\documentclass[pra,aps,twocolumn,superscriptaddress,nofootinbib,longbibliography,floatfix]{revtex4-2}
\usepackage{amsmath,amssymb,bm,graphicx,xcolor,booktabs,dsfont}
\usepackage{tikz}
\usetikzlibrary{quantikz}
\usetikzlibrary{positioning,arrows.meta,decorations.pathreplacing,calc,fit,shapes.geometric}
\graphicspath{{figures/}}
\usepackage[colorlinks=true,linkcolor=blue,citecolor=blue,urlcolor=blue]{hyperref}
\newcommand{\Cone}{\hyperref[fig:circuitapp]{$\mathcal C$}}
\usepackage{orcidlink}

\newcommand{\Tr}{\operatorname{Tr}}

\providecommand{\ket}{}\renewcommand{\ket}[1]{\lvert #1\rangle}
\providecommand{\bra}{}\renewcommand{\bra}[1]{\langle #1\rvert}
\newcommand{\Lame}{\mathcal{L}_{\rm AME}}

\newcommand{\tauA}{\tau_0}
\newcommand{\hX}{\hat{X}}
\newcommand{\hY}{\hat{Y}}
\newcommand{\hZ}{\hat{Z}}

\newcommand{\hH}{\hat{H}}

\newcommand{\hU}{\hat{U}}
\newcommand{\hrho}{\hat{\rho}}
\newcommand{\hone}{\hat{\mathds{1}}}
\newcommand{\hcU}{\hat{\mathcal U}}
\newcommand{\hSw}{\hat{\Pi}}

\usepackage{placeins}
\definecolor{cplus}{HTML}{7F2704}\definecolor{cminus}{HTML}{8E63C7}\definecolor{cmone}{HTML}{2D004B}\definecolor{cphalf}{HTML}{F28E2B}
\definecolor{pA}{HTML}{4C72B0}\definecolor{pB}{HTML}{DD8452}\definecolor{pC}{HTML}{55A868}\definecolor{pD}{HTML}{C44E52}
\begin{document}

\title{Analog preparation of absolutely maximally entangled states}

\author{Marcin P\l{}odzie\'n\,\orcidlink{0000-0002-0835-1644}}
\affiliation{Instytut Fizyki Teoretycznej, Wydzia\l{} Fizyki, Astronomii i Informatyki Stosowanej,
Uniwersytet Jagiello\'nski, \L{}ojasiewicza 11, PL-30-348 Krak\'ow, Poland}
\author{Grzegorz Rajchel-Mieldzio\'c\,\orcidlink{0000-0003-2652-9482}}
\affiliation{BEIT, ul. Wadowicka 8a, 30-415 Krak\'ow, Poland}
\author{Karol \.Zyczkowski\,\orcidlink{0000-0002-0653-3639}}
\affiliation{Instytut Fizyki Teoretycznej, Wydzia\l{} Fizyki, Astronomii i Informatyki Stosowanej,
Uniwersytet Jagiello\'nski, \L{}ojasiewicza 11, PL-30-348 Krak\'ow, Poland}
\affiliation{Center for Theoretical Physics, Polish Academy of Sciences, al. Lotnik\'ow 32/46, 02-668 Warszawa, Poland}

\date{November 27, 2026}

\begin{abstract}
In an absolutely maximally entangled (AME) state, every group of at most half of the parties is maximally entangled with the rest, which makes these states a resource for quantum error correction and multiparty teleportation. They are usually prepared by sequences of quantum gates. Here we show that a single evolution under a time-independent Hamiltonian of pairwise Ising couplings, applied to a product state of qubits that encode the parties, drives all bipartitions of the parties to maximal entanglement at once. Such evolutions exist for four, five and six parties of dimension four and for four parties of dimension eight, including states that are not locally equivalent to any stabilizer state, and an analogous evolution prepares AME states of qutrits. A sequence of local drives and coupling evolutions prepares the real AME state built from a quantum Latin square of order six. When the hardware drives all couplings at once, the state is ready in the time of a single two-qubit gate, and in simulations with decoherence it loses less fidelity than the known gate circuit for the same state.
\end{abstract}

\maketitle

\section{Introduction}
Absolutely maximally entangled (AME) states are multipartite states in which every choice of half of the parties is maximally entangled with the other half~\cite{Helwig2012,Goyeneche2015}. For $n$ parties of dimension $d$ such states are denoted AME$(n,d)$. They enable quantum secret sharing and parallel teleportation~\cite{Helwig2012}, and they are equivalent to perfect tensors and to optimal quantum error-correcting codes~\cite{Scott2004,Pastawski2015}. Qubit AME states exist only for two, three, five and six parties~\cite{Higuchi2000,Scott2004,Huber2017}. For four parties of dimension four there are infinitely many AME states that are not related by local unitaries, including states that are not equivalent to stabilizer states~\cite{Rather2023}. Infinitely many inequivalent AME states are also known for five parties~\cite{Ramadas2024}.

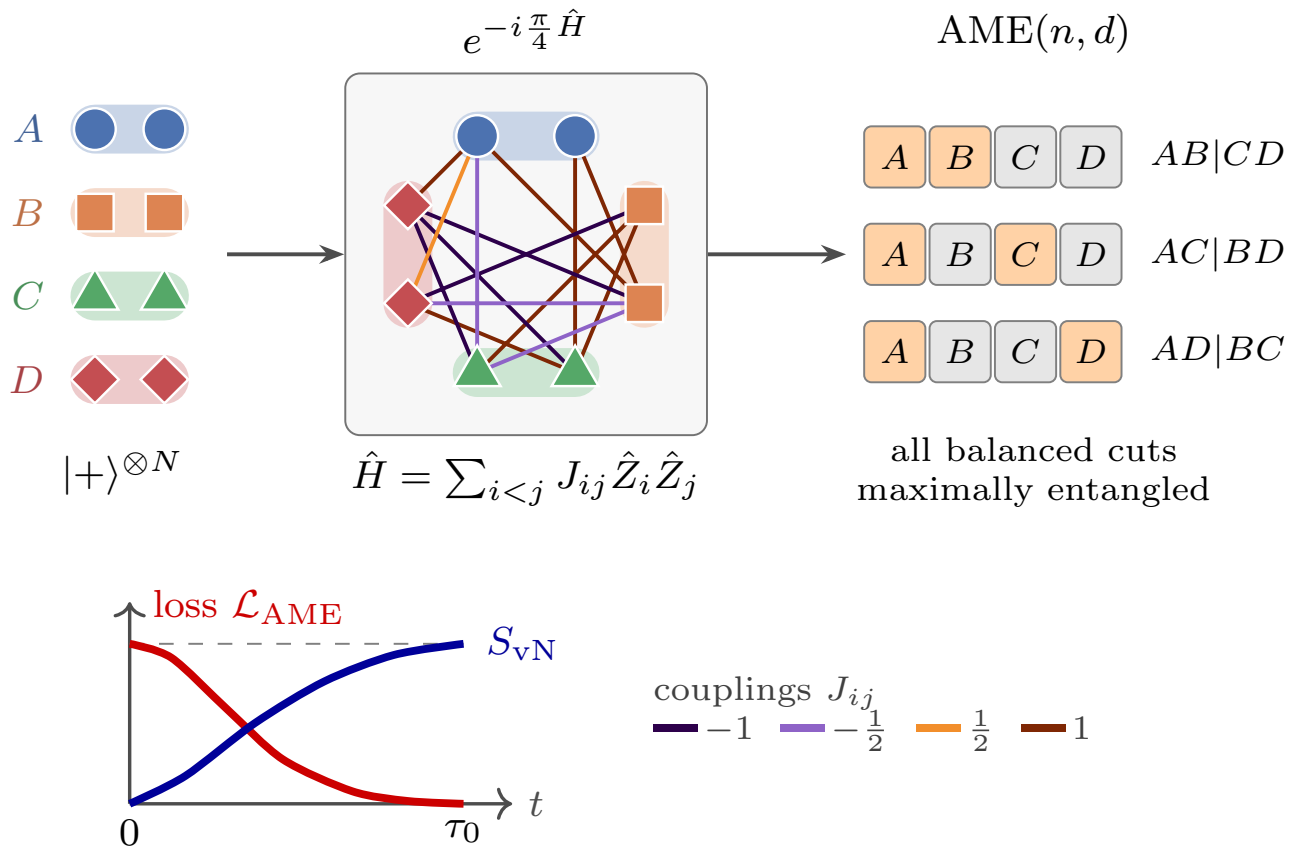
\begin{figure}[t]
\centering

\resizebox{\columnwidth}{!}{%
\begin{tikzpicture}[
  font=\footnotesize,
  qb/.style={circle,draw=white,line width=0.5pt,minimum size=3mm,inner sep=0pt},
  qA/.style={qb}, qB/.style={rectangle,draw=white,line width=0.5pt,minimum size=2.7mm,inner sep=0pt},
  qC/.style={regular polygon,regular polygon sides=3,draw=white,line width=0.5pt,minimum size=3.9mm,inner sep=0pt},
  qD/.style={diamond,draw=white,line width=0.5pt,minimum size=3.5mm,inner sep=0pt},
  arr/.style={-{Stealth[length=1.8mm]},thick,black!70}]
\foreach \nm/\y/\col in {A/0.9/pA,B/0.3/pB,C/-0.3/pC,D/-0.9/pD}{
  \draw[\col!30,line width=10pt,line cap=round] (0,\y) -- (0.5,\y);
  \node[q\nm,fill=\col] at (0,\y) {}; \node[q\nm,fill=\col] at (0.5,\y) {};
  \node[text=\col!85!black,anchor=east] at (-0.25,\y) {$\nm$};
}
\node[anchor=north] at (0.2,-1.3) {$\ket{+}^{\otimes N}$};
\begin{scope}[shift={(3.1,0)}]
  \node[draw=black!55,rounded corners=3pt,fill=black!3,minimum width=2.6cm,minimum height=2.6cm] (box) at (0,0) {};
  \foreach \k in {0,...,7}{\coordinate (v\k) at ({112.5-45*\k}:0.92);}
  \foreach \a/\b/\col in {0/1/pA,2/3/pB,4/5/pC,6/7/pD}{\draw[\col!30,line width=10pt,line cap=round] (v\a) -- (v\b);}
  \draw[cplus,thick] (v0)--(v3) (v0)--(v7) (v1)--(v3) (v1)--(v4) (v2)--(v4) (v2)--(v5) (v4)--(v6);
  \draw[cmone,thick] (v2)--(v6) (v3)--(v7) (v4)--(v7) (v5)--(v7);
  \draw[cphalf,thick] (v0)--(v6);
  \draw[cminus,thick] (v0)--(v5) (v3)--(v5) (v3)--(v6);
  \foreach \k/\col/\sh in {0/pA/A,1/pA/A,2/pB/B,3/pB/B,4/pC/C,5/pC/C,6/pD/D,7/pD/D}{\node[q\sh,fill=\col] at (v\k) {};}
  \node[anchor=south] at (0,1.33) {$e^{-i\frac{\pi}{4}\hH}$};
  \node[anchor=north] at (0,-1.3) {$\hH=\sum_{i<j}J_{ij}\hZ_i\hZ_j$};
\end{scope}
\draw[arr] (0.95,0) -- (box.west);
\begin{scope}[shift={(5.75,0)}]
  \foreach \row/\lab/\ca/\cb/\cc/\cd in {0.7/{$AB|CD$}/1/1/0/0,0/{$AC|BD$}/1/0/1/0,-0.7/{$AD|BC$}/1/0/0/1}{
    \foreach \i/\nm/\s in {0/A/\ca,1/B/\cb,2/C/\cc,3/D/\cd}{
      \ifnum\s=1 \node[draw=black!50,rounded corners=1.5pt,fill=orange!35,minimum width=4.4mm,minimum height=4.4mm,inner sep=0pt,font=\scriptsize] at (\i*0.47,\row) {$\nm$};
      \else \node[draw=black!50,rounded corners=1.5pt,fill=black!10,minimum width=4.4mm,minimum height=4.4mm,inner sep=0pt,font=\scriptsize] at (\i*0.47,\row) {$\nm$};\fi}
    \node[font=\scriptsize,anchor=west] at (1.72,\row) {\lab};
  }
  \node[anchor=south] at (1.0,1.33) {AME$(n,d)$};
  \node[anchor=north,font=\scriptsize,align=center] at (1.0,-1.2) {all balanced cuts\\ maximally entangled};
\end{scope}
\draw[arr] (box.east) -- (5.4,0);
\begin{scope}[shift={(0.25,-3.95)}]
  \draw[black!70,line width=0.7pt,->] (0,0) -- (2.75,0) node[right,font=\footnotesize] {$t$};
  \draw[black!70,line width=0.7pt,->] (0,0) -- (0,1.45);
  \draw[black!50,dashed] (0,1.15) -- (2.4,1.15);
  \draw[line width=1.6pt,red!80!black] plot[smooth] coordinates {(0,1.15) (0.3,1.05) (0.7,0.68) (1.1,0.32) (1.6,0.09) (2.0,0.02) (2.4,0)};
  \draw[line width=1.6pt,blue!60!black] plot[smooth] coordinates {(0,0) (0.4,0.2) (0.9,0.58) (1.4,0.88) (1.9,1.07) (2.4,1.15)};
  \node[font=\footnotesize,anchor=south west,text=red!80!black] at (0.05,1.18) {loss $\Lame$};
  \node[font=\footnotesize,anchor=west,text=blue!60!black] at (2.45,1.15) {$S_{\rm vN}$};
  \node[font=\footnotesize,anchor=north] at (2.4,0) {$\tauA$};
  \node[font=\footnotesize,anchor=north] at (0,0) {0};
\end{scope}
\node[font=\scriptsize,align=left,anchor=west,text=black!75] at (3.9,-3.3) {couplings $J_{ij}$\\ \textcolor{cmone}{\rule[1.5pt]{9pt}{1.6pt}}\,$-1$\ \ \textcolor{cminus}{\rule[1.5pt]{9pt}{1.6pt}}\,$-\tfrac12$\ \ \textcolor{cphalf}{\rule[1.5pt]{9pt}{1.6pt}}\,$\tfrac12$\ \ \textcolor{cplus}{\rule[1.5pt]{9pt}{1.6pt}}\,$1$};
\end{tikzpicture}}
\caption{A single analog evolution under one Ising Hamiltonian prepares an AME state, sketched here for four ququarts $A,\dots,D$, each encoded in two qubits. All qubits start polarised along $x$ and evolve for the time $\tauA$ under the Hamiltonian $\hH$, whose couplings connect qubits of different parties. Coloured bands join the qubits of one party, which share a node shape, and the couplings drawn are those of the AME$(4,4)$ state of Sec.~\ref{sec:ns44}. In the final AME state every bipartition into two pairs of parties, shown by the two shadings, is maximally entangled. During the evolution the loss function decreases (red) and the entanglement entropy of every such bipartition grows (blue) to its maximal value (dashed line).}
\label{fig:concept}
\end{figure}

AME states have mostly been prepared by circuits of one- and two-qudit gates~\cite{CerveraLierta2019,Casas2026,Chen2016,Rather2024,Majidy2025}, often starting from AME graph states~\cite{Helwig2013,Raissi2022}, which are stabilizer states built from controlled-phase gates~\cite{Briegel2001,Hein2004}. Qubit AME states have been prepared experimentally, the five-qubit state on trapped ions~\cite{Stricker2022} and a six-qubit perfect tensor in nuclear magnetic resonance~\cite{Li2019}. Reference~\cite{Casas2026} constructs AME states of four parties that are not equivalent to any stabilizer state under local unitaries, and its circuit for four ququarts (four-level systems) uses three-qubit controlled-controlled-phase gates.

In an interacting spin system, a time-independent Hamiltonian with pairwise couplings entangles an initial product state without any sequence of gates. In trapped ions, Ising interactions between the spins can be tuned by shaping the driving laser fields~\cite{PorrasCirac2004,Kim2009,Monroe2021}. Programmable couplings between all pairs of up to four ions, with signs chosen independently for each pair, have been applied together in a single Ising evolution~\cite{Lu2025}. Ising evolutions with unit couplings prepare graph states~\cite{Briegel2001,Dur2005,Hein2006}. Reference~\cite{Cieslinski2023} shows that a single evolution under two-body interactions sharing the symmetry of the five-qubit AME state generates this state, in a time set by the available energy range. Graph states are stabilizer states, and the symmetric interactions of Ref.~\cite{Cieslinski2023} are tailored to one target state. Choosing every coupling independently, as we do here, gives exact AME states for several numbers of parties and local dimensions, including states that are not locally equivalent to any stabilizer state.

In this work we prepare AME states, including states outside the stabilizer family, by a single evolution under a time-independent Hamiltonian of pairwise couplings, and compare it with a gate circuit for a state in the same local-unitary class. We show that pairwise Ising couplings, each chosen independently, between the qubits that encode the parties drive every balanced bipartition, into two groups of parties whose sizes are equal or differ by one, to maximal entanglement at the same time (Fig.~\ref{fig:concept}). All terms of the Hamiltonian commute, so the evolution amounts to two-qubit phase gates that act simultaneously. On hardware that drives all couplings at full strength at once, the state is prepared in the duration of a single controlled-phase gate. We find the couplings numerically, by minimising a loss function that vanishes exactly on AME states. We present them for the states AME$(4,4)$, AME$(4,8)$, AME$(5,4)$ and AME$(6,4)$ of qudits encoded in qubits and for the states AME$(4,3)$, AME$(5,3)$ and AME$(6,3)$ of qutrits. A criterion that we derive from the local-unitary invariant of Ref.~\cite{Rather2023} certifies that several of the AME states found here for four ququarts and for four parties of dimension eight, and all those found for five ququarts, lie outside the stabilizer family. The same holds for the members of the continuous families of AME states constructed below for four and six ququarts and for four parties of dimension eight, whenever the added coupling takes a non-integer value. For the real AME$(4,6)$ state of Ref.~\cite{HontarenkoLatin2026}, we prepare two maximally entangled pairs and apply a sequence of local drives and coupling evolutions to one party from each pair. We quantify the resource of each single-evolution preparation by the energy uncertainty of the initial state under the physical Hamiltonian, which is proportional to the normalised Hilbert--Schmidt norm of the dimensionless Hamiltonian. On a common hardware model the evolution is shorter than the known gate circuit for the non-stabilizer AME$(4,4)$ state~\cite{Casas2026} and loses less fidelity under decoherence.

\section{Setting}\label{sec:setting}
\subsection{AME states of qudits encoded in qubits}\label{sec:prelim}

We consider $n$ parties of dimension $d$ in a pure state $\ket\psi$. For a group $S$ of parties with complement $\bar S$, the reduced state is
\begin{equation}
\hrho_S=\Tr_{\bar S}\ket\psi\bra\psi .
\label{eq:reduced}
\end{equation}
The state is AME$(n,d)$ if the reduced state of every group of $\lfloor n/2\rfloor$ parties is maximally mixed,
\begin{equation}
\hrho_S=\frac{\hone}{d^{|S|}}\qquad\text{for all } S \text{ with } |S|=\Big\lfloor\frac n2\Big\rfloor .
\label{eq:kuniform}
\end{equation}
The entanglement of a bipartition $S|\bar S$ is measured by the von Neumann entropy of the reduced state,
\begin{equation}
S_{\rm vN}=-\Tr\big(\hrho_S\log_2\hrho_S\big) ,
\label{eq:SvN}
\end{equation}
the entanglement entropy. In an AME state every bipartition into $\lfloor n/2\rfloor$ parties and the rest, which we call a balanced bipartition or, for short, a balanced cut, carries the largest possible value, $S_{\rm vN}=\lfloor n/2\rfloor\log_2 d$ (Fig.~\ref{fig:concept}). Two states that differ only by unitaries acting on single parties,
\begin{equation}
\ket{\psi'}=\big(\hU_1\otimes\hU_2\otimes\cdots\otimes\hU_n\big)\ket\psi ,
\label{eq:lu}
\end{equation}
have the same entanglement across every cut and are called locally equivalent. The states locally equivalent to a given state form its local-unitary (LU) class.

A party of dimension $d=2^r$ is represented by $r$ qubits, so that the system consists of $N=nr$ qubits, and its basis states are the computational states of these qubits~\cite{CerveraLierta2019,Casas2026},
\begin{equation}
\ket a=\ket{x_1x_2\cdots x_r} ,\qquad a=\sum_{p=1}^{r}x_p\,2^{r-p}\in\{0,\dots,d-1\} ,
\label{eq:encoding}
\end{equation}
with $x_p\in\{0,1\}$. For four ququarts the eight qubits $q_1,\dots,q_8$ form the parties
\begin{equation}
\begin{aligned}
A&=(q_1,q_2), & B&=(q_3,q_4),\\ C&=(q_5,q_6), & D&=(q_7,q_8),
\end{aligned}
\label{eq:parties}
\end{equation}
and for four parties of dimension eight the twelve qubits are grouped in consecutive triples. The AME condition of Eq.~\eqref{eq:kuniform} refers to groups of parties, so cuts that separate the two qubits of one party are not tested. Every unitary on the qubits of one party, including an entangling one, is a single-party unitary in Eq.~\eqref{eq:lu}.

A stabilizer state of $N$ qubits is the simultaneous eigenstate, $\hat g_j\ket\psi=\ket\psi$ for $j=1,\dots,N$, of $N$ independent commuting products $\hat g_j$ of Pauli operators. It is the output of a circuit of Clifford gates, which map Pauli operators onto Pauli operators. Following Ref.~\cite{Casas2026}, we call an encoded AME state a non-stabilizer state when no single-party unitaries map it onto a stabilizer state of the encoding qubits.

\subsection{The analog evolution}\label{sec:quench}

The $N$ qubits start in the product state
\begin{equation}
\ket{\psi_0}=\ket{+}^{\otimes N} ,
\label{eq:psi0}
\end{equation}
with $\ket+=(\ket0+\ket1)/\sqrt2$, so that every qubit is polarised along $x$, and evolve under a time-independent Hamiltonian of two-body Ising couplings (Fig.~\ref{fig:concept}),
\begin{equation}
\ket{\psi_J(t)}=e^{-i\frac{\pi}{4}\frac{t}{\tauA}\hH}\ket{\psi_0},\qquad \hH=\sum_{i<j}J_{ij}\,\hZ_i\hZ_j ,
\label{eq:quench}
\end{equation}
for $0\le t\le\tauA$, where $\hZ_i$ is the Pauli $z$ operator of qubit $i$, the index $J$ stands for the symmetric matrix of the dimensionless couplings $J_{ij}=J_{ji}$ and $\ket{\psi_J}=\ket{\psi_J(\tauA)}$ denotes the final state. The physical Hamiltonian is $J_0\hH$, where the energy $J_0$ is the coupling strength that corresponds to $J_{ij}=1$, and we set $\hbar=1$, so that $\tauA=\pi/(4J_0)$.

All terms of $\hH$ commute, so the evolution equals the simultaneous action of one two-qubit phase gate for every coupling. In the computational basis $\ket x$, $x\in\{0,1\}^N$, with $z_i=1-2x_i$, the final state has the amplitudes
\begin{equation}
\psi_x=2^{-N/2}\exp\Big(-i\frac{\pi}{4}\sum_{i<j}J_{ij}\,z_iz_j\Big) .
\label{eq:amplitudes}
\end{equation}
All amplitudes have the same modulus, and the couplings enter only through the phases. States of this form are known as weighted graph states~\cite{Dur2005,Hein2006,Hartmann2007}, with the qubits as vertices and the couplings as weighted edges. A coupling $J_{ij}=1$ acts as a controlled-$Z$ gate up to single-qubit operators, and a coupling $J_{ij}=2$ as a product of single-qubit operators,
\begin{equation}
\begin{aligned}
e^{-i\frac{\pi}{4}\hZ_i\hZ_j}&=e^{-i\frac{\pi}{4}}\,\hat\Phi_i\hat\Phi_j\,\widehat{CZ}_{ij} ,\\
e^{-i\frac{\pi}{2}\hZ_i\hZ_j}&=-i\,\hZ_i\hZ_j ,
\end{aligned}
\label{eq:czid}
\end{equation}
with the phase gate $\hat\Phi={\rm diag}(1,i)$ and $\widehat{CZ}={\rm diag}(1,1,1,-1)$. Couplings $J_{ij}=1$ on the edges $E$ of a graph and zero elsewhere therefore produce, up to single-qubit operators, the graph state of that graph~\cite{Briegel2001,Hein2004},
\begin{equation}
\ket G=\prod_{(i,j)\in E}\widehat{CZ}_{ij}\,\ket+^{\otimes N} .
\label{eq:graphstate}
\end{equation}
Since couplings that differ by 2 give the same state up to single-qubit operators, we take them in the interval $[-1,1]$, in which $J_{ij}=1$ and $J_{ij}=-1$ are equivalent. We split every coupling into its nearest integer $\lfloor J_{ij}\rceil$, with half-integers rounded towards zero, and the remainder $\tilde J_{ij}=J_{ij}-\lfloor J_{ij}\rceil$, with $|\tilde J_{ij}|\le1/2$. Up to single-qubit operators, every final state is then a graph state followed by fractional two-qubit phases,
\begin{equation}
\ket{\psi_J}=\prod_{i<j}e^{-i\frac{\pi}{4}\tilde J_{ij}\hZ_i\hZ_j}\,\ket{G_{\lfloor J\rceil}} ,
\label{eq:graphcorr}
\end{equation}
where the graph of $\ket{G_{\lfloor J\rceil}}$ has an edge on every pair with odd $\lfloor J_{ij}\rceil$, i.e.\ on the pairs with $|J_{ij}|>1/2$ for couplings in $[-1,1]$. The remainders $\tilde J_{ij}$ add the two-qubit phases that Clifford gates cannot produce.

\subsection{Loss function}\label{sec:loss}

To find couplings for which $\ket{\psi_J}$ is AME, we minimise a loss that measures how far the balanced cuts are from maximal entanglement. For a group $S$ of at most half of the parties we arrange the amplitudes in a matrix $\hat M_S$ with the elements $(\hat M_S)_{s,\bar s}=\psi_x$, where $s$ and $\bar s$ are the parts of the configuration $x$ on the parties in $S$ and on the remaining parties, so that
\begin{equation}
\hrho_S=\hat M_S\hat M_S^\dagger .
\label{eq:psimatrix}
\end{equation}
With the eigenvalues $\lambda_k$ of $\hrho_S$, the purity $\Tr\hrho_S^2=\sum_k\lambda_k^2$ takes its minimum $d^{-|S|}$ only when $\hrho_S$ is maximally mixed and its maximum 1 only when the cut carries no entanglement. We rescale it to the interval $[0,1]$ by the excess purity
\begin{equation}
e_S=\frac{\Tr\hrho_S^2-d^{-|S|}}{1-d^{-|S|}} ,
\label{eq:eS}
\end{equation}
whose numerator is the squared distance of $\hrho_S$ from the maximally mixed state in the Hilbert--Schmidt norm $\|\hat A\|_{\rm HS}^2=\Tr\hat A^\dagger\hat A$,
\begin{equation}
\Tr\hrho_S^2-d^{-|S|}=\sum_k\big(\lambda_k-d^{-|S|}\big)^2=\big\|\hrho_S-\hone/d^{|S|}\big\|_{\rm HS}^2 ,
\label{eq:frob}
\end{equation}
because $\sum_k\lambda_k=1$. Averaging over the set $\mathcal S$ of balanced cuts, each represented by a group $S$ of $\lfloor n/2\rfloor$ parties and for even $n$ by one of its two halves, gives the loss
\begin{equation}
\Lame=\frac{1}{|\mathcal S|}\sum_{S\in\mathcal S}e_S .
\label{eq:L}
\end{equation}
For four parties $\mathcal S$ contains the three cuts $AB|CD$, $AC|BD$ and $AD|BC$. The loss lies between 0 and 1, does not change under single-party unitaries and vanishes if and only if the pure state $\ket{\psi_J}$ is AME. 
Averaged purities of this kind have been minimised to find highly entangled multiqubit states~\cite{Facchi2008}.

For the analog evolution the purity of a cut is an explicit function of the couplings (Appendix~\ref{app:closedform}),
\begin{equation}
\Tr\hrho_S^2=4^{-N_S}\sum_{\mathbf a\in\{0,\pm1\}^{N_S}}2^{n_0(\mathbf a)}\prod_{j\notin S}\cos^2\!\Big(\frac{\pi}{2}\sum_{i\in S}J_{ij}a_i\Big) ,
\label{eq:purityJ}
\end{equation}
where $N_S=|S|r$ is the number of qubits in $S$. The sum runs over all assignments of a value $a_i\in\{0,+1,-1\}$ to each qubit $i\in S$, and $n_0(\mathbf a)$ is the number of qubits with $a_i=0$. These variables come from writing the purity as an overlap of two copies of the state. A qubit has $a_i=0$ when its value $z_i$ is the same in both copies and $a_i=\pm1$ when the values differ.

Only the couplings between a qubit in $S$ and a qubit outside $S$ enter Eq.~\eqref{eq:purityJ}, because couplings within either side act as unitaries on that side. Without couplings across the cut the purity is 1, and a single coupling $J_{ij}=1$ across the cut lowers it to $1/2$, the value for one qubit of a maximally entangled pair.

Equation~\eqref{eq:purityJ} makes the loss a smooth, periodic function of the couplings, and its gradient is given in Appendix~\ref{app:closedform}. We minimise $\Lame$ from random initial couplings (Appendix~\ref{app:numerics}).

\subsection{Identifying the local-unitary class}\label{sec:identify}

To distinguish AME states that are not related by unitaries within the parties, we use local-unitary invariants~\cite{Rather2023,Casas2026}. For four parties $A$, $B$, $C$, $D$ we arrange the amplitudes $\psi_{\alpha\beta\gamma\delta}$ of the state in the $d^2\times d^2$ matrix
\begin{equation}
\hcU_{(\alpha\beta),(\gamma\delta)}=d\,\psi_{\alpha\beta\gamma\delta} .
\label{eq:Udef}
\end{equation}
The reduced state of $A$ and $B$ is $\hcU\hcU^\dagger/d^2$, so for an AME state $\hcU$ is unitary and acts as an evolution from the parties $C$, $D$ to the parties $A$, $B$.

On two copies of the system, the operator of Ref.~\cite{Rather2023}, which we call the echo operator, evolves both copies with $\hcU$, exchanges party $B$ between the copies, evolves both back with $\hcU^\dagger$ and exchanges party $D$,
\begin{equation}
\hat I(\hcU)=\hSw_{24}\,(\hcU^\dagger\otimes\hcU^\dagger)\,\hSw_{24}\,(\hcU\otimes\hcU) ,
\label{eq:inv}
\end{equation}
where $\hSw_{24}$ exchanges the second party between the copies, $B$ on the output side of $\hcU$ and $D$ on its input side. If $\hcU$ were a product of a map from $C$ to $A$ and a map from $D$ to $B$, the two exchanges would cancel and $\hat I(\hcU)$ would be the identity, so $\hat I(\hcU)$ measures how $\hcU$ correlates the parties. A change of basis within the parties rotates both copies in the same way, and exchanging a party between two identically rotated copies commutes with the rotation. The operator $\hat I(\hcU)$ therefore changes only by a unitary similarity transformation, and its moments
\begin{equation}
m_k=\Tr\big[\hat I(\hcU)^k\big]
\label{eq:moments}
\end{equation}
are the same for all locally equivalent states. The first moment, $m_1=d^2$, is the same for all AME states, so we compare AME states by $m_2$ and $m_4$, which obey $|m_k|\le d^4$ with equality for $\hat I(\hcU)^k=\hone$, as for graph states at $k=2$.

Each factor $\hcU$ in $\hat I(\hcU)^k$ carries one copy of the state and each factor $\hcU^\dagger$ one copy of its complex conjugate, so $m_k$ is an expectation value on $2k$ copies of the state. As for the R\'enyi moments $\Tr\hrho_S^q$, which are expectation values of a cyclic permutation of $q$ copies~\cite{Calabrese2004}, the operator in this expectation value permutes the copies, here with a different permutation for each party (Appendix~\ref{app:invariant}). For the second moment,
\begin{equation}
m_2=d^8\,\bra\psi^{\otimes4}\,\hone_A\otimes\hat\Pi^{B}_{(12)(34)}\otimes\hat\Pi^{C}_{(13)(24)}\otimes\hat\Pi^{D}_{(14)(23)}\,\ket\psi^{\otimes4} ,
\label{eq:m2copies}
\end{equation}
where $\hat\Pi^X_\sigma$ permutes the four copies of party $X$ according to $\sigma$. Together with the identity on $A$, these permutations form the Klein four-group, and $m_2/d^8$ is the Klein contraction with which Ref.~\cite{HontarenkoLatin2026} distinguishes AME states of four parties. By Eq.~\eqref{eq:m2copies}, $m_2$ can be measured on four copies of the prepared state~\cite{Ekert2002} or estimated from randomised local measurements~\cite{Brydges2019}.

Since the moments do not change under unitaries within the parties, for an AME state that is locally equivalent to a stabilizer state of the encoding qubits they equal the moments of that stabilizer state. For a stabilizer state $\hcU$ is a Clifford unitary, which maps products of Pauli operators onto products of Pauli operators. The exchange of a block of $b$ qubits between the two copies is $2^{-b}\sum_{\hat P}\hat P\otimes\hat P$, summed over the products $\hat P$ of Pauli operators on the block, so both factors of $\hat I(\hcU)$, the exchange $\hSw_{24}$ and its conjugate $(\hcU^\dagger\otimes\hcU^\dagger)\hSw_{24}(\hcU\otimes\hcU)$, are sums of operators $\hat P\otimes\hat P$. Two such operators $\hat P\otimes\hat P$ and $\hat P'\otimes\hat P'$ always commute, because the signs from exchanging $\hat P$ and $\hat P'$ on the two copies cancel. The two factors therefore commute and square to one, so $\hat I(\hcU)^2=\hone$ and
\begin{equation}
m_2=d^4
\label{eq:stabcrit}
\end{equation}
for every AME state of four parties that is locally equivalent to a stabilizer state. A second moment below $d^4$ excludes this equivalence. For five and six parties we merge the parties into four blocks, $(A)(B)|(C)(DE)$ and $(AB)(C)|(DE)(F)$, which defines grouped moments with the same invariance. For six parties both sides of the grouped $\hcU$ have dimension 64, and Eq.~\eqref{eq:stabcrit} becomes $m_2=4096$. For five parties, where $\hcU$ is rectangular, the overlaps of stabilizer states~\cite{Aaronson2004,Garcia2012} give
\begin{equation}
\big|m_k\big|^2\in\{0\}\cup\{2^j,\ j\in\mathbb{Z}\} .
\label{eq:stabcritgroup}
\end{equation}

\section{Results}\label{sec:results}
We apply the protocol to AME$(4,4)$ states (Sec.~\ref{sec:ns44}), AME$(4,8)$ states (Sec.~\ref{sec:ame48}), and AME$(5,4)$, AME$(6,4)$ and the qutrit states AME$(4,3)$, AME$(5,3)$ and AME$(6,3)$ (Sec.~\ref{sec:new48}). We then compare the norms of these Hamiltonians (Sec.~\ref{sec:norm}), prepare the AME$(4,6)$ state of a quantum Latin square by a sequence of evolutions (Sec.~\ref{sec:ame46}), and study coupling errors and decoherence (Secs.~\ref{sec:robust} and~\ref{sec:decoherence}).

\begin{figure}[t]
\includegraphics[width=\columnwidth]{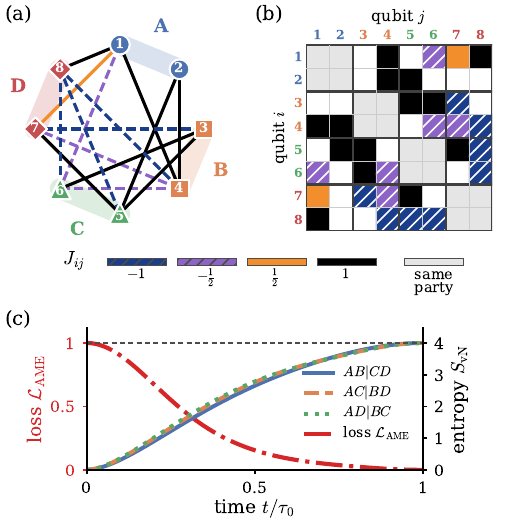}
\caption{Fifteen two-qubit couplings, switched on together, prepare a non-stabilizer AME$(4,4)$ state in a single evolution. (a) Graph of the couplings $J_{ij}$, with a colour and a node shape for each of the parties $A=(q_1,q_2)$, $B=(q_3,q_4)$, $C=(q_5,q_6)$ and $D=(q_7,q_8)$. Negative couplings are dashed in the graph and hatched in the matrix. (b) The symmetric coupling matrix on the same colour scale. Grey blocks are pairs of qubits within a party and white cells are vanishing couplings. (c) Entanglement entropy $S_{\rm vN}$ of the three balanced cuts (right axis) and the loss $\Lame$ (red, left axis) during the evolution. The entropy of all three cuts reaches its maximum $S_{\rm vN}=4$ at the time $\tauA$, where the loss vanishes and the state is AME.}
\label{fig:graphs}
\end{figure}

\subsection{AME$(4,4)$ states}\label{sec:ns44}\label{sec:bonds}

For AME$(4,4)$ states, with $n=4$ parties of dimension $d=4$, each party is encoded in $r=2$ qubits, so the analog evolution of Eq.~\eqref{eq:quench} acts on eight qubits. We minimise the loss of Eq.~\eqref{eq:L} over all couplings between qubits of different parties, starting from random values. Rounding the couplings of one minimum to multiples of $\tfrac12$ and removing a redundant coupling gives the fifteen couplings of Fig.~\ref{fig:graphs} (Appendix~\ref{app:numerics}). Eleven of them equal $\pm1$ and act as controlled-$Z$ gates, and the remaining four, of strength $\pm\tfrac12$, couple $q_1$ and $q_4$ to $q_6$ and $q_7$.

During the evolution the entanglement entropy of the three balanced cuts grows from zero and reaches its maximum, $S_{\rm vN}=2\log_24=4$, on all three cuts at $t=\tauA$, where the loss vanishes [Fig.~\ref{fig:graphs}(c)]. The eleven couplings $\pm1$ alone give the graph state of Eq.~\eqref{eq:graphcorr} whose cut $AB|CD$ has $S_{\rm vN}=3$, and the four couplings of half strength raise it to 4.

The second and fourth moments of the state are
\begin{equation}
m_2=64,\qquad m_4=160 .
\label{eq:inv64}
\end{equation}
Since $m_2$ lies below the value 256 of Eq.~\eqref{eq:stabcrit}, the state is not locally equivalent to any stabilizer state of the eight qubits, whereas for the AME$(4,4)$ graph states both moments equal 256 by the argument leading to Eq.~\eqref{eq:stabcrit}.

The state found is the non-stabilizer AME state of four ququarts constructed in Ref.~\cite{Casas2026} by a gate circuit,
\begin{equation}
\begin{aligned}
\ket{\mathrm{AME}(4,4)}&=\big(\hcU_{\rm AME}\otimes\hone_{CD}\big)\ket\Omega_{AC}\ket\Omega_{BD} ,\\
\ket\Omega&=\frac12\sum_{\alpha=0}^{3}\ket\alpha\ket\alpha ,
\end{aligned}
\label{eq:casasstate}
\end{equation}
in which two maximally entangled pairs of ququarts, $A$ with $C$ and $B$ with $D$, are followed by the two-ququart gate
\begin{equation}
\hcU_{\rm AME}=\hat C\,\mathsf H^{\otimes4}\,\hat D_\Lambda\,\mathsf H^{\otimes4}\,\hat C
\label{eq:casasU}
\end{equation}
on $A$ and $B$. Here $\mathsf H^{\otimes4}$ applies the Hadamard gate $\mathsf H$ to each of the four qubits of $A$ and $B$. The gate $\hat C$ applies controlled-$Z$ gates between corresponding qubits of $A$ and $B$, with $\bra{kl}\hat C\ket{kl}=(-1)^{k_1l_1+k_2l_2}$ for the qubit values $k=(k_1,k_2)$ of $A$ and $l=(l_1,l_2)$ of $B$. The diagonal phase gate $\hat D_\Lambda$, which is not a Clifford gate, has the entries $\Lambda=(1,1,-i,i,i,i,1,-1,i,-1,1,i,i,-1,-1,-i)$ in the computational basis of $A$ and $B$. The amplitudes of $\ket{\mathrm{AME}(4,4)}$ are $\hcU_{\rm AME}/4$, and its moments are those of Eq.~\eqref{eq:inv64}. The analog evolution prepares this state up to unitaries within each party,
\begin{equation}
\max_{\hU_A,\hU_B,\hU_C,\hU_D}\big|\bra{\mathrm{AME}(4,4)}\,\hU_A\otimes\hU_B\otimes\hU_C\otimes\hU_D\,\ket{\psi_J}\big|^2=1
\label{eq:overlap}
\end{equation}
(Appendix~\ref{app:numerics}), although the minimisation used no information about this state.

\begin{figure}[t]
\includegraphics[width=\columnwidth]{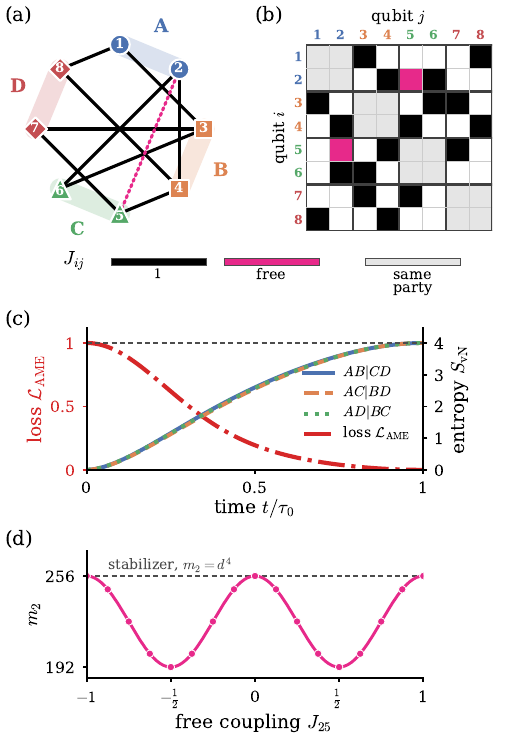}
\caption{Tuning one added coupling $J_{25}$ sweeps a continuum of AME$(4,4)$ states, which are non-stabilizer states for every non-integer $J_{25}$. (a) Coupling graph of the AME$(4,4)$ graph state (black edges) with the free coupling $J_{25}$ (pink, dotted) of arbitrary strength. (b) The coupling matrix. (c) Entanglement entropy $S_{\rm vN}$ of the three balanced cuts (right axis) and the loss $\Lame$ (red, left axis) during the analog evolution from $\ket+^{\otimes8}$ with all couplings switched on, for $J_{25}=-\tfrac12$. (d) Second moment along the family, computed from the states (dots) and from Eq.~\eqref{eq:curve} (line). It lies below the value $d^4=256$ of every stabilizer state [Eq.~\eqref{eq:stabcrit}, dashed line] for every non-integer $J_{25}$, reaches this value for the graph state at integer $J_{25}$, and is smallest at half-integer $J_{25}$.}
\label{fig:families}
\end{figure}

To obtain continuous families of AME$(4,4)$ states, we add one coupling of variable strength to an AME graph state. Sets of nine couplings $J_{ij}=1$ give such graph states, and we use the one shown by the black edges in Fig.~\ref{fig:families}(a). We place the added coupling on each uncoupled pair of qubits in different parties and compute the loss as a function of its strength (Appendix~\ref{app:further}). Twelve of these couplings, which we call free, keep the state AME for every strength. Six of them act on the graph state as a unitary within one party and leave its LU class unchanged, and the other six give new states. For one of these, the coupling $J_{25}$ between the qubits $q_2$ and $q_5$ [dotted edge in Fig.~\ref{fig:families}(a)], the states are
\begin{equation}
\ket{\psi(J_{25})}=e^{-i\frac{\pi}{4}J_{25}\hZ_2\hZ_5}\,e^{-i\frac{\pi}{4}\sum_{i<j}J_{ij}\hZ_i\hZ_j}\ket+^{\otimes8} ,
\label{eq:bond}
\end{equation}
where the sum runs over the nine couplings of the graph state. The second moment of the states $\ket{\psi(J_{25})}$,
\begin{equation}
m_2=256-64\sin^2(\pi J_{25}),
\label{eq:curve}
\end{equation}
changes continuously with $J_{25}$, so the family passes through a continuum of LU classes. The states with $J_{25}$, $-J_{25}$ and $1-J_{25}$ have the same $m_2$ and are related by unitaries within the parties. The product of single-qubit operators $\hat X_2\hZ_4\hZ_6$, a stabilizer of the graph state that anticommutes with $\hZ_2\hZ_5$, maps $\ket{\psi(J_{25})}$ onto $\ket{\psi(-J_{25})}$, and maximising the overlap as in Eq.~\eqref{eq:overlap} gives the same for $1-J_{25}$. Hence $m_2$, which follows Eq.~\eqref{eq:curve} numerically, labels the LU classes of the family one to one, and the fourth moment, $m_4=256-64\sin^2(2\pi J_{25})$, is fixed by $m_2$.

The other five free couplings that give new states produce the same family up to a relabelling of the parties. For a non-integer $J_{25}$ the second moment lies below 256, so by Eq.~\eqref{eq:stabcrit} every such member is a non-stabilizer state [Fig.~\ref{fig:families}(d)]. This family adds to the known continuous family of AME states of four ququarts~\cite{Rather2023} and is swept by tuning one Ising coupling. Since the state stays AME for every strength of the added coupling, an error in this coupling moves the state along the family without reducing the entanglement of any balanced cut.

\begin{figure}[t]
\includegraphics[width=\columnwidth]{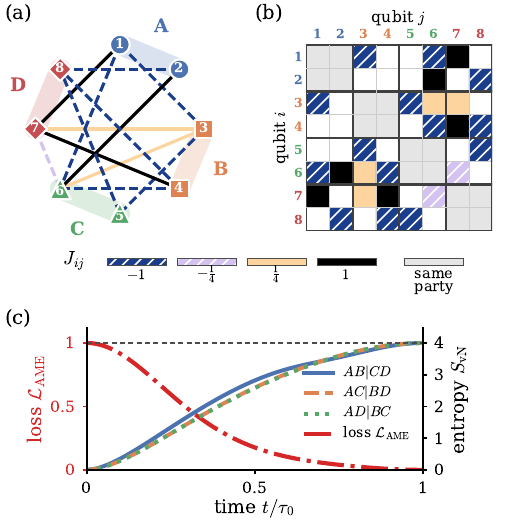}
\caption{Thirteen two-qubit couplings, switched on together, prepare the Latin-square AME$(4,4)$ state, up to unitaries within the parties, in a single evolution. (a) Graph of the couplings $J_{ij}$, drawn as in Fig.~\ref{fig:graphs}. (b) The symmetric coupling matrix on the same colour scale. (c) Entanglement entropy $S_{\rm vN}$ of the three balanced cuts (right axis) and the loss $\Lame$ (red, left axis) during the evolution. The entropy of all three cuts reaches its maximum $S_{\rm vN}=4$ at the time $\tauA$, where the loss vanishes.}
\label{fig:latin44}
\end{figure}

The evolution also reaches the AME$(4,4)$ state that Ref.~\cite{HontarenkoLatin2026} constructs from a quantum orthogonal Latin square, a $4\times4$ array of sixteen orthonormal states of two ququarts [Eqs.~\eqref{eq:latin} and~\eqref{eq:latincells} of Appendix~\ref{app:ame46}], whose cells follow a SudoQ pattern of maximal cardinality~\cite{Paczos2021}. Its moments, $m_2=208$ and $m_4=256$, differ from those of the state of Fig.~\ref{fig:graphs}, and $m_2<256$ makes it a non-stabilizer state by Eq.~\eqref{eq:stabcrit}. The family of Eq.~\eqref{eq:bond} takes the value $m_2=208$ at $J_{25}=\tfrac13$, where its fourth moment is $m_4=208$ for every pairing of the parties, so no value of $J_{25}$ gives the Latin-square state. Minimising the loss together with a penalty that steers $m_2$ and $m_4$ towards these values gives thirteen couplings, ten of strength $\pm1$ and three of strength $\pm\tfrac14$, shown in Fig.~\ref{fig:latin44} and listed in Table~\ref{tab:couplings} (Appendix~\ref{app:numerics}). Switched on together, they prepare the Latin-square state, up to unitaries within the parties, in a single evolution of $\ket+^{\otimes8}$ for the time $\tauA$ as in Eq.~\eqref{eq:quench}. The maximised overlap of Eq.~\eqref{eq:overlap} between the final state and the Latin-square state equals one. Appendix~\ref{app:classes} lists the other LU classes reached by the minimisation.

An exhaustive search over all sets of couplings $J_{ij}\in\{0,1\}$ between qubits of different parties shows that the smallest AME graph states have eight couplings, such as the ring of eight qubits listed in Table~\ref{tab:couplings}, which we call the ring graph. Adding one free coupling to such a graph state gives non-stabilizer AME states with nine couplings. For the ring, a coupling between $q_1$ and $q_3$ gives the family of Eq.~\eqref{eq:bond} itself, up to unitaries within the parties. The graph of Fig.~\ref{fig:families} uses ten couplings for the same family, the Latin-square state thirteen and the state of Fig.~\ref{fig:graphs} fifteen. With all couplings driven at once each of these states is ready after the time $\tauA$. Appendix~\ref{app:recipe} lists the steps that prepare the family of Eq.~\eqref{eq:bond}.

\subsection{AME$(4,8)$ states}\label{sec:ame48}

\begin{figure}[t]
\centering
\includegraphics[width=\columnwidth]{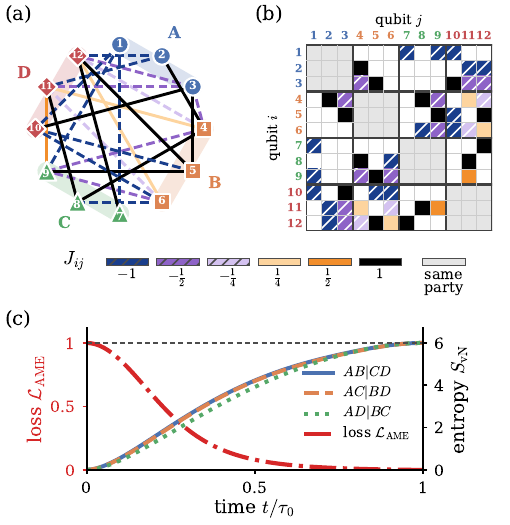}
\caption{A single evolution of twelve qubits under two-qubit couplings prepares a non-stabilizer AME$(4,8)$ state, whose second moment places it in a local-unitary class different from those of the two non-stabilizer AME$(4,8)$ states of Ref.~\cite{Casas2026}. (a) Graph of the couplings $J_{ij}$ of the state with $m_2=832$ and $m_4=1648$, with a colour and a node shape for each of the parties $A=(q_1,q_2,q_3)$, $B=(q_4,q_5,q_6)$, $C=(q_7,q_8,q_9)$ and $D=(q_{10},q_{11},q_{12})$. (b) The symmetric coupling matrix on the same colour scale. Negative couplings are drawn as in Fig.~\ref{fig:graphs}. Grey blocks are pairs of qubits within a party and white cells are vanishing couplings. (c) Entanglement entropy $S_{\rm vN}$ of the three balanced cuts (right axis) and the loss $\Lame$ (red, left axis) during the evolution. The entropy of all three cuts reaches its maximum $S_{\rm vN}=6$ at the time $\tauA$, where the loss vanishes.}
\label{fig:ising832}
\end{figure}

\begin{figure}[t]
\includegraphics[width=\columnwidth]{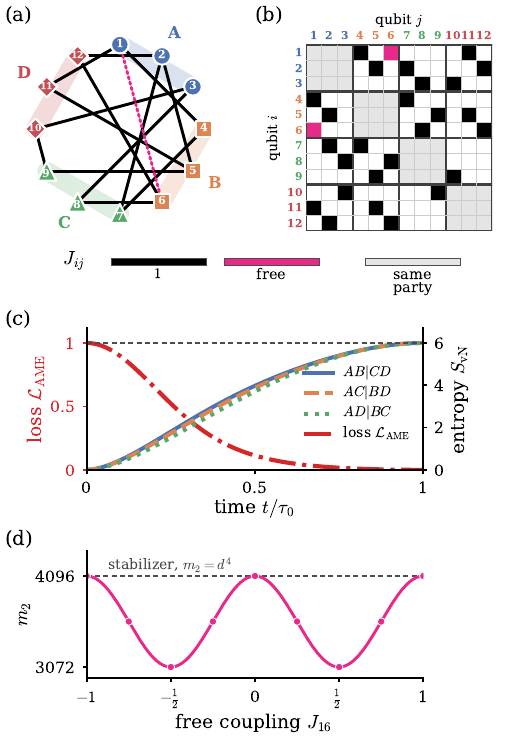}
\caption{One added coupling $J_{16}$ gives a continuous family of AME$(4,8)$ states, which are non-stabilizer states for every non-integer $J_{16}$, as for the AME$(4,4)$ family of Fig.~\ref{fig:families}. The panels are drawn as in Fig.~\ref{fig:families}. (a) Coupling graph of the AME$(4,8)$ graph state (black edges) with the free coupling $J_{16}$ (pink, dotted). (b) The coupling matrix. (c) Entanglement entropy and loss during the analog evolution from $\ket+^{\otimes12}$ with all couplings switched on, for $J_{16}=-\tfrac12$. (d) Second moment along the family, with the stabilizer value $4096$ (dashed line).}
\label{fig:family48}
\end{figure}

For AME$(4,8)$ states, with $n=4$ parties of dimension $d=8$, each party is encoded in $r=3$ qubits, so the analog evolution of Eq.~\eqref{eq:quench} acts on twelve qubits. We minimise the loss over the couplings between qubits of different parties, starting from the couplings between parties obtained by a minimisation over Hamiltonians acting on at most two parties (Appendix~\ref{app:numerics}), and round the result to multiples of $\tfrac14$. The minimisation reaches the 26 couplings of Fig.~\ref{fig:ising832}, which take the values $\pm1$, $\pm\tfrac12$ and $\pm\tfrac14$. During the evolution the entanglement entropy of the three balanced cuts grows to its maximum, $S_{\rm vN}=6$, at $t=\tauA$, where the loss vanishes [Fig.~\ref{fig:ising832}(c)].

The second and fourth moments of this state are $m_2=832$ and $m_4=1648$. The second moment lies below the value 4096 of Eq.~\eqref{eq:stabcrit}, so the state is not locally equivalent to any stabilizer state. The second moment also differs from the values 314 and 1744 of the two non-stabilizer AME$(4,8)$ states of Ref.~\cite{Casas2026}, so the analog evolution reaches a different LU class.

An added coupling gives a continuous family of AME$(4,8)$ states in the same way. A search over sets of couplings $J_{ij}\in\{0,1\}$, guided by the loss (Appendix~\ref{app:numerics}), gives an AME graph state with thirteen couplings $J_{ij}=1$ [black edges in Fig.~\ref{fig:family48}(a)], and of its thirty free couplings twenty give new states. For one of them, the coupling $J_{16}$ between the qubits $q_1$ and $q_6$ [dotted edge in Fig.~\ref{fig:family48}(a)], the states are
\begin{equation}
\ket{\psi(J_{16})}=e^{-i\frac{\pi}{4}J_{16}\hZ_1\hZ_6}\,e^{-i\frac{\pi}{4}\sum_{i<j}J_{ij}\hZ_i\hZ_j}\ket+^{\otimes12} ,
\label{eq:bond48}
\end{equation}
which remain AME for every value of $J_{16}$. Their second moment, $m_2=4096-1024\sin^2(\pi J_{16})$ (Appendix~\ref{app:further}), changes continuously with $J_{16}$ and stays above the value 832, and every member with a non-integer $J_{16}$ is a non-stabilizer state [Fig.~\ref{fig:family48}(d)].

\subsection{AME$(5,4)$, AME$(6,4)$, and qutrit AME$(4,3)$, AME$(5,3)$, AME$(6,3)$ states}\label{sec:new48}

For $n=5$ and $n=6$ parties of dimension $d=4$, each encoded in $r=2$ qubits, the analog evolution acts on ten and twelve qubits, and the same minimisation reaches AME states in both cases (Appendix~\ref{app:further}). The couplings that prepare the AME$(5,4)$ states include several of fractional strength, and the grouped second moments of these states violate Eq.~\eqref{eq:stabcritgroup}, so they are non-stabilizer AME$(5,4)$ states. The AME$(6,4)$ states are prepared by graph-state couplings, in some cases with one or two additional free couplings. For one of these states the grouped second moment lies below 4096 whenever a free coupling takes a non-integer value, so the evolution also prepares non-stabilizer AME$(6,4)$ states.

For AME$(n,3)$ states we take $n$ qutrits, one per party, with the basis states $\ket0$, $\ket1$ and $\ket2$. Every qutrit starts in the equal superposition of its three basis states,
\begin{equation}
\ket{\psi_0}=\Big(\frac{\ket0+\ket1+\ket2}{\sqrt3}\Big)^{\otimes n} ,
\label{eq:psi0qutrit}
\end{equation}
and the qutrits evolve under a Hamiltonian that couples the operators $\hat n_i$ of every pair, which number the basis states, with $\hat n_i\ket{x_i}=x_i\ket{x_i}$ and $x_i\in\{0,1,2\}$,
\begin{equation}
\ket{\psi(t)}=e^{-i\frac{t}{\tau_3}\hH_3}\ket{\psi_0},\qquad \hH_3=\frac{2\pi}{3}\sum_{i<j}K_{ij}\,\hat n_i\hat n_j .
\label{eq:nn}
\end{equation}
The physical Hamiltonian is $K_0\sum_{i<j}K_{ij}\hat n_i\hat n_j$ with the coupling energy $K_0$, so the evolution takes the time $\tau_3=2\pi/(3K_0)$. This evolution is the qutrit counterpart of the Ising evolution, and the dimensionless couplings $K_{ij}$ play the role of the $J_{ij}$. They matter modulo 3, and $K_{ij}=\pm1$ acts as the controlled-$Z$ gate of two qutrits or its inverse, so integer couplings produce qutrit graph states~\cite{Hostens2005}.

Minimising the same loss over the couplings $K_{ij}$ gives AME states of four, five and six qutrits (Appendix~\ref{app:further}). The AME$(6,3)$ state is a qutrit graph state, and the AME$(4,3)$ and AME$(5,3)$ states come from qutrit graph states with one added free coupling for four qutrits and one or two for five. For five qutrits such a coupling gives a continuous family that passes through a continuum of LU classes. For three qutrits the two couplings $K_{12}=K_{23}=1$ give an AME$(3,3)$ state.

The AME$(3,3)$ and AME$(4,3)$ evolutions use the fewest couplings possible. If every coupling across a cut touches one and the same qutrit on one side, the Schmidt rank across the cut is at most three, so every balanced cut of four qutrits must be crossed by two couplings without a common qutrit. Each coupling crosses two of the three balanced cuts, and no graph of three couplings on four qutrits satisfies this condition for all of them, whereas the ring of four couplings in Table~\ref{tab:couplings} does. For three qutrits, two couplings that connect all of them are necessary and suffice.

\subsection{Norm of the Hamiltonian and duration of the preparation}\label{sec:norm}

To compare the Hamiltonians by the resource needed to run them, we characterise each by its normalised Hilbert--Schmidt norm,
\begin{equation}
\|\hH\|_2=\Big(\frac{\Tr\hH^2}{2^N}\Big)^{1/2}=\Big(\sum_{i<j}J_{ij}^2\Big)^{1/2} ,
\label{eq:norm2}
\end{equation}
where the second equality holds because the products $\hZ_i\hZ_j$ are traceless and orthogonal in the trace inner product. Every product of $\hZ$ operators other than the identity has zero expectation value in $\ket+^{\otimes N}$, so the initial state has $\langle\hH\rangle=0$ and $\langle\hH^2\rangle=\sum_{i<j}J_{ij}^2$. The energy uncertainty of the initial state under the physical Hamiltonian $J_0\hH$ is $\Delta E=J_0\|\hH\|_2$ and is conserved during the evolution. The product of the energy uncertainty and the duration, which we call the action,
\begin{equation}
\mathcal A=\tauA\,\Delta E=\frac{\pi}{4}\,\|\hH\|_2 ,
\label{eq:action}
\end{equation}
does not depend on the coupling energy $J_0$. If the hardware caps the energy uncertainty at $E$, rescaling the Hamiltonian to saturate that cap gives a duration $\mathcal A/E$.

For a sequence of time-independent Hamiltonians the action is the sum $\mathcal A=\sum_k\tau_k\Delta E_k$ over the pieces, with $\Delta E_k$ the energy uncertainty of the state that enters piece $k$. The quantum speed limit~\cite{MandelstamTamm1945,AnandanAharonov1990,Deffner2017}
\begin{equation}
\int_0^{\tau}\Delta E\,dt\ \ge\ \arccos\big|\langle\psi_0|\psi(\tau)\rangle\big|
\label{eq:qsl}
\end{equation}
bounds the action from below by $\arccos|\langle\psi_0|\psi_J\rangle|$. For the qubit states of Table~\ref{tab:couplings}, the largest ratio of the action to this bound is about 2.6, for AME$(6,4)$.

Among the AME$(4,4)$ preparations compared here, the ring graph of Table~\ref{tab:couplings} has the smallest norm. Because the states with $J_{25}$, $-J_{25}$ and $1-J_{25}$ belong to the same LU class, every class of the family of Eq.~\eqref{eq:bond} is reached with $|J_{25}|\le\tfrac12$, so leaving the stabilizer family raises $\|\hH\|_2^2$ of its nine-coupling graph state by at most $\tfrac14$. The coupling between $q_1$ and $q_3$ added to the ring gives the same family (Sec.~\ref{sec:ns44}) and likewise raises the squared norm of the ring, 8, by at most $\tfrac14$. The Hamiltonian of the Latin-square state (Fig.~\ref{fig:latin44}) has a smaller norm than that of Fig.~\ref{fig:graphs}, because it has fewer couplings and three of them have strength $\pm\tfrac14$.

For the qutrit evolution of Sec.~\ref{sec:new48} we write $\hat n_i\hat n_j=\hat s_i\hat s_j+\hat s_i+\hat s_j+\hone$ with $\hat s_i=\hat n_i-\hone$. The single-qutrit terms commute with $\hH_3$ and add only single-qutrit phases, which do not change the LU class. For the action comparison we cancel the single-qutrit terms during the evolution and use the two-body Hamiltonian $\sum_{i<j}K_{ij}\hat s_i\hat s_j$, whose normalised norm is $\frac23(\sum_{i<j}K_{ij}^2)^{1/2}$. Diagonal operators have the same expectation value in the equal superposition of the qutrit levels as in the maximally mixed state, so this norm is again the energy uncertainty of the initial state in units of $K_0$, and the action is $\tau_3\Delta E=\frac{2\pi}{3}\cdot\frac23(\sum_{i<j}K_{ij}^2)^{1/2}$. For this Hamiltonian the largest ratio of the action to the bound of Eq.~\eqref{eq:qsl} is about 3.1, for AME$(6,3)$. The AME$(3,3)$ and AME$(4,3)$ evolutions of Sec.~\ref{sec:new48} use the fewest couplings, all of unit modulus, so among evolutions with integer couplings they have the smallest norm.

\subsection{AME$(4,6)$ state from a quantum Latin square}\label{sec:ame46}

Reference~\cite{HontarenkoLatin2026} constructs a real AME$(4,6)$ state from a quantum Latin square of order six. Its nonzero amplitudes take two different moduli, whereas a diagonal evolution of the equal superposition gives amplitudes of equal modulus, so a single evolution could reach only a locally equivalent state with amplitudes of equal modulus. Minimising the loss of Eq.~\eqref{eq:L} over all diagonal phases between pairs of six-level parties, which include every diagonal two-body coupling of a qubit and a qutrit, did not reach an AME state. In Appendix~\ref{app:ame46} the state is prepared from two maximally entangled pairs by a sequence of single-party drives alternating with coupling evolutions on two of the parties, for a six-level qudit, for a qubit and a qutrit, and for three qubits per party (Fig.~\ref{fig:ame46protocol}).

\subsection{Coupling errors}\label{sec:robust}

\begin{figure}[t]
\includegraphics[width=\columnwidth]{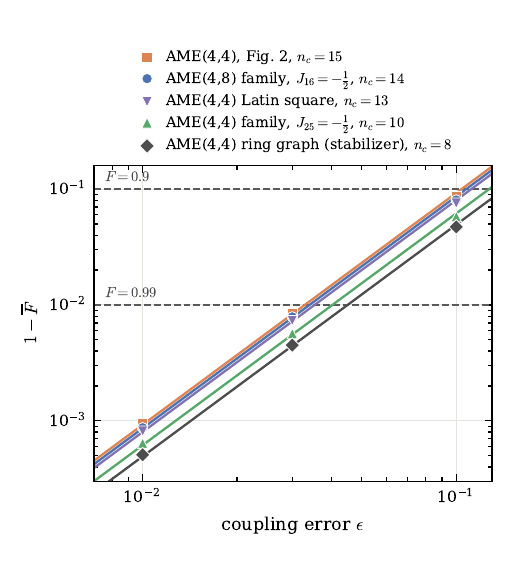}
\caption{For weak errors the sensitivity of a preparation to coupling errors is set by the number of couplings $n_c$ alone, so the stabilizer AME$(4,4)$ ring graph state, which has the fewest couplings, is the least sensitive. Mean infidelity with the target state at the time $\tauA$ under the coupling errors of Eq.~\eqref{eq:noise}, as a function of the error strength $\epsilon$, for the non-stabilizer AME$(4,4)$ state of Fig.~\ref{fig:graphs} (orange squares), the members of the AME$(4,4)$ and AME$(4,8)$ families of Figs.~\ref{fig:families} and~\ref{fig:family48} with $J_{25}=J_{16}=-\tfrac12$ (green and blue), the Latin-square AME$(4,4)$ state of Fig.~\ref{fig:latin44} (violet) and the stabilizer AME$(4,4)$ ring graph state of Table~\ref{tab:couplings} (grey). The dashed lines mark $\overline F=0.99$ and $0.9$. Markers show simulations averaged over 400 realisations and lines show Eq.~\eqref{eq:infid} with the number of couplings $n_c$ given in the legend.}
\label{fig:robust}
\end{figure}

To model the finite precision of the couplings, we replace every nonzero coupling by
\begin{equation}
J_{ij}\ \to\ J_{ij}+\epsilon\,\xi_{ij},
\label{eq:noise}
\end{equation}
with independent standard normal random numbers $\xi_{ij}$, so that $\epsilon$ is the standard deviation of the error in units of the controlled-$Z$ coupling $J_{ij}=1$. The fidelity of the perturbed state with the target is $F=|\langle\psi_J|\psi_{J'}\rangle|^2$, with $J'_{ij}=J_{ij}+\epsilon\,\xi_{ij}$. The errors add a random phase to every configuration, and to second order in $\epsilon$ the variance of these phases gives the mean infidelity
\begin{equation}
1-\overline F\simeq\Big(\frac{\pi}{4}\Big)^2 n_c\,\epsilon^2
\label{eq:infid}
\end{equation}
for $n_c$ nonzero couplings. It does not depend on the values of the couplings, because all states of Eq.~\eqref{eq:quench} have amplitudes of equal modulus, and for weak errors it agrees with the numerical averages in Fig.~\ref{fig:robust}. The infidelity grows with the number of couplings. Among the AME$(4,4)$ states of Fig.~\ref{fig:robust} it increases from the stabilizer ring graph state, with eight couplings, through the non-stabilizer family member with $J_{25}=-\tfrac12$, with ten, and the Latin-square state, with thirteen, to the non-stabilizer state with fifteen couplings. By Eq.~\eqref{eq:infid}, the nine-coupling preparations of Sec.~\ref{sec:ns44} (the stabilizer graph state of Fig.~\ref{fig:families} with $J_{25}=0$ and the non-stabilizer states of the ring with one added coupling) fall between the ring and the ten-coupling member. The ring with one added coupling is therefore the simplest non-stabilizer AME$(4,4)$ preparation among those found. A systematic error in the duration of the evolution rescales all couplings by $1+\eta$ and gives
\begin{equation}
1-F\simeq\Big(\frac{\pi}{4}\Big)^2\eta^2\sum_{i<j}J_{ij}^2=\eta^2\mathcal A^2 ,
\label{eq:timing}
\end{equation}
which depends on the coupling strengths through the action of Eq.~\eqref{eq:action}.

\subsection{Decoherence}\label{sec:decoherence}

We characterise the mixed state $\hat\rho$ prepared under noise by the entanglement that remains across the balanced cuts and by its fidelity with the target AME state $\ket{\psi_J}$,
\begin{equation}
F=\bra{\psi_J}\hat\rho\ket{\psi_J} .
\label{eq:fid}
\end{equation}
For Markovian noise at the rate $\kappa$ the state obeys the master equation~\cite{Lindblad1976,Gorini1976}
\begin{equation}
\frac{d\hat\rho}{dt}=-i\,\frac{\pi}{4\tauA}\big[\hH,\hat\rho\big]+\kappa\sum_{k}g_k\Big(\hat L_k\hat\rho\hat L_k^\dagger-\tfrac12\big\{\hat L_k^\dagger\hat L_k,\hat\rho\big\}\Big) ,
\label{eq:lindblad}
\end{equation}
with the jump operators $\hat L_k$ and their relative rates $g_k$ of Table~\ref{tab:noise}. For local noise the index $k$ runs over the qubits $i=1,\dots,N$ and over the jump operators of one qubit, with the lowering operator $\hat\sigma^-_i=\ket{0}\bra{1}_i$. The rates make $\kappa$ the decay rate of the coherence of a qubit under dephasing, of its excited-state population under amplitude damping and of its Bloch vector under depolarizing noise. Collective dephasing has a single jump operator, the total magnetisation, and describes a fluctuating field common to all qubits. Since the evolution operator depends on $J_0t$ only, the noise enters through the dimensionless strength $\kappa\tauA=\pi\kappa/(4J_0)$, the duration of the evolution in units of the coherence time $1/\kappa$.

\begin{table}[t]
\caption{Noise models of Eq.~\eqref{eq:lindblad}. The local models act on every qubit $i$ with the listed jump operators $\hat L$ and relative rates $g$.}
\label{tab:noise}
\begin{ruledtabular}
\begin{tabular}{lll}
model & $\hat L$ & $g$ \\ \hline
dephasing & $\hZ_i$ & $1/2$ \\
depolarizing & $\hX_i$, $\hY_i$, $\hZ_i$ & $1/4$ each \\
amplitude damping & $\hat\sigma^-_i$ & $1$ \\
bit flip & $\hX_i$ & $1/2$ \\
collective dephasing & $\sum_{i}\hZ_i$ & $1/2$
\end{tabular}
\end{ruledtabular}
\end{table}

\begin{figure}[t]
\includegraphics[width=\columnwidth]{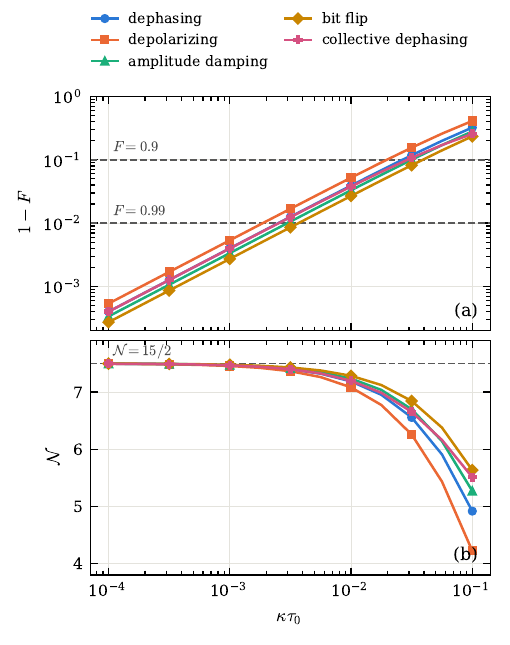}
\caption{Under all five noise models of Table~\ref{tab:noise} the infidelity grows in proportion to the noise strength for weak noise, so a high fidelity requires an evolution much shorter than the coherence time. Simulated analog evolution to the non-stabilizer AME$(4,4)$ state as a function of the dimensionless noise strength $\kappa\tauA$, the duration $\tauA$ of the evolution, equal to one controlled-$Z$ gate time, in units of the coherence time $1/\kappa$. (a) Infidelity $1-F$ with the AME state. The dashed lines mark $F=0.99$ and $F=0.9$. (b) Negativity $\mathcal{N}$, Eq.~\eqref{eq:EN}, at the end of the evolution, averaged over the three balanced cuts. The dashed line marks the value $15/2$ of the AME state. The relative decrease of $\mathcal{N}$ follows the infidelity of panel (a).}
\label{fig:fidelity}
\end{figure}

Dephasing commutes with $\hH$, so $F=[(1+e^{-\kappa\tauA})/2]^8$ for every set of couplings. The other models of Table~\ref{tab:noise} reach a given fidelity above 0.9 at noise strengths that differ from that of dephasing by less than a factor of two [Fig.~\ref{fig:fidelity}(a)].

We measure the entanglement of the mixed state across a balanced cut $S|\bar S$ by the negativity~\cite{Vidal2002}
\begin{equation}
\mathcal{N}=\sum_{\nu_j<0}|\nu_j| ,
\label{eq:EN}
\end{equation}
where $\nu_j$ are the eigenvalues of the partial transpose of $\hat\rho$ with respect to the parties in $S$. Every balanced cut of an AME$(4,4)$ state has $\mathcal{N}=15/2$. Stronger noise lowers the negativity at the end of the evolution for each noise model in Table~\ref{tab:noise} [Fig.~\ref{fig:fidelity}(b)].

\section{Analog versus gate-based preparation}\label{sec:analogvsgates}
Because the infidelity depends only on $\kappa/J_0$ (Sec.~\ref{sec:decoherence}), we simulate all protocols on one hardware model, in which every pair of qubits has an Ising coupling of strength at most $J_0$ and single-qubit gates are instantaneous and noiseless. On this hardware $\tauA$ is the duration of a controlled-$Z$ gate, a ZZ phase with $|J_{ij}|=1$ at full strength. The noise strength $\kappa\tauA$ is of the order of the error that decoherence alone causes in such a gate, comparable to the two-qubit gate errors of present devices~\cite{Moses2023}. All qubits decohere during the whole preparation according to Eq.~\eqref{eq:lindblad}, with the Hamiltonian that the protocol applies at each time.

\begin{table}[t]
\caption{The analog evolution with all couplings at once is the shortest protocol, fourteen times shorter than the compiled gate circuit \Cone{}. Simulated protocols, with the number of two-body phases, the duration in units of $\tauA$, and the spectral width $w$ integrated over the protocol in units of $\tauA$. The spectral width is the difference between the largest and smallest eigenvalue of the dimensionless Hamiltonian, and for a gate circuit its integral is the sum of $2|J|$ times the duration of each two-body phase. The analog rows assume couplings up to $J_0$ on all pairs at once (all couplings) or a summed coupling of at most $J_0$ per qubit (qubit budget). The row ``gates on disjoint pairs'' applies the same fifteen couplings as two-qubit phases at full strength, scheduled on disjoint pairs. The gate circuit \Cone{} of Fig.~12 of Ref.~\cite{Casas2026} is drawn in Fig.~\ref{fig:circuitapp}, and ``native CCZ'' counts its two controlled-controlled-$Z$ gates, the ``$+2$'', as single three-body operations of duration $\tauA$, for which the hardware model defines no two-body spectral width, hence no entry in the last column.}
\label{tab:protocols}
\begin{ruledtabular}
\begin{tabular}{lccc}
protocol & two-body phases & duration & $w$ \\ \hline
\multicolumn{4}{l}{{\bfseries\itshape non-stabilizer AME$(4,4)$ state}}\\
analog, all couplings & 15 & 1 & 18 \\
analog, qubit budget & 15 & 4 & 18 \\
gates on disjoint pairs & 15 & 4 & 26 \\
\Cone{}, compiled [Fig.~\ref{fig:circuitapp}(a)] & 23 & 14 & 44 \\
\Cone{}, native CCZ & $11+2$ & 5.5 & --
\end{tabular}
\end{ruledtabular}
\end{table}

The gate circuit \Cone{} of Ref.~\cite{Casas2026} [Fig.~\ref{fig:circuitapp}(a)] prepares the state of the fifteen couplings of Fig.~\ref{fig:graphs} up to unitaries within the parties. Its compilation for two-body hardware, which gives the durations of Table~\ref{tab:protocols}, is described in Appendix~\ref{app:circuit}.

\begin{figure}[t]
\includegraphics[width=\columnwidth]{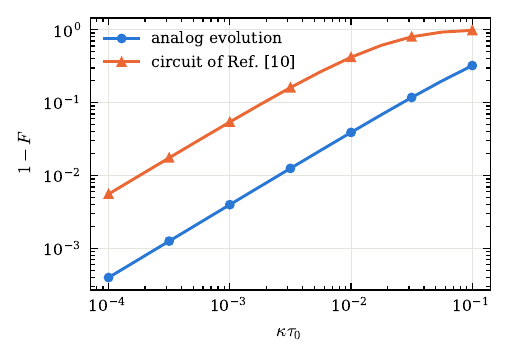}
\caption{Infidelity $1-F$ under dephasing as a function of the dimensionless noise strength: duration $\tauA$ of a controlled-$Z$ gate in units of the coherence time $1/\kappa$. Analog evolution with all couplings at once proposed in this work (blue circles) prepares the non-stabilizer AME$(4,4)$ state with a smaller infidelity than the compiled gate circuit \Cone{} of Ref.~\cite{Casas2026} (orange triangles). For weak noise the infidelity is smaller by the ratio of the durations, because the evolution lasts $\tauA$ whereas the circuit lasts $14\tauA$ (Table~\ref{tab:protocols}). }
\label{fig:analogvsgates}
\end{figure}

Under decoherence the infidelity of every protocol grows with its duration, in proportion for weak dephasing [Fig.~\ref{fig:analogvsgates}], whereas under the other local noise models the compiled circuit loses more fidelity than its duration alone predicts (Table~\ref{tab:ratios}). If the hardware realises the whole coupling matrix with its largest coupling at $J_0$, the analog evolution lasts a single $\tauA$ and is faster than the compiled circuit \Cone{} by more than an order of magnitude. If the summed coupling of every qubit is limited to $J_0$ (budget per qubit), it lasts $\tauA$ times the largest summed coupling $\sum_j|J_{ij}|$ of a qubit, $4\tauA$ (Table~\ref{tab:protocols}), and remains several times faster than the compiled circuit. Applied as two-qubit gates on disjoint pairs, the same couplings also take $4\tauA$. Both schedules tolerate about four times less noise than the analog evolution with all couplings at once, still more than the circuit. The gain over \Cone{} has two factors of comparable size, one from the couplings found, which replace the three-qubit gates of \Cone{} by fifteen commuting two-qubit phases, and one from driving these phases simultaneously. If the spectral width of the Hamiltonian is fixed~\cite{Cieslinski2023}, every protocol lasts in proportion to its width $w$, and the analog evolution is again the fastest (Table~\ref{tab:protocols}).

Under the coupling errors of Eq.~\eqref{eq:noise} every two-body phase adds a term to the infidelity of Eq.~\eqref{eq:infid}. Simulations show that for weak errors Eq.~\eqref{eq:infid} also describes the compiled circuit \Cone{} with its twenty-three two-body phases, so the analog evolution, with fifteen couplings, loses less fidelity.

\section{Discussion}\label{sec:discussion}
A single evolution under pairwise Ising couplings, applied to a product state, prepares AME states of qudits encoded in qubits and of qutrits, including states that no unitaries within the parties map onto a stabilizer state. The evolution reaches two previously known LU classes of non-stabilizer AME states of four ququarts, represented by the output of a gate circuit~\cite{Casas2026} and by a quantum orthogonal Latin square~\cite{HontarenkoLatin2026}. It also reaches an AME state of four parties of dimension eight that is not locally equivalent to any stabilizer state or to the AME$(4,8)$ states of Ref.~\cite{Casas2026}. One coupling of arbitrary strength added to an AME graph state gives continuous families of AME states, which pass through a continuum of LU classes and stay AME under errors in that coupling. Leaving the stabilizer family therefore costs a single coupling.

On hardware that drives every coupling at full strength simultaneously, the state is ready after the duration of a single two-qubit gate. The known gate circuit for the non-stabilizer state~\cite{Casas2026}, compiled into two-qubit gates, lasts more than an order of magnitude longer on the same hardware and loses at least correspondingly more fidelity under decoherence. With a fixed coupling budget per qubit, the single evolution and a schedule on disjoint pairs~\cite{Figgatt2019,Grzesiak2020,Evered2023} both take four controlled-$Z$ gate times. Programmable simultaneous couplings with independently chosen signs have been applied to up to four trapped ions~\cite{Lu2025}, and in larger systems programmability and drive power constrain the available couplings~\cite{Kyprianidis2024,Peleg2026}.

For four parties of dimension six~\cite{Rather2022,Bruzda2024,Casas2026,HontarenkoLatin2026,BruzdaGeomagic2026,Pandya2026}, the quantum Latin-square state can be prepared by alternating local drives with diagonal coupling evolutions after creating two maximally entangled pairs. The same construction works with coupled six-level qudits, with separate qubits and qutrits, and with qubits alone. Splitting each party into a qubit and a qutrit, or into qubits alone, restricts the available couplings and increases the number of evolutions in these sequences.

\section*{Data availability}
The code and data supporting the findings of this article are openly available in the repository of Ref.~\cite{AMEcode}, including a notebook with the complete parameters of the AME$(4,6)$ sequence of Appendix~\ref{app:latin46}.

\begin{acknowledgments}
We thank Mykhailo Hontarenko and Suhail Ahmad Rather for numerous recent discussions on AME states. 
K{\.Z} acknowledges financial support from the European Union through ERC Advanced Grant TAtypic,  Project No. 101142236.
\end{acknowledgments}

\FloatBarrier
\appendix
\section{Numerical methods}\label{app:numerics}
\subsection{Explicit loss function and gradient}\label{app:closedform}

We write a configuration of the qubits as $x=(s,\bar s)$, with $s$ the values of the qubits in $S$ and $\bar s$ those of the remaining qubits, and the amplitudes of Eq.~\eqref{eq:amplitudes} as $\psi_{s\bar s}=2^{-N/2}e^{i\varphi(s,\bar s)}$. The reduced state $\hrho_S=\hat M_S\hat M_S^\dagger$ of Eq.~\eqref{eq:psimatrix} and its purity are
\begin{equation}
(\hrho_S)_{ss'}=\sum_{\bar s}\psi_{s\bar s}\,\psi^*_{s'\bar s} ,\qquad \Tr\hrho_S^2=\sum_{s,s'}(\hrho_S)_{ss'}(\hrho_S)_{s's} ,
\label{eq:rhoelements}
\end{equation}
so the purity is a sum of products of four amplitudes,
\begin{equation}
\Tr\hrho_S^2=\sum_{s,s',\bar s,\bar s'}\psi_{s\bar s}\,\psi^*_{s'\bar s}\,\psi_{s'\bar s'}\,\psi^*_{s\bar s'}
=4^{-N}\sum_{s,s',\bar s,\bar s'}e^{i\Theta} ,
\label{eq:purity4}
\end{equation}
with the phase
\begin{equation}
\Theta=\varphi(s,\bar s)-\varphi(s',\bar s)+\varphi(s',\bar s')-\varphi(s,\bar s') .
\label{eq:Phi}
\end{equation} The sum runs over two configurations of the whole system, so the purity is an expectation value in two copies of the state, that of the operator exchanging the qubits of $S$ between the copies~\cite{Ekert2002}. Its negative logarithm, $-\log_2\Tr\hrho_S^2$, is the second R\'enyi entropy measured in quantum simulators~\cite{Islam2015,Brydges2019}.

The phase splits into the couplings within $S$, within $\bar S$ and across the cut,
\begin{equation}
\varphi(s,\bar s)=\varphi_S(s)+\varphi_{\bar S}(\bar s)-\frac{\pi}{4}\sum_{i\in S}\sum_{j\notin S}J_{ij}z_iz_j .
\label{eq:phase}
\end{equation}
In $\Theta$ the term $\varphi_S$ enters as $\varphi_S(s)-\varphi_S(s')+\varphi_S(s')-\varphi_S(s)=0$ and $\varphi_{\bar S}$ as $\varphi_{\bar S}(\bar s)-\varphi_{\bar S}(\bar s)+\varphi_{\bar S}(\bar s')-\varphi_{\bar S}(\bar s')=0$. Denoting by $z_i,z_i'$ the values of a qubit $i\in S$ in $s,s'$ and by $z_j,z_j'$ those of a qubit $j\notin S$ in $\bar s,\bar s'$, the couplings across the cut give
\begin{equation}
\begin{aligned}
\Theta&=-\frac{\pi}{4}\sum_{i\in S}\sum_{j\notin S}J_{ij}\,(z_iz_j-z_i'z_j+z_i'z_j'-z_iz_j')\\
&=-\pi\sum_{i\in S}\sum_{j\notin S}J_{ij}\,a_ib_j ,
\end{aligned}
\label{eq:exponent}
\end{equation}
with $a_i=(z_i-z_i')/2$ and $b_j=(z_j-z_j')/2$.

Of the four pairs $(z_i,z_i')$, the pairs $(1,1)$ and $(-1,-1)$ give the difference spin $a_i=0$, the pair $(1,-1)$ gives $a_i=1$ and the pair $(-1,1)$ gives $a_i=-1$. A sum over the values of a qubit in both copies is therefore a sum over $a_i\in\{0,\pm1\}$ with the weight $2$ for $a_i=0$ and $1$ otherwise, and Eq.~\eqref{eq:purity4} becomes the partition function of a classical model of difference spins with imaginary couplings across the cut,
\begin{equation}
\Tr\hrho_S^2=4^{-N}\sum_{\mathbf a,\mathbf b}2^{n_0(\mathbf a)+n_0(\mathbf b)}\exp\Big(-i\pi\sum_{i\in S}\sum_{j\notin S}J_{ij}\,a_ib_j\Big) ,
\label{eq:spinmodel}
\end{equation}
where $n_0$ counts the zero entries. The spins $b_j$ are not coupled to each other, so each is summed out separately in the field $\pi c_j$ of the spins in $S$, with $c_j=\sum_{i\in S}J_{ij}a_i$,
\begin{equation}
\begin{aligned}
\sum_{b_j\in\{0,\pm1\}}2^{1-|b_j|}\,e^{-i\pi c_jb_j}&=2+e^{-i\pi c_j}+e^{i\pi c_j}\\
&=4\cos^2\!\Big(\frac{\pi c_j}{2}\Big) .
\end{aligned}
\label{eq:factor}
\end{equation}
The product of these factors over the $N-N_S$ qubits outside $S$ gives Eq.~\eqref{eq:purityJ}, with the prefactor $4^{-N}4^{N-N_S}=4^{-N_S}$.

A coupling $J_{ij}=J_{ji}$ across the cut, with $i\in S$ and $j\notin S$, enters Eq.~\eqref{eq:purityJ} only through the factor of the qubit $j$, so the derivative of the purity with respect to this coupling is
\begin{equation}
\begin{aligned}
\frac{\partial\Tr\hrho_S^2}{\partial J_{ij}}=-\frac{\pi}{2}\,4^{-N_S}\sum_{\mathbf a}\,&2^{n_0(\mathbf a)}\,a_i\sin(\pi c_j)\\
&\times\prod_{k\notin S,\,k\ne j}\cos^2\!\Big(\frac{\pi c_k}{2}\Big) .
\end{aligned}
\label{eq:gradpurity}
\end{equation}
The gradient of the loss follows from Eqs.~\eqref{eq:eS} and~\eqref{eq:L},
\begin{equation}
\frac{\partial\Lame}{\partial J_{ij}}=\frac{1}{|\mathcal S|}\sum_{S\in\mathcal S}\frac{1}{1-d^{-|S|}}\,\frac{\partial\Tr\hrho_S^2}{\partial J_{ij}} ,
\label{eq:gradL}
\end{equation}
where only the cuts that separate qubit $i$ from qubit $j$ contribute. The sum over $\mathbf a$ has $3^{N_S}$ terms, 81 for a balanced cut of four ququarts. For random couplings Eqs.~\eqref{eq:purityJ} and~\eqref{eq:gradL} agree to numerical precision with the purity of Eq.~\eqref{eq:psimatrix} and with the gradient obtained by automatic differentiation in JAX~\cite{Bradbury2018}, which the minimisation uses.

\subsection{Minimisation}\label{app:ame46methods}

We minimise the loss of Eq.~\eqref{eq:L} over the couplings $J_{ij}$ between qubits of different parties from random initial values, except for the AME$(4,8)$ state of Fig.~\ref{fig:ising832}. For the AME$(4,4)$ state of Fig.~\ref{fig:graphs}, the AME$(5,4)$ and AME$(6,4)$ states and, with the couplings $K_{ij}$ as variables, the qutrit states AME$(4,3)$, AME$(5,3)$ and AME$(6,3)$, we use the Adam method~\cite{Kingma2015} of Optax~\cite{DeepMind2020} followed by the BFGS method~\cite{Broyden1970,Fletcher1970,Goldfarb1970,Shanno1970} of SciPy~\cite{Virtanen2020}. For the AME$(4,8)$ state of Fig.~\ref{fig:ising832} we start from a Hamiltonian of Pauli strings acting on at most two parties, found with the Adam method followed by the L-BFGS method~\cite{Liu1989} of Optax. We minimise the loss over its sixty largest terms with the L-BFGS method~\cite{Liu1989,Byrd1995} of SciPy and take the two-body $\hZ\hZ$ couplings between parties as initial values of the Ising minimisation, done with the same method, whose result is reduced modulo 2 and rounded to multiples of $\tfrac14$. A state counts as AME when
\begin{equation}
\Delta=\max_{S\in\mathcal S}\big\|\hrho_S-\hone/d^{|S|}\big\|_{\rm HS}
\label{eq:residual}
\end{equation}
is below $10^{-6}$. To reach a given class of AME$(4,4)$ states we add the penalty $\frac{1}{3}\sum[(m_2-m_2^\ast)^2+(m_4-m_4^\ast)^2]/256^2$, summed over the three pairings of the parties. The minimisations of Appendix~\ref{app:classes}, with and without this penalty, use the L-BFGS method of SciPy. For the Latin-square state we round the couplings found to multiples of $\tfrac14$ and drop two couplings without changing the AME property or the moments. For AME$(4,4)$ we also test every set of at most eight unit couplings. The search for the AME$(4,8)$ graph state adds one unit coupling at a time and keeps the sets of couplings with the smallest loss. A coupling added to the graph states of Figs.~\ref{fig:families} and~\ref{fig:family48} is free when the loss vanishes on a grid of 201 strengths in $[-1,1]$. Shifting a coupling by 2 multiplies the state by $\hZ_i\hZ_j$, so this interval contains every distinct strength. For one varying coupling every purity has the form $a+b\cos\pi J+c\sin\pi J$, so its values on the grid fix it for every strength.

For the states of Figs.~\ref{fig:graphs} and~\ref{fig:ising832} the sum $\sum_{i<j}J_{ij}z_iz_j$ is an integer for every configuration, and for the Latin-square state, whose three couplings $\pm\tfrac14$ form a triangle, it is an integer up to a constant. Every amplitude is then $2^{-N/2}$ times a power of $e^{i\pi/4}$, so the reduced states, evaluated in exact integer arithmetic, are exactly maximally mixed. The maximum in Eq.~\eqref{eq:overlap} is found by updating one single-party unitary at a time to the unitary factor of the polar decomposition of its overlap matrix.

For the AME$(4,6)$ state the variables are the drive and coupling strengths of Eq.~\eqref{eq:ame46sequence}. The sequence $\hat U$ acting on the pairs of Eq.~\eqref{eq:ame46pairs} prepares $\ket\psi=(\hat U\otimes\hone_{CD})\ket{\Omega_6}_{AC}\ket{\Omega_6}_{BD}$, and we minimise one minus its overlap with the target,
\begin{equation}
\ell=1-|\langle\psi_{96}|\psi\rangle|=1-\frac{|\Tr(\hcU_{96}^\dagger\hat U)|}{36},
\label{eq:ame46fit}
\end{equation}
where the second form follows from Eq.~\eqref{eq:ame46pairs}. The trace is the sum of the 36 eigenvalues of the unitary $\hcU_{96}^\dagger\hat U$, so $\ell$ vanishes only when all of them are equal, that is when $\hat U=e^{i\chi}\hcU_{96}$. We minimise $\ell$ with the L-BFGS method of SciPy and refine the result with the Levenberg--Marquardt method~\cite{Levenberg1944,Marquardt1963} applied to the matrix elements of $\hat U-e^{i\chi}\hcU_{96}$. The minimisation over diagonal phases of Sec.~\ref{sec:ame46} also uses the L-BFGS method of SciPy. Each rotation of a six-level party is then written as a product of rotations between neighbouring levels~\cite{Reck1994}.

\section{Local-unitary invariants}\label{app:invariant}
\subsection{Echo operator on two copies}

We take two copies of the system,
\begin{equation}
\ket\psi\otimes\ket\psi=\sum\psi_{\alpha\beta\gamma\delta}\,\psi_{\alpha'\beta'\gamma'\delta'}\,\ket{\alpha,\beta,\gamma,\delta}\ket{\alpha',\beta',\gamma',\delta'} ,
\label{eq:copies}
\end{equation}
with primed indices for the second copy. By Eq.~\eqref{eq:Udef} the amplitudes of $\ket\psi\otimes\ket\psi$ are the matrix elements of $\hcU\otimes\hcU$ divided by $d^2$, so two copies of the state correspond to two copies of the evolution operator. A single copy evolves as
\begin{equation}
\hcU\ket{\gamma,\delta}=\sum_{\alpha,\beta}\hcU_{(\alpha\beta),(\gamma\delta)}\ket{\alpha,\beta} ,
\label{eq:Uone}
\end{equation}
and the two copies evolve independently, $(\hcU\otimes\hcU)\ket{\gamma,\delta}\ket{\gamma',\delta'}=(\hcU\ket{\gamma,\delta})(\hcU\ket{\gamma',\delta'})$. The swap of the second and the fourth index,
\begin{equation}
\hSw_{24}\ket{\alpha,\beta,\alpha',\beta'}=\ket{\alpha,\beta',\alpha',\beta} ,
\label{eq:swap}
\end{equation}
exchanges party $B$ between the copies after the evolution, and acting on $\ket{\gamma,\delta,\gamma',\delta'}$ it exchanges party $D$ before the evolution. These are the two exchanges of the echo operator of Eq.~\eqref{eq:inv}, which we write as
\begin{equation}
\hat I(\hcU)=\hSw_{24}\,\hat W ,\qquad \hat W=(\hcU^\dagger\otimes\hcU^\dagger)\,\hSw_{24}\,(\hcU\otimes\hcU) ,
\label{eq:W}
\end{equation}
where $\hat W$ is the exchange of $B$ carried back through the evolution.

If $\hcU$ factorises into two unitary $d\times d$ matrices,
\begin{equation}
\hcU_{(\alpha\beta),(\gamma\delta)}=(w_1)_{\alpha\gamma}\,(w_2)_{\beta\delta} ,\qquad \hcU=\hat w_1\otimes\hat w_2 ,
\label{eq:uv}
\end{equation}
where $\hat w_1$ connects party $C$ with party $A$ and $\hat w_2$ connects party $D$ with party $B$, the state is a product of a state of $A$ and $C$ and a state of $B$ and $D$. On two copies $\hcU\otimes\hcU=\hat w_1\otimes\hat w_2\otimes\hat w_1\otimes\hat w_2$, and $\hSw_{24}$ exchanges the two factors $\hat w_2$ and commutes with $\hcU\otimes\hcU$, so that $\hat W=\hSw_{24}$ and $\hat I(\hcU)=\hSw_{24}^2=\hone$.

\subsection{Invariance of the moments}

A change of basis within the parties, Eq.~\eqref{eq:lu}, inserted into Eq.~\eqref{eq:Udef}, gives
\begin{equation}
\hcU'=\hU_{AB}\,\hcU\,\hU_{CD}^{T} ,\quad \hU_{AB}=\hU_A\otimes\hU_B ,\quad \hU_{CD}=\hU_C\otimes\hU_D .
\label{eq:Uprime}
\end{equation}
The operators $\hU_{AB}\otimes\hU_{AB}$ and $\hat Q=\hU_{CD}^{T}\otimes\hU_{CD}^{T}$ carry the same unitary on the second and on the fourth factor and therefore commute with $\hSw_{24}$, so that
\begin{equation}
\hat I(\hcU')=\hSw_{24}\,\hat Q^\dagger(\hcU^\dagger\otimes\hcU^\dagger)\,\hSw_{24}\,(\hcU\otimes\hcU)\hat Q=\hat Q^\dagger\,\hat I(\hcU)\,\hat Q .
\label{eq:Iconj}
\end{equation}
By the cyclic property of the trace, $\Tr[\hat I(\hcU')^k]=\Tr[\hat I(\hcU)^k]$, so the moments of Eq.~\eqref{eq:moments} are the same for all locally equivalent states. They are real, because $\hSw_{24}$ and $\hat W$ are Hermitian, so the trace of $(\hSw_{24}\hat W)^k$ equals the trace of its adjoint $(\hat W\hSw_{24})^k$.

\subsection{Moments as observables on copies of the state}

Writing out the matrix products in $\hat I(\hcU)^k$ with Eq.~\eqref{eq:Udef}, each factor $\hcU$ contributes an amplitude $d\,\psi$ and each factor $\hcU^\dagger$ a conjugate amplitude $d\,\psi^*$. The $k$th moment is therefore an expectation value on $2k$ copies of the state,
\begin{equation}
m_k=d^{4k}\,\bra{\psi}^{\otimes 2k}\,\hat\Sigma_A\otimes\hat\Sigma_B\otimes\hat\Sigma_C\otimes\hat\Sigma_D\,\ket{\psi}^{\otimes 2k} ,
\label{eq:replica}
\end{equation}
where $\hat\Sigma_X$ permutes the copies of party $X$ among themselves. We label the copies by $(j,c)$, with $j=1,\dots,k$ the factor of $\hat I(\hcU)^k$ and $c=1,2$ the copy in Eq.~\eqref{eq:copies}. The swap of $B$ after the evolution exchanges the two copies $c$ of party $B$, the swap of $D$ before the evolution links consecutive factors, and the trace closes the chain. Party $A$ is therefore not permuted, party $B$ swaps the two copies of each factor, party $C$ moves every copy to the next factor, and party $D$ does both. Each $\hat\Sigma_X$ is a product of swaps, so $m_k$ can be measured on $2k$ copies by the methods used for R\'enyi entropies, by interfering the copies~\cite{Ekert2002} or by randomised local measurements~\cite{Brydges2019}.

For the second moment we number the four copies $(1,1),(1,2),(2,1),(2,2)$ as $1,\dots,4$. The permutations are then those of Eq.~\eqref{eq:m2copies}, and in components
\begin{equation}
\begin{aligned}
m_2=d^8\sum&\psi_{\alpha_1\beta_1\gamma_1\delta_1}\psi_{\alpha_2\beta_2\gamma_2\delta_2}\psi_{\alpha_3\beta_3\gamma_3\delta_3}\psi_{\alpha_4\beta_4\gamma_4\delta_4}\\
\times\,&\psi^*_{\alpha_1\beta_2\gamma_3\delta_4}\,\psi^*_{\alpha_2\beta_1\gamma_4\delta_3}\,\psi^*_{\alpha_3\beta_4\gamma_1\delta_2}\,\psi^*_{\alpha_4\beta_3\gamma_2\delta_1} ,
\end{aligned}
\label{eq:kleincomp}
\end{equation}
summed over all indices, with the subscript $1,\dots,4$ of an index labelling the copy. Without the factor $d^8$ the sum in Eq.~\eqref{eq:kleincomp} is the Klein contraction $K(\psi)$ of Ref.~\cite{HontarenkoLatin2026}. Because these permutations and the identity form the Klein four-group, $m_2$ is the same for every ordering of the parties, as shown in Ref.~\cite{HontarenkoLatin2026}, whereas $m_4$ in general is not. The AME$(4,8)$ state of Fig.~\ref{fig:ising832} has $m_2=832$ for all orderings and $m_4=1648$, $1600$ and $2240$ when $A$ is paired with $B$, $C$ and $D$. The stabilizer AME$(4,4)$ state $\frac14\sum_{x,y}\ket{x,y,x+y,x+\eta y}$ of Ref.~\cite{HontarenkoLatin2026}, with $x,y$ in the field of four elements and $\eta^2+\eta+1=0$, has $m_2=256$. We verified Eq.~\eqref{eq:replica} numerically against Eq.~\eqref{eq:moments} for $k=2$ and $k=3$.

\subsection{Classes of AME$(4,4)$ states reached by the evolution}\label{app:classes}

Minimising the loss of Eq.~\eqref{eq:L} ends in AME$(4,4)$ states of several LU classes, which the moments $m_2$ and $m_4$ of the cut $AB|CD$ separate. Besides the graph states, with $(m_2,m_4)=(256,256)$, the members of the family of Eq.~\eqref{eq:bond}, and the state of Fig.~\ref{fig:graphs}, with $(64,160)$, the evolution reaches classes whose moments lie below and above the family curve. The Latin-square state of Appendix~\ref{app:latin44}, with $(208,256)$, was found only in the minimisations with a penalty that steers the moments towards these values. Equal moments are necessary but not sufficient for local equivalence, so the equivalence with the state of Fig.~\ref{fig:graphs} and with the Latin-square state is confirmed by the maximised overlap of Eq.~\eqref{eq:overlap}.

\subsection{Grouped moments for five and six parties}

For five and six parties we merge the parties into four blocks, $(A)(B)|(C)(DE)$ for five parties and $(AB)(C)|(DE)(F)$ for six, with the block dimensions $D_1,\dots,D_4$, and define
\begin{equation}
\hcU_{(b_1b_2),(b_3b_4)}=\sqrt{D_1D_2}\,\psi_{b_1b_2b_3b_4} ,
\label{eq:Ugroup}
\end{equation}
a matrix with $D_1D_2$ rows and $D_3D_4$ columns, where $b_1,\dots,b_4$ label the basis states of the blocks. The echo operator of Eq.~\eqref{eq:inv}, with the exchanges acting on the second and the fourth block, gives the grouped moments $m_k$. A change of basis within a party acts within one block, so the grouped moments do not depend on the basis of the parties, and Eq.~\eqref{eq:replica} holds with $d^{4k}$ replaced by $(D_1D_2)^{2k}$ and the permutations acting on the blocks. A permutation of qubits is a Clifford operator, so for a stabilizer state both $\ket\psi^{\otimes 2k}$ and $\hat\Sigma_A\otimes\cdots\otimes\hat\Sigma_D\ket\psi^{\otimes 2k}$ are stabilizer states, and the squared modulus of their overlap is zero or a power of two~\cite{Aaronson2004,Garcia2012}. With $D_1D_2$ a power of two this gives Eq.~\eqref{eq:stabcritgroup}.

\section{Gate circuits of the comparison}\label{app:circuit}
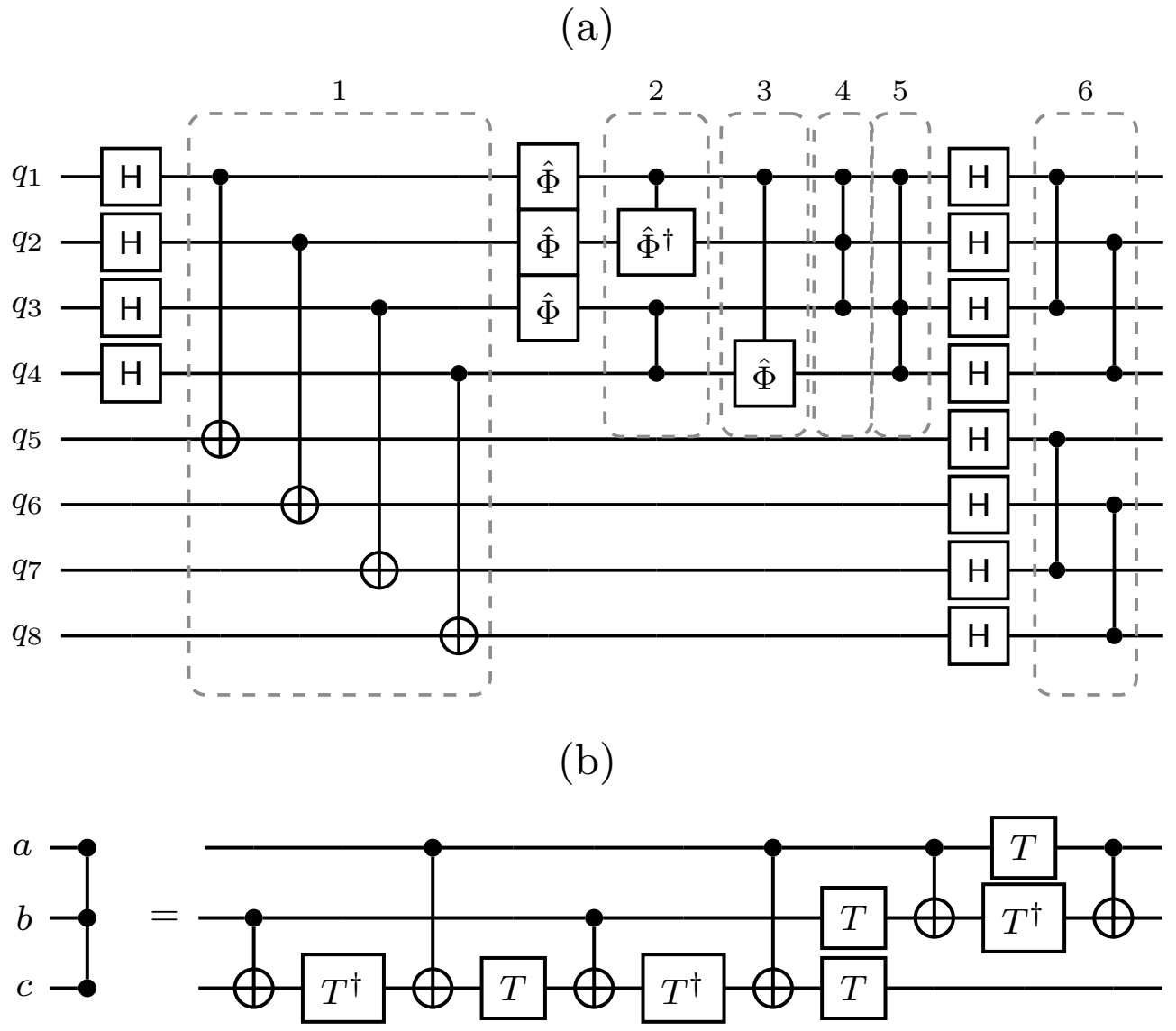
\begin{figure}[t]
\centering
{\small (a)}\\[1mm]
\resizebox{\columnwidth}{!}{%
\begin{quantikz}[row sep={0.55cm,between origins}, column sep=0.34cm, font=\footnotesize]
\lstick{$q_1$} & \gate{\mathsf H} & \ctrl{4}\gategroup[8,steps=4,style={dashed,rounded corners,inner xsep=0pt,draw=black!45},label style={label position=above,yshift=-1pt}]{\scriptsize 1} & \qw & \qw & \qw & \gate{\hat\Phi} & \ctrl{1}\gategroup[4,steps=1,style={dashed,rounded corners,inner xsep=0pt,draw=black!45},label style={label position=above,yshift=-1pt}]{\scriptsize 2} & \ctrl{3}\gategroup[4,steps=1,style={dashed,rounded corners,inner xsep=0pt,draw=black!45},label style={label position=above,yshift=-1pt}]{\scriptsize 3} & \ctrl{2}\gategroup[4,steps=1,style={dashed,rounded corners,inner xsep=0pt,draw=black!45},label style={label position=above,yshift=-1pt}]{\scriptsize 4} & \ctrl{3}\gategroup[4,steps=1,style={dashed,rounded corners,inner xsep=0pt,draw=black!45},label style={label position=above,yshift=-1pt}]{\scriptsize 5} & \gate{\mathsf H} & \ctrl{2}\gategroup[8,steps=2,style={dashed,rounded corners,inner xsep=0pt,draw=black!45},label style={label position=above,yshift=-1pt}]{\scriptsize 6} & \qw & \qw \\
\lstick{$q_2$} & \gate{\mathsf H} & \qw & \ctrl{4} & \qw & \qw & \gate{\hat\Phi} & \gate{\hat\Phi^\dagger} & \qw & \control{} & \qw & \gate{\mathsf H} & \qw & \ctrl{2} & \qw \\
\lstick{$q_3$} & \gate{\mathsf H} & \qw & \qw & \ctrl{4} & \qw & \gate{\hat\Phi} & \ctrl{1} & \qw & \control{} & \control{} & \gate{\mathsf H} & \control{} & \qw & \qw \\
\lstick{$q_4$} & \gate{\mathsf H} & \qw & \qw & \qw & \ctrl{4} & \qw & \control{} & \gate{\hat\Phi} & \qw & \control{} & \gate{\mathsf H} & \qw & \control{} & \qw \\
\lstick{$q_5$} & \qw & \targ{} & \qw & \qw & \qw & \qw & \qw & \qw & \qw & \qw & \gate{\mathsf H} & \ctrl{2} & \qw & \qw \\
\lstick{$q_6$} & \qw & \qw & \targ{} & \qw & \qw & \qw & \qw & \qw & \qw & \qw & \gate{\mathsf H} & \qw & \ctrl{2} & \qw \\
\lstick{$q_7$} & \qw & \qw & \qw & \targ{} & \qw & \qw & \qw & \qw & \qw & \qw & \gate{\mathsf H} & \control{} & \qw & \qw \\
\lstick{$q_8$} & \qw & \qw & \qw & \qw & \targ{} & \qw & \qw & \qw & \qw & \qw & \gate{\mathsf H} & \qw & \control{} & \qw 
\end{quantikz}%
}

\vspace{3mm}
{\small (b)}\\[1mm]
\resizebox{\columnwidth}{!}{%
\begin{quantikz}[row sep={0.55cm,between origins}, column sep=0.22cm, font=\footnotesize]
\lstick{$a$} & \ctrl{2} & \midstick[3,brackets=none]{=} & \qw & \qw & \ctrl{2} & \qw & \qw & \qw & \ctrl{2} & \qw & \ctrl{1} & \gate{T} & \ctrl{1} & \qw \\
\lstick{$b$} & \control{} & & \ctrl{1} & \qw & \qw & \qw & \ctrl{1} & \qw & \qw & \gate{T} & \targ{} & \gate{T^\dagger} & \targ{} & \qw \\
\lstick{$c$} & \control{} & & \targ{} & \gate{T^\dagger} & \targ{} & \gate{T} & \targ{} & \gate{T^\dagger} & \targ{} & \gate{T} & \qw & \qw & \qw & \qw 
\end{quantikz}%
}
\caption{(a) Circuit \Cone{}, Fig.~12 of Ref.~\cite{Casas2026}, for the non-stabilizer AME$(4,4)$ state, starting from $\ket{0}^{\otimes8}$. Dashed boxes mark the six entangling layers when every gate, including the controlled-controlled-$Z$ gate, is native. (b) Compilation of the controlled-controlled-$Z$ gate on the qubits $a$, $b$ and $c$ into six CNOT gates and seven $T$ gates, $\hat T={\rm diag}(1,e^{i\pi/4})$, used for \Cone{} on two-body hardware.}
\label{fig:circuitapp}
\end{figure}

The gate circuit \Cone{} for the non-stabilizer AME$(4,4)$ state is that of Fig.~12 of Ref.~\cite{Casas2026}, reproduced gate by gate in Fig.~\ref{fig:circuitapp} with the parties of Eq.~\eqref{eq:parties}. It starts from $\ket{0}^{\otimes8}$, creates Bell pairs between the qubits $q_i$ and $q_{i+4}$, applies to $q_1,\dots,q_4$ the diagonal non-Clifford unitary $\hat D_\Lambda$ of Eq.~\eqref{eq:casasU}, built from controlled-$\hat\Phi$ and controlled-controlled-$Z$ gates, and ends with Hadamard and controlled-$Z$ gates. Its diagonal stage equals $\hat D_\Lambda$ times $\hZ$ on $q_3$, so it prepares the state of Eq.~\eqref{eq:casasstate} up to the single-qubit operator $\hZ_1\hat X_3$, which leaves the moments $m_2=64$ and $m_4=160$ unchanged. The fidelities of Sec.~\ref{sec:analogvsgates} are taken with the noiseless output of each protocol.

On the hardware model of Sec.~\ref{sec:analogvsgates} a controlled-$\hat\Phi$ gate is a ZZ phase of duration $\tauA/2$ together with single-qubit phases, and a controlled-controlled-$Z$ gate is compiled into six CNOT gates and seven $T$ gates [Fig.~\ref{fig:circuitapp}(b)]~\cite{NielsenChuang2010}. The single-qubit phases of each controlled-phase gate act before its ZZ phase, which fixes the direction of the transverse Bloch vectors during the evolution and thereby the effect of bit-flip noise. Every gate starts as soon as the preceding gates on its qubits have ended, which gives the durations of Table~\ref{tab:protocols}. With native controlled-controlled-$Z$ gates each of these gates lasts $\tauA$.

\begin{table}[t]
\caption{For every noise model the analog evolution tolerates noise more than ten times stronger than the compiled circuit. Noise strength $\kappa\tauA$, in units of $10^{-4}$, at which the fidelity of the prepared non-stabilizer AME$(4,4)$ state falls to $F=0.99$, for the analog evolution with all couplings at once (duration $\tauA$) and for the compiled circuit \Cone{} of Ref.~\cite{Casas2026} (duration $14\tauA$).}
\label{tab:ratios}
\begin{ruledtabular}
\begin{tabular}{lcc}
noise & analog evolution & circuit of Ref.~\cite{Casas2026} \\ \hline
dephasing & 25 & 1.8 \\
depolarizing & 19 & 1.2 \\
amplitude damping & 30 & 1.8 \\
bit flip & 37 & 1.8 \\
collective dephasing & 25 & 1.8
\end{tabular}
\end{ruledtabular}
\end{table}

\FloatBarrier
\section{Preparing the AME$(4,4)$ family}\label{app:recipe}
A member $\ket{\psi(J_{25})}$ of the family of Eq.~\eqref{eq:bond} is prepared on eight qubits $q_1,\dots,q_8$, grouped into the parties of Eq.~\eqref{eq:parties}, by rotating every qubit from $\ket0$ to $\ket+$ and applying the Ising Hamiltonian of Eq.~\eqref{eq:quench} for the time $\tauA=\pi/(4J_0)$. The coupling is $J_0$ on the nine pairs $(1,3)$, $(1,8)$, $(2,4)$, $(2,6)$, $(3,6)$, $(3,7)$, $(4,5)$, $(4,8)$, $(5,7)$, $J_{25}J_0$ on the pair $(2,5)$, and zero on the other pairs. Each term contributes the phase $e^{-i\frac{\pi}{4}J_{ij}\hZ_i\hZ_j}$ and no single-qubit correction follows. The choice $J_{25}=\tfrac12$ gives the smallest second moment of the family, $m_2=192$.

All ten terms commute, so they can also act in three layers of duration $\tauA$, each made of couplings on disjoint pairs,
\begin{equation}
\begin{aligned}
&\{(1,3),(2,4),(5,7)\},\quad \{(3,6),(4,8),(2,5)\},\\
&\{(1,8),(2,6),(3,7),(4,5)\},
\end{aligned}
\label{eq:layers}
\end{equation}
in any order. The same family is also prepared by nine couplings, the eight couplings of the ring graph and a coupling between $q_1$ and $q_3$ (Sec.~\ref{sec:ns44}), in two layers of four disjoint pairs followed by that coupling. For the ten couplings of Eq.~\eqref{eq:bond}, a summed coupling of at most $J_0$ per qubit gives the same duration $3\tauA$, set by the qubits $q_3$ and $q_4$, whose three couplings of unit strength give the largest summed coupling.

Independent Gaussian errors of standard deviation $\epsilon$ in the ten couplings give $1-\overline F\simeq(\pi/4)^2\,10\,\epsilon^2$ by Eq.~\eqref{eq:infid}, so a mean fidelity of $0.99$ requires $\epsilon\approx0.04$. Under dephasing of rate $\kappa$ on every qubit the fidelity after the single evolution is $[(1+e^{-\kappa\tauA})/2]^8$, so a fidelity of $0.99$ requires $\kappa\tauA\le2.5\times10^{-3}$, and a three times smaller value for the layered version.

\section{Further AME states}\label{app:further}
\subsection{AME graph states}

With couplings $J_{ij}=1$ on the edges listed in Table~\ref{tab:couplings} the analog evolution prepares AME graph states. The AME$(5,2)$ and AME$(6,2)$ states are the known graph states of five and six qubits~\cite{Helwig2013}, and the AME$(4,4)$ ring graph is the graph of Refs.~\cite{Helwig2013,Casas2026}. The AME$(4,8)$ graph state is the graph of Fig.~\ref{fig:family48} without the added coupling. During the evolution the loss decreases to zero and the entanglement entropy of every balanced cut reaches its maximum at the time $\tauA$.

\subsection{Added coupling and the AME$(4,8)$ family}

When $\hZ_k\hZ_l$ acts on the graph state as a Pauli operator of a single party, the added coupling is a unitary within that party and the state stays in the LU class of the graph state. Half of the free couplings of the nine-edge AME$(4,4)$ graph state and a third of those of the AME$(4,8)$ graph state are of this kind. The other free couplings change the second moment, for the AME$(4,4)$ graph state according to Eq.~\eqref{eq:curve}, and for the AME$(4,8)$ graph state the added coupling $J_{16}$ between $q_1$ and $q_6$ gives
\begin{equation}
m_2=4096-1024\sin^2(\pi J_{16}) ,
\label{eq:family48}
\end{equation}
so by Eq.~\eqref{eq:stabcrit} every member with a non-integer $J_{16}$ is a non-stabilizer state. Equations~\eqref{eq:curve} and~\eqref{eq:family48} are numerical fits to the second moment computed from the states [Figs.~\ref{fig:families}(d) and~\ref{fig:family48}(d)].

\begin{figure*}[t]
\begin{minipage}[t]{\columnwidth}
\vspace{0pt}
\includegraphics[width=\columnwidth]{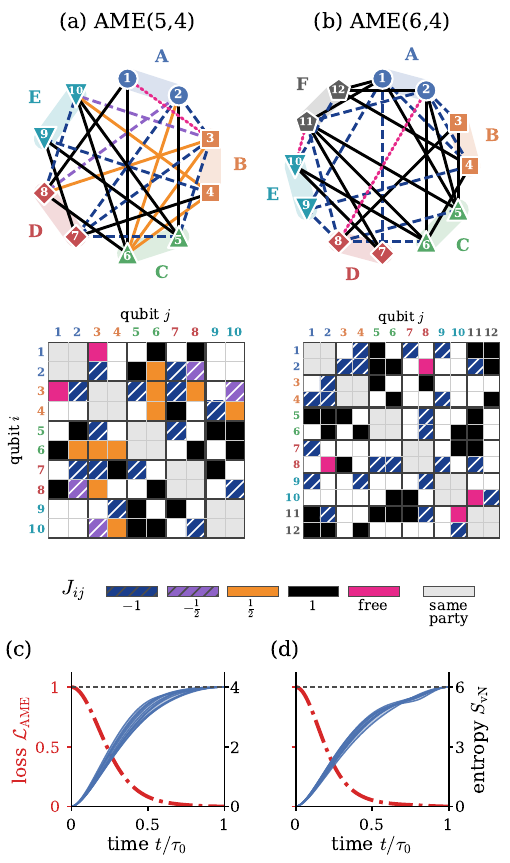}
\caption{Single evolutions under two-qubit couplings prepare AME states of five and six ququarts. Couplings of (a) an AME$(5,4)$ state, with five ququarts $A,\dots,E$, and (b) an AME$(6,4)$ state, with six ququarts $A,\dots,F$, each ququart made of two qubits, drawn as a graph (top), with negative couplings as in Fig.~\ref{fig:graphs}, and as the symmetric coupling matrix (middle). The couplings of arbitrary strength (pink, dotted), $J_{13}$ in (a) and $J_{28}$ and $J_{10,11}$ in (b), are set to zero for the curves below. (c), (d) Loss $\Lame$ (red, left axis) and entanglement entropy $S_{\rm vN}$ of the ten balanced cuts (thin lines, right axis) during the evolution. The shaded band spans the range of the ten entropies and closes at the time $\tauA$, where all balanced cuts are maximally entangled.}
\label{fig:graphs5664}
\end{minipage}\hfill
\begin{minipage}[t]{\columnwidth}
\vspace{0pt}
\includegraphics[width=\columnwidth]{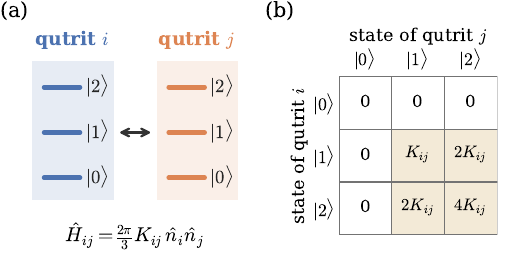}
\caption{Qutrit evolution of Eq.~\eqref{eq:nn}. (a) Two qutrits $i$ and $j$ with the basis states $\ket0$, $\ket1$ and $\ket2$. (b) Integrated interaction energy of their nine configurations in units of $2\pi/3$.}
\label{fig:qutritmodel}

\vspace{4mm}
\includegraphics[width=\columnwidth]{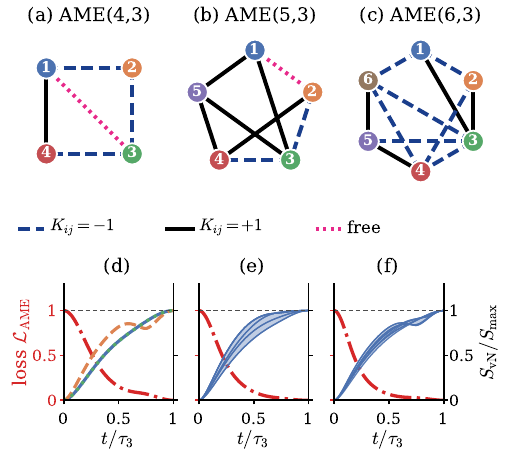}
\caption{The same loss function also finds AME states of four, five and six qutrits, each prepared by a single evolution under two-qutrit couplings. (a)--(c) Couplings $K_{ij}$ that prepare AME$(4,3)$, AME$(5,3)$ and AME$(6,3)$ states, with one node per qutrit. Couplings $K_{ij}=\pm1$ are controlled-$Z$ couplings of two qutrits, and the dotted pink coupling can take any value. (d)--(f) Loss (red, left axis) and entanglement entropy of every balanced cut in units of its maximum $S_{\max}=\lfloor n/2\rfloor\log_23$ (right axis) during the evolution for the states of (a)--(c), with the coupling of arbitrary strength set to zero. In (d) the cuts $AB|CD$, $AC|BD$ and $AD|BC$ are drawn solid, dashed and dotted, and in (e) and (f) the thin lines are the ten balanced cuts and the shaded band spans their range. Every balanced cut reaches its maximal entropy at the end of the evolution.}
\label{fig:qutrits}
\end{minipage}
\end{figure*}

\subsection{AME$(5,4)$ and AME$(6,4)$ states}

Figure~\ref{fig:graphs5664} shows the couplings of an AME$(5,4)$ state, encoded in ten qubits, and of an AME$(6,4)$ state, encoded in twelve qubits. The AME$(5,4)$ couplings are $\pm1$ and $\pm\tfrac12$ with the free coupling $J_{13}$, and the AME$(6,4)$ couplings are $\pm1$ with the free couplings $J_{28}$ and $J_{10,11}$.

With the free couplings set to zero the AME$(6,4)$ state is a graph state, and its grouped second moment for $(AB)(C)|(DE)(F)$ equals 4096. The free couplings lower it, and a numerical fit gives $m_2=4096\,[1-\tfrac14\sin^2(\pi J_{28})][1-\tfrac14\sin^2(\pi J_{10,11})]$, so every member with a non-integer free coupling is a non-stabilizer AME$(6,4)$ state.

For the grouping $(A)(B)|(C)(DE)$ the grouped second moments of the AME$(5,4)$ states found all differ and all violate Eq.~\eqref{eq:stabcritgroup}, as does the value $m_2=21$ of the state of Fig.~\ref{fig:graphs5664}(a) at $J_{13}=0$. For the displayed state the sum $\sum_{i<j}J_{ij}z_iz_j$ is an integer plus $\tfrac12$ for every configuration, so, as for the states of Figs.~\ref{fig:graphs} and~\ref{fig:ising832} (Appendix~\ref{app:numerics}), its balanced reductions are exactly maximally mixed.

\subsection{Qutrit evolution: AME$(4,3)$, AME$(5,3)$ and AME$(6,3)$}\label{sec:qutrits}

In the qutrit evolution of Eq.~\eqref{eq:nn}, the integrated interaction energy of two qutrits in the basis states $\ket{x_i}$ and $\ket{x_j}$ [Fig.~\ref{fig:qutritmodel}(a)] is $(2\pi/3)K_{ij}x_ix_j$, which vanishes when either qutrit is in $\ket0$ [Fig.~\ref{fig:qutritmodel}(b)]. As for qubits, the final state has amplitudes of equal modulus,
\begin{equation}
\psi_x=3^{-n/2}\exp\Big(-i\frac{2\pi}{3}\sum_{i<j}K_{ij}\,x_ix_j\Big) ,
\label{eq:ampqutrit}
\end{equation}
and the AME property is tested in the same way, through the reduced states of all groups of $\lfloor n/2\rfloor$ qutrits and the loss of Eq.~\eqref{eq:L}.

Minimising this loss gives AME$(4,3)$, AME$(5,3)$ and AME$(6,3)$ states, with the couplings of Fig.~\ref{fig:qutrits}(a)--(c). The free couplings of the AME$(4,3)$ and AME$(5,3)$ states are the qutrit counterparts of the added coupling of Sec.~\ref{sec:bonds}. For four qutrits the free couplings do not change the LU class, because the AME$(4,3)$ state is unique up to local unitaries~\cite{Rather2023}. For five qutrits the grouping $(A)(B)|(C)(DE)$ of Sec.~\ref{sec:identify} gives the same moments along the family, so we group the parties as $(A)(C)|(B)(DE)$. The grouped second moment then stays at 9 along the family, whereas the grouped fourth moment changes continuously with the free coupling $K_{12}$, so the family passes through a continuum of LU classes.

\FloatBarrier
\section{AME states from quantum Latin squares}\label{app:ame46}

In a Latin square of order $d$ every row and every column contains each of $d$ symbols once. In a quantum Latin square the symbols are replaced by quantum states~\cite{Musto2016}, and a quantum orthogonal Latin square of order $d$, a quantum Latin square whose $d^2$ cells are orthonormal states of two parties of dimension $d$, defines an AME$(4,d)$ state~\cite{Goyeneche2018,HontarenkoLatin2026}. Reference~\cite{HontarenkoLatin2026} constructs from such squares an AME$(4,4)$ state, which a single Ising evolution prepares (Sec.~\ref{sec:ns44}), and a real AME$(4,6)$ state, which is prepared here by a sequence of evolutions.

\subsection{Non-stabilizer AME$(4,4)$ state}\label{app:latin44}

The Latin-square state of Ref.~\cite{HontarenkoLatin2026} is
\begin{equation}
\ket{\psi_{\rm Latin}}=\frac14\sum_{i,j=0}^{3}\ket{i,j}\otimes\ket{\Phi_{ij}} ,
\label{eq:latin}
\end{equation}
where $\ket{i,j}$ are basis states of the parties $A$, $B$ and the cell $\ket{\Phi_{ij}}$ is a state of the parties $C$, $D$. 
This non-stabilizer state is based on the SudoQ design of maximal cardinality constructed in Ref. \cite{Paczos2021}.

With $\ket{k,l}$ the basis states of $C$ and $D$, the sixteen cells are
\begin{equation}
\begin{aligned}
&\ket{\Phi_{00}}=\ket{0,0},\quad \ket{\Phi_{01}}=\ket{1,2},\\
&\ket{\Phi_{10}}=\ket{2,3},\quad \ket{\Phi_{11}}=\ket{3,1},\\
&\ket{\Phi_{02}},\ket{\Phi_{03}}=\tfrac{1}{\sqrt2}\big(\ket{2,1}\pm\ket{3,3}\big),\\
&\ket{\Phi_{12}},\ket{\Phi_{13}}=\tfrac{1}{\sqrt2}\big(\ket{0,2}\mp\ket{1,0}\big),\\
&\ket{\Phi_{20}},\ket{\Phi_{30}}=\tfrac{1}{\sqrt2}\big(\ket{1,1}\pm\ket{3,2}\big),\\
&\ket{\Phi_{21}},\ket{\Phi_{31}}=\tfrac{1}{\sqrt2}\big(\ket{0,3}\mp\ket{2,0}\big),\\
&\ket{\Phi_{22}}=\tfrac12\big(-\ket{0,1}-\ket{1,3}-\ket{2,2}+\ket{3,0}\big),\\
&\ket{\Phi_{23}}=\tfrac12\big(\ket{0,1}-\ket{1,3}+\ket{2,2}+\ket{3,0}\big),\\
&\ket{\Phi_{32}}=\tfrac12\big(\ket{0,1}+\ket{1,3}-\ket{2,2}+\ket{3,0}\big),\\
&\ket{\Phi_{33}}=\tfrac12\big(-\ket{0,1}+\ket{1,3}+\ket{2,2}+\ket{3,0}\big),
\end{aligned}
\label{eq:latincells}
\end{equation}
where the upper sign belongs to the first cell of each pair. These are the cells of Fig.~2 of Ref.~\cite{HontarenkoLatin2026}, whose letters $A,B,C,D$ for the levels of the fourth party we write as $0,1,2,3$. The sixteen cells are orthonormal, so the reduced state of $A$ and $B$ is maximally mixed, and the reduced states of the other two pairs of parties are maximally mixed as well, which makes $\ket{\psi_{\rm Latin}}$ an AME state with 36 nonzero amplitudes.

\subsection{AME$(4,6)$ state}\label{app:latin46}

AME$(4,6)$ states were first constructed with complex amplitudes as solutions of the quantum problem of 36 officers~\cite{Rather2022}, and two-unitary complex Hadamard matrices of order 36 give further examples~\cite{Bruzda2024}. Real AME$(4,6)$ states have been found recently~\cite{BruzdaGeomagic2026,Pandya2026,HontarenkoLatin2026}. Reference~\cite{HontarenkoLatin2026} constructs one of them from a quantum Latin square of order six,
\begin{equation}
\ket{\psi_{96}}=\frac16\sum_{i,j=0}^{5}\ket{i,j}\otimes\ket{\Phi_{ij}} ,
\label{eq:latin46}
\end{equation}
where the 36 orthonormal cells $\ket{\Phi_{ij}}$ are real states of $C$ and $D$ listed in Table~2 of Ref.~\cite{HontarenkoLatin2026}. Each has two terms of amplitude $\pm1/\sqrt2$ or four of amplitude $\pm\tfrac12$. The second moment $m_2=108$ is the same for every ordering of the parties and equals $6^8$ times the Klein contraction of Ref.~\cite{HontarenkoLatin2026}. The fourth moment is $m_4=264$, $2511/8$ and $2487/8$ when $A$ is paired with $B$, $C$ and $D$.

A single diagonal evolution of the equal superposition cannot give $\ket{\psi_{96}}$ itself (Sec.~\ref{sec:ame46}). A diagonal evolution can reach a different AME$(4,6)$ state if its amplitudes give a two-unitary complex Hadamard matrix of order $36$~\cite{Bruzda2024}.

\begin{figure}[t]
\centering
{\small (a)}\\[0.5mm]
\includegraphics[width=\columnwidth]{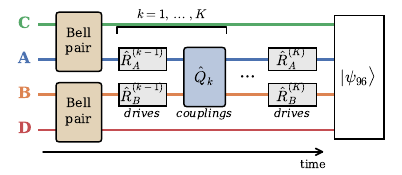}\\[1.5mm]
{\small (b)}\\[0.5mm]
\includegraphics[width=\columnwidth]{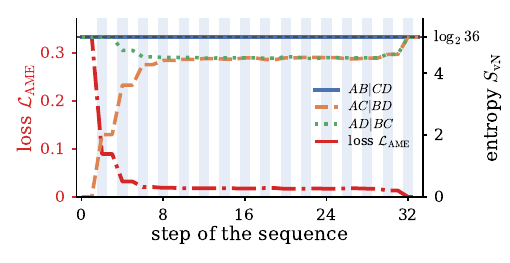}
\caption{The Latin-square AME$(4,6)$ state is prepared by a sequence of evolutions, and of the sequences found for the three encodings of a party, the one for a single six-level qudit uses the fewest coupling evolutions. The maximally entangled pairs $\ket{\Omega_6}_{AC}\ket{\Omega_6}_{BD}$ of Eq.~\eqref{eq:ame46pairs} are prepared first, by the evolution of Eq.~\eqref{eq:ame46bell} followed by a basis rotation of $C$ and $D$, and for three qubits per party by the qubit sequence of Appendix~\ref{app:ame46qubits}. On $A$ and $B$, drive layers $\hat R^{(k)}_A\otimes\hat R^{(k)}_B$ alternate with coupling evolutions $\hat Q_k$ as in Eq.~\eqref{eq:ame46sequence}. The number of coupling evolutions is $K=16$ for a six-level qudit, $K=48$ for a qubit and a qutrit, and $K=195$ for three qubits per party. (a) Scheme of the preparation. (b) Loss $\Lame$ (red, left axis) and entanglement entropy $S_{\rm vN}$ of the three balanced cuts (right axis) after each step of the six-level sequence, starting from the maximally entangled pairs. Shaded steps are coupling evolutions, the others drive layers, which act on each party separately and leave the entropies unchanged. The cut $AB|CD$ stays maximally entangled throughout, and the cuts $AC|BD$ and $AD|BC$ reach the maximum $\log_2 36$ together at the last step, where the loss vanishes.}
\label{fig:ame46protocol}
\end{figure}

For $\ket{\psi_{96}}$ we first create maximally entangled pairs $AC$ and $BD$. Equation~\eqref{eq:Udef} gives the unitary $\hcU_{96}$ on $A$ and $B$ that converts their product into the target,
\begin{equation}
\begin{aligned}
\ket{\psi_{96}}&=(\hcU_{96}\otimes\hone_{CD})\ket{\Omega_6}_{AC}\ket{\Omega_6}_{BD},\\
\ket{\Omega_6}&=\frac{1}{\sqrt 6}\sum_{a=0}^{5}\ket a\ket a .
\end{aligned}
\label{eq:ame46pairs}
\end{equation}
Equation~\eqref{eq:ame46pairs} is the dimension-six version of Eq.~\eqref{eq:casasstate}. Figure~\ref{fig:ame46protocol} shows the two pair preparations and the controls that implement $\hcU_{96}$.

We label the six basis levels by $a=3x+y$, where $x\in\{0,1\}$ is the qubit index and $y\in\{0,1,2\}$ is the qutrit index. The operators $\hZ$ and $\hat n$ have eigenvalues $(-1)^x$ and $y$ on level $a$. In the qubit-and-qutrit implementation each pair starts in the equal superposition of its levels, and $AC$ and $BD$ evolve simultaneously under
\begin{equation}
\tau_{\rm p}\hat H_{\rm p}
 =\sum_{(P,Q)\in\{(A,C),(B,D)\}}
 \left(\frac{\pi}{4}\hZ_P\hZ_Q+\frac{2\pi}{3}\hat n_P\hat n_Q\right),
\label{eq:ame46bell}
\end{equation}
where $\tau_{\rm p}$ is the evolution duration and we set $\hbar=1$. The amplitudes on each pair factor into the Ising phase of the qubits and the density--density phase of the qutrits. A basis rotation on $C$ and $D$ brings the pairs to the form of Eq.~\eqref{eq:ame46pairs} (Sec.~\ref{app:ame46qubits}).

In the qubit-and-qutrit implementation the qubit and the qutrit of each party are separate sites. The operation on $A$ and $B$ alternates two kinds of time-independent evolution. A drive layer applies simultaneous single-site drives,
\begin{equation}
\begin{aligned}
\hat R^{(k)}&=e^{-i\hat h^{(k)}},\\
\hat h^{(k)}&=\sum_{P=A,B}\Big(\sum_{\alpha=x,y,z}b^{(k)}_{P\alpha}\,\hat\sigma^{\alpha}_{P}+\sum_{j=1}^{8}c^{(k)}_{Pj}\,\hat\lambda^{j}_{P}\Big),
\end{aligned}
\label{eq:ame46drive}
\end{equation}
where $\hat\sigma^\alpha_P$ are the Pauli operators of the qubit and $\hat\lambda^j_P$ the Gell-Mann operators of the qutrit of party $P$, and the coefficients are integrated drive strengths. A coupling evolution applies diagonal two-site couplings,
\begin{equation}
\begin{aligned}
\hat Q_k&=e^{-i\hat G_k},\qquad \hat O_\mu\in\{\hZ,\hat n,\hat n^2\},\\
\hat G_k&=\sum_{\mu,\nu}J^{(k)}_{\mu\nu}\hat O_{\mu,A}\hat O_{\nu,B}
          +\sum_{P=A,B}\sum_{p=1}^{2}g_{Pp}^{(k)}\hZ_P\hat n_P^p ,
\end{aligned}
\label{eq:ame46quench}
\end{equation}
where $J^{(k)}_{\mu\nu}$ couple sites of different parties and $g_{Pp}^{(k)}$ couple the qubit and the qutrit of one party. These are Ising, dispersive and density--density couplings of the same type as in Eqs.~\eqref{eq:quench} and \eqref{eq:nn}, including products with the squared level number $\hat n^2$. For an evolution of duration $\tau_k$ under a physical Hamiltonian $\hat H_k$, the integrated couplings are $\hat G_k=\tau_k\hat H_k$. Starting from the pairs of Eq.~\eqref{eq:ame46pairs}, the operation on $A$ and $B$ is the product of $2K+1$ evolutions,
\begin{equation}
\begin{aligned}
\hat U&=e^{-i\hat h^{(K)}}e^{-i\hat G_K}\,e^{-i\hat h^{(K-1)}}\cdots e^{-i\hat h^{(1)}}e^{-i\hat G_1}\,e^{-i\hat h^{(0)}}\\
&=e^{i\chi}\hcU_{96},
\end{aligned}
\label{eq:ame46sequence}
\end{equation}
where $\chi$ is a global phase. With $K=48$ evolutions the qubit-and-qutrit sequence reaches the target to numerical precision (Sec.~\ref{app:ame46qubits}).

For one six-level qudit per party, a resonant drive couples neighbouring levels $a$ and $a+1$ through the evolution
\begin{equation}
\hat P_a(\theta,\varphi)
=\exp\!\left[-i\theta\left(e^{i\varphi}\ket{a+1}\!\bra a+\mathrm{H.c.}\right)\right],
\label{eq:ame46pulse}
\end{equation}
where $\theta$ is the integrated drive strength and $\varphi$ is the drive phase. The drive layer on party $P$ is then a product of at most fifteen such two-level rotations after a layer of level phases,
\begin{equation}
\hat R^{(k)}_P=\hat P_{a_L}(\theta_L,\varphi_L)\cdots\hat P_{a_1}(\theta_1,\varphi_1)\,
\exp\!\Big(i\sum_{a=0}^{5}\delta^{(k)}_{P,a}\ket a\!\bra a\Big),
\label{eq:ame46quditrotation}
\end{equation}
and $\hat R^{(k)}=\hat R^{(k)}_A\otimes\hat R^{(k)}_B$ replaces $e^{-i\hat h^{(k)}}$ in Eq.~\eqref{eq:ame46sequence}. The evolutions have the form of Eq.~\eqref{eq:ame46quench} with $g_{Pp}^{(k)}=0$ and the larger operator set $\{\hZ,\hat n,\hat n^2,\hZ\hat n,\hZ\hat n^2\}$ in the products $\hat O_{\mu,A}\hat O_{\nu,B}$. For two six-level devices these are level-dependent energy shifts. If the devices are split into qubits and qutrits, a product such as $\hZ_A\hat n_A\hat n_B$ involves three sites. With $K=16$ evolutions the qudit sequence reaches the target to numerical precision. Figure~\ref{fig:ame46protocol}(b) follows the entanglement of the three balanced cuts through this sequence. All parameters of the sequence are given in the repository listed under Data availability.

The $K$ coupling evolutions of Eq.~\eqref{eq:ame46sequence} alternate with $K+1$ drive layers, 17 for the six-level qudit sequence and 49 for the qubit-and-qutrit sequence, which restricts every term to two physical sites. With three qubits per party the qubit-and-qutrit sequence is written in single-qubit drives and Ising evolutions (Sec.~\ref{app:ame46qubits}). Encoding the qutrit in two qubits adds evolutions for the rotations between qutrit levels, and the qubit sequence uses 195 evolutions on $A$ and $B$ after two for the maximally entangled pairs.

\begin{table*}[t]
\caption{Couplings $J_{ij}$ of Eq.~\eqref{eq:quench} of the states of this paper, with the norms of Sec.~\ref{sec:norm}. For the qutrit states AME$(n,3)$ the column gives the couplings $K_{ij}$ of Eq.~\eqref{eq:nn}. Qubits and qutrits are numbered as in the figures. Every qubit or qutrit is a party for dimension two and three, and the parties are consecutive pairs of qubits for dimension four and consecutive triples for dimension eight. All couplings that are not listed vanish, and a free coupling can take any value. The AME$(4,8)$ graph state is that of Fig.~\ref{fig:family48} without the free coupling. The column $\|\hH\|_2$ gives the normalised Hilbert--Schmidt norm of Eq.~\eqref{eq:norm2} of $\hH=\sum_{i<j}J_{ij}\hZ_i\hZ_j$, in units of $J_0$, with free couplings set to zero. A free coupling $J$ adds $J^2$ to $\|\hH\|_2^2$. For the qutrit states it refers to the two-body part $\sum_{i<j}K_{ij}\hat s_i\hat s_j$ of Sec.~\ref{sec:norm}, in units of $K_0$. The action of Eq.~\eqref{eq:action} is $\mathcal A=\frac{\pi}{4}\|\hH\|_2$ for qubits and $\frac{2\pi}{3}\|\hH\|_2$ for qutrits.}
\label{tab:couplings}
\begin{ruledtabular}
\begin{tabular}{lclp{0.45\textwidth}}
state & $\|\hH\|_2$ & coupling $J_{ij}$ & pairs $(i,j)$ \\
\colrule
AME$(4,4)$, Fig.~\ref{fig:graphs} & $3.46$ & $1$ & $(1,4)$, $(1,8)$, $(2,4)$, $(2,5)$, $(3,5)$, $(3,6)$, $(5,7)$ \\
 & & $-1$ & $(3,7)$, $(4,8)$, $(5,8)$, $(6,8)$ \\
 & & $\tfrac12$ & $(1,7)$ \\
 & & $-\tfrac12$ & $(1,6)$, $(4,6)$, $(4,7)$ \\
\colrule
AME$(4,4)$, Fig.~\ref{fig:latin44} & $3.19$ & $1$ & $(1,7)$, $(2,6)$, $(4,7)$ \\
 & & $-1$ & $(1,3)$, $(1,6)$, $(2,8)$, $(3,5)$, $(4,6)$, $(4,8)$, $(5,8)$ \\
 & & $\tfrac14$ & $(3,6)$, $(3,7)$ \\
 & & $-\tfrac14$ & $(6,7)$ \\
\colrule
AME$(4,4)$, Fig.~\ref{fig:families} & $3.00$ & $1$ & $(1,3)$, $(1,8)$, $(2,4)$, $(2,6)$, $(3,6)$, $(3,7)$, $(4,5)$, $(4,8)$, $(5,7)$ \\
 & & free & $(2,5)$ \\
\colrule
AME$(4,8)$, Fig.~\ref{fig:family48} & $3.61$ & $1$ & $(1,4)$, $(1,11)$, $(2,5)$, $(2,7)$, $(2,12)$, $(3,8)$, $(3,10)$, $(4,7)$, $(5,9)$, $(5,11)$, $(6,8)$, $(6,12)$, $(9,10)$ \\
 & & free & $(1,6)$ \\
\colrule
AME$(4,8)$, Fig.~\ref{fig:ising832} & $4.21$ & $1$ & $(2,4)$, $(3,5)$, $(3,10)$, $(4,8)$, $(5,9)$, $(5,12)$, $(7,12)$, $(8,11)$ \\
 & & $-1$ & $(1,7)$, $(1,9)$, $(1,10)$, $(2,11)$, $(2,12)$, $(5,10)$, $(6,8)$, $(6,10)$ \\
 & & $\tfrac12$ & $(9,11)$ \\
 & & $-\tfrac12$ & $(3,4)$, $(3,11)$, $(3,12)$, $(4,9)$, $(6,9)$ \\
 & & $\tfrac14$ & $(4,11)$, $(6,12)$ \\
 & & $-\tfrac14$ & $(4,12)$, $(6,11)$ \\
\colrule
AME$(5,2)$ graph state & $2.24$ & $1$ & $(1,2)$, $(1,5)$, $(2,3)$, $(3,4)$, $(4,5)$ \\
\colrule
AME$(6,2)$ graph state & $3.00$ & $1$ & $(1,3)$, $(1,4)$, $(1,5)$, $(2,4)$, $(2,5)$, $(2,6)$, $(3,5)$, $(3,6)$, $(4,6)$ \\
\colrule
AME$(4,4)$ ring graph & $2.83$ & $1$ & $(1,4)$, $(1,5)$, $(2,3)$, $(2,7)$, $(3,6)$, $(4,8)$, $(5,7)$, $(6,8)$ \\
\colrule
AME$(5,4)$, Fig.~\ref{fig:graphs5664}(a) & $4.21$ & $1$ & $(1,6)$, $(1,8)$, $(2,5)$, $(4,7)$, $(5,9)$, $(5,10)$, $(6,8)$, $(6,10)$, $(7,9)$ \\
 & & $-1$ & $(2,3)$, $(2,7)$, $(3,5)$, $(3,7)$, $(4,9)$, $(5,7)$, $(8,10)$ \\
 & & $\tfrac12$ & $(2,6)$, $(3,6)$, $(3,8)$, $(4,6)$, $(4,10)$ \\
 & & $-\tfrac12$ & $(2,8)$, $(3,10)$ \\
 & & free & $(1,3)$ \\
\colrule
AME$(6,4)$, Fig.~\ref{fig:graphs5664}(b) & $5.20$ & $1$ & $(1,5)$, $(1,11)$, $(1,12)$, $(2,5)$, $(2,6)$, $(2,12)$, $(3,5)$, $(3,8)$, $(4,6)$, $(4,12)$, $(5,11)$, $(6,10)$, $(6,11)$, $(7,10)$, $(7,11)$ \\
 & & $-1$ & $(1,4)$, $(1,7)$, $(1,9)$, $(2,3)$, $(2,4)$, $(2,11)$, $(4,9)$, $(5,8)$, $(6,8)$, $(8,9)$, $(8,11)$, $(10,12)$ \\
 & & free & $(2,8)$, $(10,11)$ \\
\colrule
AME$(3,3)$ & $0.94$ & $1$ & $(1,2)$, $(2,3)$ \\
\colrule
AME$(4,3)$, Fig.~\ref{fig:qutrits}(a) & $1.33$ & $1$ & $(1,4)$ \\
 & & $-1$ & $(1,2)$, $(2,3)$, $(3,4)$ \\
 & & free & $(1,3)$ \\
\colrule
AME$(5,3)$, Fig.~\ref{fig:qutrits}(b) & $1.76$ & $1$ & $(1,3)$, $(1,5)$, $(2,4)$, $(3,5)$, $(4,5)$ \\
 & & $-1$ & $(2,3)$, $(3,4)$ \\
 & & free & $(1,2)$ \\
\colrule
AME$(6,3)$, Fig.~\ref{fig:qutrits}(c) & $2.21$ & $1$ & $(1,3)$, $(2,3)$, $(4,5)$, $(5,6)$ \\
 & & $-1$ & $(1,2)$, $(1,6)$, $(2,4)$, $(3,4)$, $(3,5)$, $(3,6)$, $(4,6)$ \\
\end{tabular}
\end{ruledtabular}
\end{table*}

\subsection{Maximally entangled pairs and qubit encoding}\label{app:ame46qubits}

After the pair evolution of Eq.~\eqref{eq:ame46bell}, each pair is maximally entangled up to a rotation of the partner party. A drive along $\hX$ of area $\pi/4$ on the partner qubit and the discrete Fourier transform on the partner qutrit bring the pair to the form of Eq.~\eqref{eq:ame46pairs}.

With three qubits per party, the qutrit levels $0$, $1$ and $2$ are stored as $00$, $01$ and $10$ of two qubits. The operators $\hat n$ and $\hat n^2$ are then linear in the $\hZ$ operators of these qubits, so every diagonal evolution of the qubit-and-qutrit sequence becomes an Ising evolution together with single-qubit $\hZ$ phases. Every qutrit rotation is a product of rotations between level 0 and level 1 or 2, and each of these flips one of the two qubits only when the other one is in a given state. The generator of each such rotation is a drive on the flipped qubit multiplied by $(\hone\pm\hZ)/2$ of the other qubit, which requires one Ising evolution and single-qubit rotations. The pairs are prepared from $\ket{000}$ by single-qubit rotations, one such conditional rotation and controlled flips onto the partner qubits, each again one Ising evolution with single-qubit rotations. The complete qubit sequence has 197 Ising evolutions. Multiplied out, every sequence of this appendix reproduces $\hcU_{96}$ up to a global phase and gives $\ket{\psi_{96}}$ with all balanced reductions maximally mixed to numerical precision, with no population outside the encoded levels.

\section{Couplings of the states}\label{app:couplings}
Table~\ref{tab:couplings} lists the couplings of the single-evolution states shown in the figures and of the graph states, so that each state can be rebuilt from Eq.~\eqref{eq:quench}, or from Eq.~\eqref{eq:nn} for qutrits.

\bibliography{references}

\begin{thebibliography}{70}%
\makeatletter
\providecommand \@ifxundefined [1]{%
 \@ifx{#1\undefined}
}%
\providecommand \@ifnum [1]{%
 \ifnum #1\expandafter \@firstoftwo
 \else \expandafter \@secondoftwo
 \fi
}%
\providecommand \@ifx [1]{%
 \ifx #1\expandafter \@firstoftwo
 \else \expandafter \@secondoftwo
 \fi
}%
\providecommand \natexlab [1]{#1}%
\providecommand \enquote  [1]{``#1''}%
\providecommand \bibnamefont  [1]{#1}%
\providecommand \bibfnamefont [1]{#1}%
\providecommand \citenamefont [1]{#1}%
\providecommand \href@noop [0]{\@secondoftwo}%
\providecommand \href [0]{\begingroup \@sanitize@url \@href}%
\providecommand \@href[1]{\@@startlink{#1}\@@href}%
\providecommand \@@href[1]{\endgroup#1\@@endlink}%
\providecommand \@sanitize@url [0]{\catcode `\\12\catcode `\$12\catcode
  `\&12\catcode `\#12\catcode `\^12\catcode `\_12\catcode `\%12\relax}%
\providecommand \@@startlink[1]{}%
\providecommand \@@endlink[0]{}%
\providecommand \url  [0]{\begingroup\@sanitize@url \@url }%
\providecommand \@url [1]{\endgroup\@href {#1}{\urlprefix }}%
\providecommand \urlprefix  [0]{URL }%
\providecommand \Eprint [0]{\href }%
\providecommand \doibase [0]{https://doi.org/}%
\providecommand \selectlanguage [0]{\@gobble}%
\providecommand \bibinfo  [0]{\@secondoftwo}%
\providecommand \bibfield  [0]{\@secondoftwo}%
\providecommand \translation [1]{[#1]}%
\providecommand \BibitemOpen [0]{}%
\providecommand \bibitemStop [0]{}%
\providecommand \bibitemNoStop [0]{.\EOS\space}%
\providecommand \EOS [0]{\spacefactor3000\relax}%
\providecommand \BibitemShut  [1]{\csname bibitem#1\endcsname}%
\let\auto@bib@innerbib\@empty
\bibitem [{\citenamefont {Helwig}\ \emph {et~al.}(2012)\citenamefont {Helwig},
  \citenamefont {Cui}, \citenamefont {Latorre}, \citenamefont {Riera},\ and\
  \citenamefont {Lo}}]{Helwig2012}%
  \BibitemOpen
  \bibfield  {author} {\bibinfo {author} {\bibfnamefont {W.}~\bibnamefont
  {Helwig}}, \bibinfo {author} {\bibfnamefont {W.}~\bibnamefont {Cui}},
  \bibinfo {author} {\bibfnamefont {J.~I.}\ \bibnamefont {Latorre}}, \bibinfo
  {author} {\bibfnamefont {A.}~\bibnamefont {Riera}},\ and\ \bibinfo {author}
  {\bibfnamefont {H.-K.}\ \bibnamefont {Lo}},\ }\bibfield  {title} {\bibinfo
  {title} {Absolute maximal entanglement and quantum secret sharing},\ }\href
  {https://doi.org/10.1103/PhysRevA.86.052335} {\bibfield  {journal} {\bibinfo
  {journal} {Phys. Rev. A}\ }\textbf {\bibinfo {volume} {86}},\ \bibinfo
  {pages} {052335} (\bibinfo {year} {2012})}\BibitemShut {NoStop}%
\bibitem [{\citenamefont {Goyeneche}\ \emph {et~al.}(2015)\citenamefont
  {Goyeneche}, \citenamefont {Alsina}, \citenamefont {Latorre}, \citenamefont
  {Riera},\ and\ \citenamefont {{\.Z}yczkowski}}]{Goyeneche2015}%
  \BibitemOpen
  \bibfield  {author} {\bibinfo {author} {\bibfnamefont {D.}~\bibnamefont
  {Goyeneche}}, \bibinfo {author} {\bibfnamefont {D.}~\bibnamefont {Alsina}},
  \bibinfo {author} {\bibfnamefont {J.~I.}\ \bibnamefont {Latorre}}, \bibinfo
  {author} {\bibfnamefont {A.}~\bibnamefont {Riera}},\ and\ \bibinfo {author}
  {\bibfnamefont {K.}~\bibnamefont {{\.Z}yczkowski}},\ }\bibfield  {title}
  {\bibinfo {title} {Absolutely maximally entangled states, combinatorial
  designs, and multiunitary matrices},\ }\href
  {https://doi.org/10.1103/PhysRevA.92.032316} {\bibfield  {journal} {\bibinfo
  {journal} {Phys. Rev. A}\ }\textbf {\bibinfo {volume} {92}},\ \bibinfo
  {pages} {032316} (\bibinfo {year} {2015})}\BibitemShut {NoStop}%
\bibitem [{\citenamefont {Scott}(2004)}]{Scott2004}%
  \BibitemOpen
  \bibfield  {author} {\bibinfo {author} {\bibfnamefont {A.~J.}\ \bibnamefont
  {Scott}},\ }\bibfield  {title} {\bibinfo {title} {Multipartite entanglement,
  quantum-error-correcting codes, and entangling power of quantum evolutions},\
  }\href {https://doi.org/10.1103/PhysRevA.69.052330} {\bibfield  {journal}
  {\bibinfo  {journal} {Phys. Rev. A}\ }\textbf {\bibinfo {volume} {69}},\
  \bibinfo {pages} {052330} (\bibinfo {year} {2004})}\BibitemShut {NoStop}%
\bibitem [{\citenamefont {Pastawski}\ \emph {et~al.}(2015)\citenamefont
  {Pastawski}, \citenamefont {Yoshida}, \citenamefont {Harlow},\ and\
  \citenamefont {Preskill}}]{Pastawski2015}%
  \BibitemOpen
  \bibfield  {author} {\bibinfo {author} {\bibfnamefont {F.}~\bibnamefont
  {Pastawski}}, \bibinfo {author} {\bibfnamefont {B.}~\bibnamefont {Yoshida}},
  \bibinfo {author} {\bibfnamefont {D.}~\bibnamefont {Harlow}},\ and\ \bibinfo
  {author} {\bibfnamefont {J.}~\bibnamefont {Preskill}},\ }\bibfield  {title}
  {\bibinfo {title} {Holographic quantum error-correcting codes: toy models for
  the bulk/boundary correspondence},\ }\href
  {https://doi.org/10.1007/JHEP06(2015)149} {\bibfield  {journal} {\bibinfo
  {journal} {J. High Energy Phys.}\ }\textbf {\bibinfo {volume} {2015}}\bibinfo
   {number} { (6)},\ \bibinfo {pages} {149}}\BibitemShut {NoStop}%
\bibitem [{\citenamefont {Higuchi}\ and\ \citenamefont
  {Sudbery}(2000)}]{Higuchi2000}%
  \BibitemOpen
\bibfield  {number} {  }\bibfield  {author} {\bibinfo {author} {\bibfnamefont
  {A.}~\bibnamefont {Higuchi}}\ and\ \bibinfo {author} {\bibfnamefont
  {A.}~\bibnamefont {Sudbery}},\ }\bibfield  {title} {\bibinfo {title} {How
  entangled can two couples get?},\ }\href
  {https://doi.org/10.1016/S0375-9601(00)00480-1} {\bibfield  {journal}
  {\bibinfo  {journal} {Phys. Lett. A}\ }\textbf {\bibinfo {volume} {273}},\
  \bibinfo {pages} {213} (\bibinfo {year} {2000})}\BibitemShut {NoStop}%
\bibitem [{\citenamefont {Huber}\ \emph {et~al.}(2017)\citenamefont {Huber},
  \citenamefont {G{\"u}hne},\ and\ \citenamefont {Siewert}}]{Huber2017}%
  \BibitemOpen
  \bibfield  {author} {\bibinfo {author} {\bibfnamefont {F.}~\bibnamefont
  {Huber}}, \bibinfo {author} {\bibfnamefont {O.}~\bibnamefont {G{\"u}hne}},\
  and\ \bibinfo {author} {\bibfnamefont {J.}~\bibnamefont {Siewert}},\
  }\bibfield  {title} {\bibinfo {title} {Absolutely maximally entangled states
  of seven qubits do not exist},\ }\href
  {https://doi.org/10.1103/PhysRevLett.118.200502} {\bibfield  {journal}
  {\bibinfo  {journal} {Phys. Rev. Lett.}\ }\textbf {\bibinfo {volume} {118}},\
  \bibinfo {pages} {200502} (\bibinfo {year} {2017})}\BibitemShut {NoStop}%
\bibitem [{\citenamefont {Rather}\ \emph {et~al.}(2023)\citenamefont {Rather},
  \citenamefont {Ramadas}, \citenamefont {Kodiyalam},\ and\ \citenamefont
  {Lakshminarayan}}]{Rather2023}%
  \BibitemOpen
  \bibfield  {author} {\bibinfo {author} {\bibfnamefont {S.~A.}\ \bibnamefont
  {Rather}}, \bibinfo {author} {\bibfnamefont {N.}~\bibnamefont {Ramadas}},
  \bibinfo {author} {\bibfnamefont {V.}~\bibnamefont {Kodiyalam}},\ and\
  \bibinfo {author} {\bibfnamefont {A.}~\bibnamefont {Lakshminarayan}},\
  }\bibfield  {title} {\bibinfo {title} {Absolutely maximally entangled state
  equivalence and the construction of infinite quantum solutions to the problem
  of 36 officers of {E}uler},\ }\href
  {https://doi.org/10.1103/PhysRevA.108.032412} {\bibfield  {journal} {\bibinfo
   {journal} {Phys. Rev. A}\ }\textbf {\bibinfo {volume} {108}},\ \bibinfo
  {pages} {032412} (\bibinfo {year} {2023})}\BibitemShut {NoStop}%
\bibitem [{\citenamefont {Ramadas}\ and\ \citenamefont
  {Lakshminarayan}(2025)}]{Ramadas2024}%
  \BibitemOpen
  \bibfield  {author} {\bibinfo {author} {\bibfnamefont {N.}~\bibnamefont
  {Ramadas}}\ and\ \bibinfo {author} {\bibfnamefont {A.}~\bibnamefont
  {Lakshminarayan}},\ }\bibfield  {title} {\bibinfo {title} {Local unitary
  equivalence of absolutely maximally entangled states constructed from
  orthogonal arrays},\ }\href {https://doi.org/10.1088/1751-8121/adbf75}
  {\bibfield  {journal} {\bibinfo  {journal} {J. Phys. A: Math. Theor.}\
  }\textbf {\bibinfo {volume} {58}},\ \bibinfo {pages} {125301} (\bibinfo
  {year} {2025})}\BibitemShut {NoStop}%
\bibitem [{\citenamefont {Cervera-Lierta}\ \emph {et~al.}(2019)\citenamefont
  {Cervera-Lierta}, \citenamefont {Latorre},\ and\ \citenamefont
  {Goyeneche}}]{CerveraLierta2019}%
  \BibitemOpen
  \bibfield  {author} {\bibinfo {author} {\bibfnamefont {A.}~\bibnamefont
  {Cervera-Lierta}}, \bibinfo {author} {\bibfnamefont {J.~I.}\ \bibnamefont
  {Latorre}},\ and\ \bibinfo {author} {\bibfnamefont {D.}~\bibnamefont
  {Goyeneche}},\ }\bibfield  {title} {\bibinfo {title} {Quantum circuits for
  maximally entangled states},\ }\href
  {https://doi.org/10.1103/PhysRevA.100.022342} {\bibfield  {journal} {\bibinfo
   {journal} {Phys. Rev. A}\ }\textbf {\bibinfo {volume} {100}},\ \bibinfo
  {pages} {022342} (\bibinfo {year} {2019})}\BibitemShut {NoStop}%
\bibitem [{\citenamefont {Casas}\ \emph {et~al.}(2026)\citenamefont {Casas},
  \citenamefont {Rajchel-Mieldzio{\'c}}, \citenamefont {Rather}, \citenamefont
  {P{\l}odzie{\'n}}, \citenamefont {Bruzda}, \citenamefont {Cervera-Lierta},\
  and\ \citenamefont {{\.Z}yczkowski}}]{Casas2026}%
  \BibitemOpen
  \bibfield  {author} {\bibinfo {author} {\bibfnamefont {B.}~\bibnamefont
  {Casas}}, \bibinfo {author} {\bibfnamefont {G.}~\bibnamefont
  {Rajchel-Mieldzio{\'c}}}, \bibinfo {author} {\bibfnamefont {S.~A.}\
  \bibnamefont {Rather}}, \bibinfo {author} {\bibfnamefont {M.}~\bibnamefont
  {P{\l}odzie{\'n}}}, \bibinfo {author} {\bibfnamefont {W.}~\bibnamefont
  {Bruzda}}, \bibinfo {author} {\bibfnamefont {A.}~\bibnamefont
  {Cervera-Lierta}},\ and\ \bibinfo {author} {\bibfnamefont {K.}~\bibnamefont
  {{\.Z}yczkowski}},\ }\bibfield  {title} {\bibinfo {title} {Quantum circuits
  for high-dimensional absolutely maximally entangled states},\ }\href
  {https://doi.org/10.1088/2058-9565/ae46d6} {\bibfield  {journal} {\bibinfo
  {journal} {Quantum Science and Technology}\ }\textbf {\bibinfo {volume}
  {11}},\ \bibinfo {pages} {025011} (\bibinfo {year} {2026})}\BibitemShut
  {NoStop}%
\bibitem [{\citenamefont {Chen}\ and\ \citenamefont {Zhou}(2016)}]{Chen2016}%
  \BibitemOpen
  \bibfield  {author} {\bibinfo {author} {\bibfnamefont {L.}~\bibnamefont
  {Chen}}\ and\ \bibinfo {author} {\bibfnamefont {D.~L.}\ \bibnamefont
  {Zhou}},\ }\bibfield  {title} {\bibinfo {title} {Graph states of prime-power
  dimension from generalized {CNOT} quantum circuit},\ }\href
  {https://doi.org/10.1038/srep27135} {\bibfield  {journal} {\bibinfo
  {journal} {Sci. Rep.}\ }\textbf {\bibinfo {volume} {6}},\ \bibinfo {pages}
  {27135} (\bibinfo {year} {2016})}\BibitemShut {NoStop}%
\bibitem [{\citenamefont {Rather}(2024)}]{Rather2024}%
  \BibitemOpen
  \bibfield  {author} {\bibinfo {author} {\bibfnamefont {S.~A.}\ \bibnamefont
  {Rather}},\ }\bibfield  {title} {\bibinfo {title} {Construction of perfect
  tensors using biunimodular vectors},\ }\href
  {https://doi.org/10.22331/q-2024-11-20-1528} {\bibfield  {journal} {\bibinfo
  {journal} {Quantum}\ }\textbf {\bibinfo {volume} {8}},\ \bibinfo {pages}
  {1528} (\bibinfo {year} {2024})}\BibitemShut {NoStop}%
\bibitem [{\citenamefont {Majidy}\ \emph {et~al.}(2025)\citenamefont {Majidy},
  \citenamefont {Hangleiter},\ and\ \citenamefont {Gullans}}]{Majidy2025}%
  \BibitemOpen
  \bibfield  {author} {\bibinfo {author} {\bibfnamefont {S.}~\bibnamefont
  {Majidy}}, \bibinfo {author} {\bibfnamefont {D.}~\bibnamefont {Hangleiter}},\
  and\ \bibinfo {author} {\bibfnamefont {M.~J.}\ \bibnamefont {Gullans}},\
  }\bibfield  {title} {\bibinfo {title} {Scalable and fault-tolerant
  preparation of encoded $k$-uniform states},\ }\href
  {https://doi.org/10.1103/cl23-c3jg} {\bibfield  {journal} {\bibinfo
  {journal} {Phys. Rev. A}\ }\textbf {\bibinfo {volume} {112}},\ \bibinfo
  {pages} {042409} (\bibinfo {year} {2025})}\BibitemShut {NoStop}%
\bibitem [{\citenamefont {Helwig}(2013)}]{Helwig2013}%
  \BibitemOpen
  \bibfield  {author} {\bibinfo {author} {\bibfnamefont {W.}~\bibnamefont
  {Helwig}},\ }\href {https://doi.org/10.48550/arXiv.1306.2879} {\bibinfo
  {title} {Absolutely maximally entangled qudit graph states}} (\bibinfo {year}
  {2013}),\ \bibinfo {note} {arXiv:1306.2879}\BibitemShut {NoStop}%
\bibitem [{\citenamefont {Raissi}\ \emph {et~al.}(2022)\citenamefont {Raissi},
  \citenamefont {Burchardt},\ and\ \citenamefont {Barnes}}]{Raissi2022}%
  \BibitemOpen
  \bibfield  {author} {\bibinfo {author} {\bibfnamefont {Z.}~\bibnamefont
  {Raissi}}, \bibinfo {author} {\bibfnamefont {A.}~\bibnamefont {Burchardt}},\
  and\ \bibinfo {author} {\bibfnamefont {E.}~\bibnamefont {Barnes}},\
  }\bibfield  {title} {\bibinfo {title} {General stabilizer approach for
  constructing highly entangled graph states},\ }\href
  {https://doi.org/10.1103/PhysRevA.106.062424} {\bibfield  {journal} {\bibinfo
   {journal} {Phys. Rev. A}\ }\textbf {\bibinfo {volume} {106}},\ \bibinfo
  {pages} {062424} (\bibinfo {year} {2022})}\BibitemShut {NoStop}%
\bibitem [{\citenamefont {Briegel}\ and\ \citenamefont
  {Raussendorf}(2001)}]{Briegel2001}%
  \BibitemOpen
  \bibfield  {author} {\bibinfo {author} {\bibfnamefont {H.~J.}\ \bibnamefont
  {Briegel}}\ and\ \bibinfo {author} {\bibfnamefont {R.}~\bibnamefont
  {Raussendorf}},\ }\bibfield  {title} {\bibinfo {title} {Persistent
  entanglement in arrays of interacting particles},\ }\href
  {https://doi.org/10.1103/PhysRevLett.86.910} {\bibfield  {journal} {\bibinfo
  {journal} {Phys. Rev. Lett.}\ }\textbf {\bibinfo {volume} {86}},\ \bibinfo
  {pages} {910} (\bibinfo {year} {2001})}\BibitemShut {NoStop}%
\bibitem [{\citenamefont {Hein}\ \emph {et~al.}(2004)\citenamefont {Hein},
  \citenamefont {Eisert},\ and\ \citenamefont {Briegel}}]{Hein2004}%
  \BibitemOpen
  \bibfield  {author} {\bibinfo {author} {\bibfnamefont {M.}~\bibnamefont
  {Hein}}, \bibinfo {author} {\bibfnamefont {J.}~\bibnamefont {Eisert}},\ and\
  \bibinfo {author} {\bibfnamefont {H.~J.}\ \bibnamefont {Briegel}},\
  }\bibfield  {title} {\bibinfo {title} {Multiparty entanglement in graph
  states},\ }\href {https://doi.org/10.1103/PhysRevA.69.062311} {\bibfield
  {journal} {\bibinfo  {journal} {Phys. Rev. A}\ }\textbf {\bibinfo {volume}
  {69}},\ \bibinfo {pages} {062311} (\bibinfo {year} {2004})}\BibitemShut
  {NoStop}%
\bibitem [{\citenamefont {Stricker}\ \emph {et~al.}(2022)\citenamefont
  {Stricker}, \citenamefont {Meth}, \citenamefont {Postler}, \citenamefont
  {Edmunds}, \citenamefont {Ferrie}, \citenamefont {Blatt}, \citenamefont
  {Schindler}, \citenamefont {Monz}, \citenamefont {Kueng},\ and\ \citenamefont
  {Ringbauer}}]{Stricker2022}%
  \BibitemOpen
  \bibfield  {author} {\bibinfo {author} {\bibfnamefont {R.}~\bibnamefont
  {Stricker}}, \bibinfo {author} {\bibfnamefont {M.}~\bibnamefont {Meth}},
  \bibinfo {author} {\bibfnamefont {L.}~\bibnamefont {Postler}}, \bibinfo
  {author} {\bibfnamefont {C.}~\bibnamefont {Edmunds}}, \bibinfo {author}
  {\bibfnamefont {C.}~\bibnamefont {Ferrie}}, \bibinfo {author} {\bibfnamefont
  {R.}~\bibnamefont {Blatt}}, \bibinfo {author} {\bibfnamefont
  {P.}~\bibnamefont {Schindler}}, \bibinfo {author} {\bibfnamefont
  {T.}~\bibnamefont {Monz}}, \bibinfo {author} {\bibfnamefont {R.}~\bibnamefont
  {Kueng}},\ and\ \bibinfo {author} {\bibfnamefont {M.}~\bibnamefont
  {Ringbauer}},\ }\bibfield  {title} {\bibinfo {title} {Experimental
  single-setting quantum state tomography},\ }\href
  {https://doi.org/10.1103/PRXQuantum.3.040310} {\bibfield  {journal} {\bibinfo
   {journal} {PRX Quantum}\ }\textbf {\bibinfo {volume} {3}},\ \bibinfo {pages}
  {040310} (\bibinfo {year} {2022})}\BibitemShut {NoStop}%
\bibitem [{\citenamefont {Li}\ \emph {et~al.}(2019)\citenamefont {Li},
  \citenamefont {Han}, \citenamefont {Qu}, \citenamefont {Huang}, \citenamefont
  {Long}, \citenamefont {Wan}, \citenamefont {Lu}, \citenamefont {Zeng},\ and\
  \citenamefont {Laflamme}}]{Li2019}%
  \BibitemOpen
  \bibfield  {author} {\bibinfo {author} {\bibfnamefont {K.}~\bibnamefont
  {Li}}, \bibinfo {author} {\bibfnamefont {M.}~\bibnamefont {Han}}, \bibinfo
  {author} {\bibfnamefont {D.}~\bibnamefont {Qu}}, \bibinfo {author}
  {\bibfnamefont {Z.}~\bibnamefont {Huang}}, \bibinfo {author} {\bibfnamefont
  {G.}~\bibnamefont {Long}}, \bibinfo {author} {\bibfnamefont {Y.}~\bibnamefont
  {Wan}}, \bibinfo {author} {\bibfnamefont {D.}~\bibnamefont {Lu}}, \bibinfo
  {author} {\bibfnamefont {B.}~\bibnamefont {Zeng}},\ and\ \bibinfo {author}
  {\bibfnamefont {R.}~\bibnamefont {Laflamme}},\ }\bibfield  {title} {\bibinfo
  {title} {Measuring holographic entanglement entropy on a quantum simulator},\
  }\href {https://doi.org/10.1038/s41534-019-0145-z} {\bibfield  {journal}
  {\bibinfo  {journal} {npj Quantum Inf.}\ }\textbf {\bibinfo {volume} {5}},\
  \bibinfo {pages} {30} (\bibinfo {year} {2019})}\BibitemShut {NoStop}%
\bibitem [{\citenamefont {Porras}\ and\ \citenamefont
  {Cirac}(2004)}]{PorrasCirac2004}%
  \BibitemOpen
  \bibfield  {author} {\bibinfo {author} {\bibfnamefont {D.}~\bibnamefont
  {Porras}}\ and\ \bibinfo {author} {\bibfnamefont {J.~I.}\ \bibnamefont
  {Cirac}},\ }\bibfield  {title} {\bibinfo {title} {Effective quantum spin
  systems with trapped ions},\ }\href
  {https://doi.org/10.1103/PhysRevLett.92.207901} {\bibfield  {journal}
  {\bibinfo  {journal} {Phys. Rev. Lett.}\ }\textbf {\bibinfo {volume} {92}},\
  \bibinfo {pages} {207901} (\bibinfo {year} {2004})}\BibitemShut {NoStop}%
\bibitem [{\citenamefont {Kim}\ \emph {et~al.}(2009)\citenamefont {Kim},
  \citenamefont {Chang}, \citenamefont {Islam}, \citenamefont {Korenblit},
  \citenamefont {Duan},\ and\ \citenamefont {Monroe}}]{Kim2009}%
  \BibitemOpen
  \bibfield  {author} {\bibinfo {author} {\bibfnamefont {K.}~\bibnamefont
  {Kim}}, \bibinfo {author} {\bibfnamefont {M.-S.}\ \bibnamefont {Chang}},
  \bibinfo {author} {\bibfnamefont {R.}~\bibnamefont {Islam}}, \bibinfo
  {author} {\bibfnamefont {S.}~\bibnamefont {Korenblit}}, \bibinfo {author}
  {\bibfnamefont {L.-M.}\ \bibnamefont {Duan}},\ and\ \bibinfo {author}
  {\bibfnamefont {C.}~\bibnamefont {Monroe}},\ }\bibfield  {title} {\bibinfo
  {title} {Entanglement and tunable spin-spin couplings between trapped ions
  using multiple transverse modes},\ }\href
  {https://doi.org/10.1103/PhysRevLett.103.120502} {\bibfield  {journal}
  {\bibinfo  {journal} {Phys. Rev. Lett.}\ }\textbf {\bibinfo {volume} {103}},\
  \bibinfo {pages} {120502} (\bibinfo {year} {2009})}\BibitemShut {NoStop}%
\bibitem [{\citenamefont {Monroe}\ \emph {et~al.}(2021)\citenamefont {Monroe},
  \citenamefont {Campbell}, \citenamefont {Duan}, \citenamefont {Gong},
  \citenamefont {Gorshkov}, \citenamefont {Hess}, \citenamefont {Islam},
  \citenamefont {Kim}, \citenamefont {Linke}, \citenamefont {Pagano},
  \citenamefont {Richerme}, \citenamefont {Senko},\ and\ \citenamefont
  {Yao}}]{Monroe2021}%
  \BibitemOpen
  \bibfield  {author} {\bibinfo {author} {\bibfnamefont {C.}~\bibnamefont
  {Monroe}}, \bibinfo {author} {\bibfnamefont {W.~C.}\ \bibnamefont
  {Campbell}}, \bibinfo {author} {\bibfnamefont {L.-M.}\ \bibnamefont {Duan}},
  \bibinfo {author} {\bibfnamefont {Z.-X.}\ \bibnamefont {Gong}}, \bibinfo
  {author} {\bibfnamefont {A.~V.}\ \bibnamefont {Gorshkov}}, \bibinfo {author}
  {\bibfnamefont {P.~W.}\ \bibnamefont {Hess}}, \bibinfo {author}
  {\bibfnamefont {R.}~\bibnamefont {Islam}}, \bibinfo {author} {\bibfnamefont
  {K.}~\bibnamefont {Kim}}, \bibinfo {author} {\bibfnamefont {N.~M.}\
  \bibnamefont {Linke}}, \bibinfo {author} {\bibfnamefont {G.}~\bibnamefont
  {Pagano}}, \bibinfo {author} {\bibfnamefont {P.}~\bibnamefont {Richerme}},
  \bibinfo {author} {\bibfnamefont {C.}~\bibnamefont {Senko}},\ and\ \bibinfo
  {author} {\bibfnamefont {N.~Y.}\ \bibnamefont {Yao}},\ }\bibfield  {title}
  {\bibinfo {title} {Programmable quantum simulations of spin systems with
  trapped ions},\ }\href {https://doi.org/10.1103/RevModPhys.93.025001}
  {\bibfield  {journal} {\bibinfo  {journal} {Rev. Mod. Phys.}\ }\textbf
  {\bibinfo {volume} {93}},\ \bibinfo {pages} {025001} (\bibinfo {year}
  {2021})}\BibitemShut {NoStop}%
\bibitem [{\citenamefont {Lu}\ \emph {et~al.}(2025)\citenamefont {Lu},
  \citenamefont {Chen}, \citenamefont {Zhang}, \citenamefont {Zhang},
  \citenamefont {Zhang}, \citenamefont {Zhang},\ and\ \citenamefont
  {Kim}}]{Lu2025}%
  \BibitemOpen
  \bibfield  {author} {\bibinfo {author} {\bibfnamefont {Y.}~\bibnamefont
  {Lu}}, \bibinfo {author} {\bibfnamefont {W.}~\bibnamefont {Chen}}, \bibinfo
  {author} {\bibfnamefont {S.}~\bibnamefont {Zhang}}, \bibinfo {author}
  {\bibfnamefont {K.}~\bibnamefont {Zhang}}, \bibinfo {author} {\bibfnamefont
  {J.}~\bibnamefont {Zhang}}, \bibinfo {author} {\bibfnamefont {J.-N.}\
  \bibnamefont {Zhang}},\ and\ \bibinfo {author} {\bibfnamefont
  {K.}~\bibnamefont {Kim}},\ }\bibfield  {title} {\bibinfo {title}
  {Implementing arbitrary {I}sing models with a trapped-ion quantum
  processor},\ }\href {https://doi.org/10.1103/PhysRevLett.134.050602}
  {\bibfield  {journal} {\bibinfo  {journal} {Phys. Rev. Lett.}\ }\textbf
  {\bibinfo {volume} {134}},\ \bibinfo {pages} {050602} (\bibinfo {year}
  {2025})}\BibitemShut {NoStop}%
\bibitem [{\citenamefont {D{\"u}r}\ \emph {et~al.}(2005)\citenamefont
  {D{\"u}r}, \citenamefont {Hartmann}, \citenamefont {Hein}, \citenamefont
  {Lewenstein},\ and\ \citenamefont {Briegel}}]{Dur2005}%
  \BibitemOpen
  \bibfield  {author} {\bibinfo {author} {\bibfnamefont {W.}~\bibnamefont
  {D{\"u}r}}, \bibinfo {author} {\bibfnamefont {L.}~\bibnamefont {Hartmann}},
  \bibinfo {author} {\bibfnamefont {M.}~\bibnamefont {Hein}}, \bibinfo {author}
  {\bibfnamefont {M.}~\bibnamefont {Lewenstein}},\ and\ \bibinfo {author}
  {\bibfnamefont {H.-J.}\ \bibnamefont {Briegel}},\ }\bibfield  {title}
  {\bibinfo {title} {Entanglement in spin chains and lattices with long-range
  {I}sing-type interactions},\ }\href
  {https://doi.org/10.1103/PhysRevLett.94.097203} {\bibfield  {journal}
  {\bibinfo  {journal} {Phys. Rev. Lett.}\ }\textbf {\bibinfo {volume} {94}},\
  \bibinfo {pages} {097203} (\bibinfo {year} {2005})}\BibitemShut {NoStop}%
\bibitem [{\citenamefont {Hein}\ \emph {et~al.}(2006)\citenamefont {Hein},
  \citenamefont {D{\"u}r}, \citenamefont {Eisert}, \citenamefont {Raussendorf},
  \citenamefont {Van~den Nest},\ and\ \citenamefont {Briegel}}]{Hein2006}%
  \BibitemOpen
  \bibfield  {author} {\bibinfo {author} {\bibfnamefont {M.}~\bibnamefont
  {Hein}}, \bibinfo {author} {\bibfnamefont {W.}~\bibnamefont {D{\"u}r}},
  \bibinfo {author} {\bibfnamefont {J.}~\bibnamefont {Eisert}}, \bibinfo
  {author} {\bibfnamefont {R.}~\bibnamefont {Raussendorf}}, \bibinfo {author}
  {\bibfnamefont {M.}~\bibnamefont {Van~den Nest}},\ and\ \bibinfo {author}
  {\bibfnamefont {H.-J.}\ \bibnamefont {Briegel}},\ }\bibfield  {title}
  {\bibinfo {title} {Entanglement in graph states and its applications},\ }in\
  \href {https://doi.org/10.3254/978-1-61499-018-5-115} {\emph {\bibinfo
  {booktitle} {Quantum Computers, Algorithms and Chaos}}},\ \bibinfo {series}
  {Proceedings of the International School of Physics ``Enrico Fermi''}, Vol.\
  \bibinfo {volume} {162}\ (\bibinfo  {publisher} {IOS Press},\ \bibinfo
  {address} {Amsterdam},\ \bibinfo {year} {2006})\ pp.\ \bibinfo {pages}
  {115--218}\BibitemShut {NoStop}%
\bibitem [{\citenamefont {Cie{\'s}li{\'n}ski}\ \emph
  {et~al.}(2023)\citenamefont {Cie{\'s}li{\'n}ski}, \citenamefont {K{\l}obus},
  \citenamefont {Kurzy{\'n}ski}, \citenamefont {Paterek},\ and\ \citenamefont
  {Laskowski}}]{Cieslinski2023}%
  \BibitemOpen
  \bibfield  {author} {\bibinfo {author} {\bibfnamefont {P.}~\bibnamefont
  {Cie{\'s}li{\'n}ski}}, \bibinfo {author} {\bibfnamefont {W.}~\bibnamefont
  {K{\l}obus}}, \bibinfo {author} {\bibfnamefont {P.}~\bibnamefont
  {Kurzy{\'n}ski}}, \bibinfo {author} {\bibfnamefont {T.}~\bibnamefont
  {Paterek}},\ and\ \bibinfo {author} {\bibfnamefont {W.}~\bibnamefont
  {Laskowski}},\ }\bibfield  {title} {\bibinfo {title} {The fastest generation
  of multipartite entanglement with natural interactions},\ }\href
  {https://doi.org/10.1088/1367-2630/acf953} {\bibfield  {journal} {\bibinfo
  {journal} {New J. Phys.}\ }\textbf {\bibinfo {volume} {25}},\ \bibinfo
  {pages} {093040} (\bibinfo {year} {2023})}\BibitemShut {NoStop}%
\bibitem [{\citenamefont {Hontarenko}\ and\ \citenamefont
  {{\.Z}yczkowski}(2026)}]{HontarenkoLatin2026}%
  \BibitemOpen
  \bibfield  {author} {\bibinfo {author} {\bibfnamefont {M.}~\bibnamefont
  {Hontarenko}}\ and\ \bibinfo {author} {\bibfnamefont {K.}~\bibnamefont
  {{\.Z}yczkowski}},\ }\href {https://arxiv.org/abs/2610.17170} {\bibinfo
  {title} {Fractional and quantum {L}atin squares: real solutions of quantum 36
  officers problem}} (\bibinfo {year} {2026}),\ \bibinfo {note}
  {arXiv:2610.17170}\BibitemShut {NoStop}%
\bibitem [{\citenamefont {Hartmann}\ \emph {et~al.}(2007)\citenamefont
  {Hartmann}, \citenamefont {Calsamiglia}, \citenamefont {D{\"u}r},\ and\
  \citenamefont {Briegel}}]{Hartmann2007}%
  \BibitemOpen
  \bibfield  {author} {\bibinfo {author} {\bibfnamefont {L.}~\bibnamefont
  {Hartmann}}, \bibinfo {author} {\bibfnamefont {J.}~\bibnamefont
  {Calsamiglia}}, \bibinfo {author} {\bibfnamefont {W.}~\bibnamefont
  {D{\"u}r}},\ and\ \bibinfo {author} {\bibfnamefont {H.~J.}\ \bibnamefont
  {Briegel}},\ }\bibfield  {title} {\bibinfo {title} {Weighted graph states and
  applications to spin chains, lattices and gases},\ }\href
  {https://doi.org/10.1088/0953-4075/40/9/S01} {\bibfield  {journal} {\bibinfo
  {journal} {J. Phys. B: At. Mol. Opt. Phys.}\ }\textbf {\bibinfo {volume}
  {40}},\ \bibinfo {pages} {S1} (\bibinfo {year} {2007})}\BibitemShut {NoStop}%
\bibitem [{\citenamefont {Facchi}\ \emph {et~al.}(2008)\citenamefont {Facchi},
  \citenamefont {Florio}, \citenamefont {Parisi},\ and\ \citenamefont
  {Pascazio}}]{Facchi2008}%
  \BibitemOpen
  \bibfield  {author} {\bibinfo {author} {\bibfnamefont {P.}~\bibnamefont
  {Facchi}}, \bibinfo {author} {\bibfnamefont {G.}~\bibnamefont {Florio}},
  \bibinfo {author} {\bibfnamefont {G.}~\bibnamefont {Parisi}},\ and\ \bibinfo
  {author} {\bibfnamefont {S.}~\bibnamefont {Pascazio}},\ }\bibfield  {title}
  {\bibinfo {title} {Maximally multipartite entangled states},\ }\href
  {https://doi.org/10.1103/PhysRevA.77.060304} {\bibfield  {journal} {\bibinfo
  {journal} {Phys. Rev. A}\ }\textbf {\bibinfo {volume} {77}},\ \bibinfo
  {pages} {060304(R)} (\bibinfo {year} {2008})}\BibitemShut {NoStop}%
\bibitem [{\citenamefont {Calabrese}\ and\ \citenamefont
  {Cardy}(2004)}]{Calabrese2004}%
  \BibitemOpen
  \bibfield  {author} {\bibinfo {author} {\bibfnamefont {P.}~\bibnamefont
  {Calabrese}}\ and\ \bibinfo {author} {\bibfnamefont {J.}~\bibnamefont
  {Cardy}},\ }\bibfield  {title} {\bibinfo {title} {Entanglement entropy and
  quantum field theory},\ }\href
  {https://doi.org/10.1088/1742-5468/2004/06/P06002} {\bibfield  {journal}
  {\bibinfo  {journal} {J. Stat. Mech.}\ }\textbf {\bibinfo {volume} {2004}},\
  \bibinfo {pages} {P06002} (\bibinfo {year} {2004})}\BibitemShut {NoStop}%
\bibitem [{\citenamefont {Ekert}\ \emph {et~al.}(2002)\citenamefont {Ekert},
  \citenamefont {Alves}, \citenamefont {Oi}, \citenamefont {Horodecki},
  \citenamefont {Horodecki},\ and\ \citenamefont {Kwek}}]{Ekert2002}%
  \BibitemOpen
  \bibfield  {author} {\bibinfo {author} {\bibfnamefont {A.~K.}\ \bibnamefont
  {Ekert}}, \bibinfo {author} {\bibfnamefont {C.~M.}\ \bibnamefont {Alves}},
  \bibinfo {author} {\bibfnamefont {D.~K.~L.}\ \bibnamefont {Oi}}, \bibinfo
  {author} {\bibfnamefont {M.}~\bibnamefont {Horodecki}}, \bibinfo {author}
  {\bibfnamefont {P.}~\bibnamefont {Horodecki}},\ and\ \bibinfo {author}
  {\bibfnamefont {L.~C.}\ \bibnamefont {Kwek}},\ }\bibfield  {title} {\bibinfo
  {title} {Direct estimations of linear and nonlinear functionals of a quantum
  state},\ }\href {https://doi.org/10.1103/PhysRevLett.88.217901} {\bibfield
  {journal} {\bibinfo  {journal} {Phys. Rev. Lett.}\ }\textbf {\bibinfo
  {volume} {88}},\ \bibinfo {pages} {217901} (\bibinfo {year}
  {2002})}\BibitemShut {NoStop}%
\bibitem [{\citenamefont {Brydges}\ \emph {et~al.}(2019)\citenamefont
  {Brydges}, \citenamefont {Elben}, \citenamefont {Jurcevic}, \citenamefont
  {Vermersch}, \citenamefont {Maier}, \citenamefont {Lanyon}, \citenamefont
  {Zoller}, \citenamefont {Blatt},\ and\ \citenamefont {Roos}}]{Brydges2019}%
  \BibitemOpen
  \bibfield  {author} {\bibinfo {author} {\bibfnamefont {T.}~\bibnamefont
  {Brydges}}, \bibinfo {author} {\bibfnamefont {A.}~\bibnamefont {Elben}},
  \bibinfo {author} {\bibfnamefont {P.}~\bibnamefont {Jurcevic}}, \bibinfo
  {author} {\bibfnamefont {B.}~\bibnamefont {Vermersch}}, \bibinfo {author}
  {\bibfnamefont {C.}~\bibnamefont {Maier}}, \bibinfo {author} {\bibfnamefont
  {B.~P.}\ \bibnamefont {Lanyon}}, \bibinfo {author} {\bibfnamefont
  {P.}~\bibnamefont {Zoller}}, \bibinfo {author} {\bibfnamefont
  {R.}~\bibnamefont {Blatt}},\ and\ \bibinfo {author} {\bibfnamefont {C.~F.}\
  \bibnamefont {Roos}},\ }\bibfield  {title} {\bibinfo {title} {Probing
  r{\'e}nyi entanglement entropy via randomized measurements},\ }\href
  {https://doi.org/10.1126/science.aau4963} {\bibfield  {journal} {\bibinfo
  {journal} {Science}\ }\textbf {\bibinfo {volume} {364}},\ \bibinfo {pages}
  {260} (\bibinfo {year} {2019})}\BibitemShut {NoStop}%
\bibitem [{\citenamefont {Aaronson}\ and\ \citenamefont
  {Gottesman}(2004)}]{Aaronson2004}%
  \BibitemOpen
  \bibfield  {author} {\bibinfo {author} {\bibfnamefont {S.}~\bibnamefont
  {Aaronson}}\ and\ \bibinfo {author} {\bibfnamefont {D.}~\bibnamefont
  {Gottesman}},\ }\bibfield  {title} {\bibinfo {title} {Improved simulation of
  stabilizer circuits},\ }\href {https://doi.org/10.1103/PhysRevA.70.052328}
  {\bibfield  {journal} {\bibinfo  {journal} {Phys. Rev. A}\ }\textbf {\bibinfo
  {volume} {70}},\ \bibinfo {pages} {052328} (\bibinfo {year}
  {2004})}\BibitemShut {NoStop}%
\bibitem [{\citenamefont {Garc{\'i}a}\ \emph {et~al.}(2012)\citenamefont
  {Garc{\'i}a}, \citenamefont {Markov},\ and\ \citenamefont
  {Cross}}]{Garcia2012}%
  \BibitemOpen
  \bibfield  {author} {\bibinfo {author} {\bibfnamefont {H.~J.}\ \bibnamefont
  {Garc{\'i}a}}, \bibinfo {author} {\bibfnamefont {I.~L.}\ \bibnamefont
  {Markov}},\ and\ \bibinfo {author} {\bibfnamefont {A.~W.}\ \bibnamefont
  {Cross}},\ }\href {https://arxiv.org/abs/1210.6646} {\bibinfo {title}
  {Efficient inner-product algorithm for stabilizer states}} (\bibinfo {year}
  {2012}),\ \bibinfo {note} {arXiv:1210.6646}\BibitemShut {NoStop}%
\bibitem [{\citenamefont {Paczos}\ \emph {et~al.}(2021)\citenamefont {Paczos},
  \citenamefont {Wierzbi{\'n}ski}, \citenamefont {Rajchel-Mieldzio{\'c}},
  \citenamefont {Burchardt},\ and\ \citenamefont
  {{\.Z}yczkowski}}]{Paczos2021}%
  \BibitemOpen
  \bibfield  {author} {\bibinfo {author} {\bibfnamefont {J.}~\bibnamefont
  {Paczos}}, \bibinfo {author} {\bibfnamefont {M.}~\bibnamefont
  {Wierzbi{\'n}ski}}, \bibinfo {author} {\bibfnamefont {G.}~\bibnamefont
  {Rajchel-Mieldzio{\'c}}}, \bibinfo {author} {\bibfnamefont {A.}~\bibnamefont
  {Burchardt}},\ and\ \bibinfo {author} {\bibfnamefont {K.}~\bibnamefont
  {{\.Z}yczkowski}},\ }\bibfield  {title} {\bibinfo {title} {Genuinely quantum
  solutions of the game {S}udoku and their cardinality},\ }\href
  {https://doi.org/10.1103/PhysRevA.104.042423} {\bibfield  {journal} {\bibinfo
   {journal} {Phys. Rev. A}\ }\textbf {\bibinfo {volume} {104}},\ \bibinfo
  {pages} {042423} (\bibinfo {year} {2021})},\ \Eprint
  {https://arxiv.org/abs/2106.02967} {arXiv:2106.02967} \BibitemShut {NoStop}%
\bibitem [{\citenamefont {Hostens}\ \emph {et~al.}(2005)\citenamefont
  {Hostens}, \citenamefont {Dehaene},\ and\ \citenamefont
  {De~Moor}}]{Hostens2005}%
  \BibitemOpen
  \bibfield  {author} {\bibinfo {author} {\bibfnamefont {E.}~\bibnamefont
  {Hostens}}, \bibinfo {author} {\bibfnamefont {J.}~\bibnamefont {Dehaene}},\
  and\ \bibinfo {author} {\bibfnamefont {B.}~\bibnamefont {De~Moor}},\
  }\bibfield  {title} {\bibinfo {title} {Stabilizer states and {C}lifford
  operations for systems of arbitrary dimensions and modular arithmetic},\
  }\href {https://doi.org/10.1103/PhysRevA.71.042315} {\bibfield  {journal}
  {\bibinfo  {journal} {Phys. Rev. A}\ }\textbf {\bibinfo {volume} {71}},\
  \bibinfo {pages} {042315} (\bibinfo {year} {2005})}\BibitemShut {NoStop}%
\bibitem [{\citenamefont {Mandelstam}\ and\ \citenamefont
  {Tamm}(1991)}]{MandelstamTamm1945}%
  \BibitemOpen
  \bibfield  {author} {\bibinfo {author} {\bibfnamefont {L.}~\bibnamefont
  {Mandelstam}}\ and\ \bibinfo {author} {\bibfnamefont {I.}~\bibnamefont
  {Tamm}},\ }\bibfield  {title} {\bibinfo {title} {The uncertainty relation
  between energy and time in non-relativistic quantum mechanics},\ }in\ \href
  {https://doi.org/10.1007/978-3-642-74626-0_8} {\emph {\bibinfo {booktitle}
  {Selected Papers}}},\ \bibinfo {editor} {edited by\ \bibinfo {editor}
  {\bibfnamefont {B.~M.}\ \bibnamefont {Bolotovskii}}, \bibinfo {editor}
  {\bibfnamefont {V.~Y.}\ \bibnamefont {Frenkel}},\ and\ \bibinfo {editor}
  {\bibfnamefont {R.}~\bibnamefont {Peierls}}}\ (\bibinfo  {publisher}
  {Springer},\ \bibinfo {address} {Berlin, Heidelberg},\ \bibinfo {year}
  {1991})\ pp.\ \bibinfo {pages} {115--123},\ \bibinfo {note} {originally
  published in J. Phys. (USSR) \textbf{9}, 249 (1945)}\BibitemShut {NoStop}%
\bibitem [{\citenamefont {Anandan}\ and\ \citenamefont
  {Aharonov}(1990)}]{AnandanAharonov1990}%
  \BibitemOpen
  \bibfield  {author} {\bibinfo {author} {\bibfnamefont {J.}~\bibnamefont
  {Anandan}}\ and\ \bibinfo {author} {\bibfnamefont {Y.}~\bibnamefont
  {Aharonov}},\ }\bibfield  {title} {\bibinfo {title} {Geometry of quantum
  evolution},\ }\href {https://doi.org/10.1103/PhysRevLett.65.1697} {\bibfield
  {journal} {\bibinfo  {journal} {Phys. Rev. Lett.}\ }\textbf {\bibinfo
  {volume} {65}},\ \bibinfo {pages} {1697} (\bibinfo {year}
  {1990})}\BibitemShut {NoStop}%
\bibitem [{\citenamefont {Deffner}\ and\ \citenamefont
  {Campbell}(2017)}]{Deffner2017}%
  \BibitemOpen
  \bibfield  {author} {\bibinfo {author} {\bibfnamefont {S.}~\bibnamefont
  {Deffner}}\ and\ \bibinfo {author} {\bibfnamefont {S.}~\bibnamefont
  {Campbell}},\ }\bibfield  {title} {\bibinfo {title} {Quantum speed limits:
  from {H}eisenberg's uncertainty principle to optimal quantum control},\
  }\href {https://doi.org/10.1088/1751-8121/aa86c6} {\bibfield  {journal}
  {\bibinfo  {journal} {J. Phys. A: Math. Theor.}\ }\textbf {\bibinfo {volume}
  {50}},\ \bibinfo {pages} {453001} (\bibinfo {year} {2017})}\BibitemShut
  {NoStop}%
\bibitem [{\citenamefont {Lindblad}(1976)}]{Lindblad1976}%
  \BibitemOpen
  \bibfield  {author} {\bibinfo {author} {\bibfnamefont {G.}~\bibnamefont
  {Lindblad}},\ }\bibfield  {title} {\bibinfo {title} {On the generators of
  quantum dynamical semigroups},\ }\href {https://doi.org/10.1007/BF01608499}
  {\bibfield  {journal} {\bibinfo  {journal} {Commun. Math. Phys.}\ }\textbf
  {\bibinfo {volume} {48}},\ \bibinfo {pages} {119} (\bibinfo {year}
  {1976})}\BibitemShut {NoStop}%
\bibitem [{\citenamefont {Gorini}\ \emph {et~al.}(1976)\citenamefont {Gorini},
  \citenamefont {Kossakowski},\ and\ \citenamefont {Sudarshan}}]{Gorini1976}%
  \BibitemOpen
  \bibfield  {author} {\bibinfo {author} {\bibfnamefont {V.}~\bibnamefont
  {Gorini}}, \bibinfo {author} {\bibfnamefont {A.}~\bibnamefont
  {Kossakowski}},\ and\ \bibinfo {author} {\bibfnamefont {E.~C.~G.}\
  \bibnamefont {Sudarshan}},\ }\bibfield  {title} {\bibinfo {title} {Completely
  positive dynamical semigroups of {N}-level systems},\ }\href
  {https://doi.org/10.1063/1.522979} {\bibfield  {journal} {\bibinfo  {journal}
  {J. Math. Phys.}\ }\textbf {\bibinfo {volume} {17}},\ \bibinfo {pages} {821}
  (\bibinfo {year} {1976})}\BibitemShut {NoStop}%
\bibitem [{\citenamefont {Vidal}\ and\ \citenamefont
  {Werner}(2002)}]{Vidal2002}%
  \BibitemOpen
  \bibfield  {author} {\bibinfo {author} {\bibfnamefont {G.}~\bibnamefont
  {Vidal}}\ and\ \bibinfo {author} {\bibfnamefont {R.~F.}\ \bibnamefont
  {Werner}},\ }\bibfield  {title} {\bibinfo {title} {Computable measure of
  entanglement},\ }\href {https://doi.org/10.1103/PhysRevA.65.032314}
  {\bibfield  {journal} {\bibinfo  {journal} {Phys. Rev. A}\ }\textbf {\bibinfo
  {volume} {65}},\ \bibinfo {pages} {032314} (\bibinfo {year}
  {2002})}\BibitemShut {NoStop}%
\bibitem [{\citenamefont {Moses}\ \emph {et~al.}(2023)\citenamefont {Moses},
  \citenamefont {Baldwin}, \citenamefont {Allman}, \citenamefont {Ancona},
  \citenamefont {Ascarrunz}, \citenamefont {Barnes}, \citenamefont
  {Bartolotta}, \citenamefont {Bjork}, \citenamefont {Blanchard}, \citenamefont
  {Bohn}, \citenamefont {Bohnet}, \citenamefont {Brown} \emph
  {et~al.}}]{Moses2023}%
  \BibitemOpen
  \bibfield  {author} {\bibinfo {author} {\bibfnamefont {S.~A.}\ \bibnamefont
  {Moses}}, \bibinfo {author} {\bibfnamefont {C.~H.}\ \bibnamefont {Baldwin}},
  \bibinfo {author} {\bibfnamefont {M.~S.}\ \bibnamefont {Allman}}, \bibinfo
  {author} {\bibfnamefont {R.}~\bibnamefont {Ancona}}, \bibinfo {author}
  {\bibfnamefont {L.}~\bibnamefont {Ascarrunz}}, \bibinfo {author}
  {\bibfnamefont {C.}~\bibnamefont {Barnes}}, \bibinfo {author} {\bibfnamefont
  {J.}~\bibnamefont {Bartolotta}}, \bibinfo {author} {\bibfnamefont
  {B.}~\bibnamefont {Bjork}}, \bibinfo {author} {\bibfnamefont
  {P.}~\bibnamefont {Blanchard}}, \bibinfo {author} {\bibfnamefont
  {M.}~\bibnamefont {Bohn}}, \bibinfo {author} {\bibfnamefont {J.~G.}\
  \bibnamefont {Bohnet}}, \bibinfo {author} {\bibfnamefont {N.~C.}\
  \bibnamefont {Brown}}, \emph {et~al.},\ }\bibfield  {title} {\bibinfo {title}
  {A race-track trapped-ion quantum processor},\ }\href
  {https://doi.org/10.1103/PhysRevX.13.041052} {\bibfield  {journal} {\bibinfo
  {journal} {Phys. Rev. X}\ }\textbf {\bibinfo {volume} {13}},\ \bibinfo
  {pages} {041052} (\bibinfo {year} {2023})}\BibitemShut {NoStop}%
\bibitem [{\citenamefont {Figgatt}\ \emph {et~al.}(2019)\citenamefont
  {Figgatt}, \citenamefont {Ostrander}, \citenamefont {Linke}, \citenamefont
  {Landsman}, \citenamefont {Zhu}, \citenamefont {Maslov},\ and\ \citenamefont
  {Monroe}}]{Figgatt2019}%
  \BibitemOpen
  \bibfield  {author} {\bibinfo {author} {\bibfnamefont {C.}~\bibnamefont
  {Figgatt}}, \bibinfo {author} {\bibfnamefont {A.}~\bibnamefont {Ostrander}},
  \bibinfo {author} {\bibfnamefont {N.~M.}\ \bibnamefont {Linke}}, \bibinfo
  {author} {\bibfnamefont {K.~A.}\ \bibnamefont {Landsman}}, \bibinfo {author}
  {\bibfnamefont {D.}~\bibnamefont {Zhu}}, \bibinfo {author} {\bibfnamefont
  {D.}~\bibnamefont {Maslov}},\ and\ \bibinfo {author} {\bibfnamefont
  {C.}~\bibnamefont {Monroe}},\ }\bibfield  {title} {\bibinfo {title} {Parallel
  entangling operations on a universal ion-trap quantum computer},\ }\href
  {https://doi.org/10.1038/s41586-019-1427-5} {\bibfield  {journal} {\bibinfo
  {journal} {Nature}\ }\textbf {\bibinfo {volume} {572}},\ \bibinfo {pages}
  {368} (\bibinfo {year} {2019})}\BibitemShut {NoStop}%
\bibitem [{\citenamefont {Grzesiak}\ \emph {et~al.}(2020)\citenamefont
  {Grzesiak}, \citenamefont {Bl{\"u}mel}, \citenamefont {Wright}, \citenamefont
  {Beck}, \citenamefont {Pisenti}, \citenamefont {Li}, \citenamefont {Chaplin},
  \citenamefont {Amini}, \citenamefont {Debnath}, \citenamefont {Chen},\ and\
  \citenamefont {Nam}}]{Grzesiak2020}%
  \BibitemOpen
  \bibfield  {author} {\bibinfo {author} {\bibfnamefont {N.}~\bibnamefont
  {Grzesiak}}, \bibinfo {author} {\bibfnamefont {R.}~\bibnamefont
  {Bl{\"u}mel}}, \bibinfo {author} {\bibfnamefont {K.}~\bibnamefont {Wright}},
  \bibinfo {author} {\bibfnamefont {K.~M.}\ \bibnamefont {Beck}}, \bibinfo
  {author} {\bibfnamefont {N.~C.}\ \bibnamefont {Pisenti}}, \bibinfo {author}
  {\bibfnamefont {M.}~\bibnamefont {Li}}, \bibinfo {author} {\bibfnamefont
  {V.}~\bibnamefont {Chaplin}}, \bibinfo {author} {\bibfnamefont {J.~M.}\
  \bibnamefont {Amini}}, \bibinfo {author} {\bibfnamefont {S.}~\bibnamefont
  {Debnath}}, \bibinfo {author} {\bibfnamefont {J.-S.}\ \bibnamefont {Chen}},\
  and\ \bibinfo {author} {\bibfnamefont {Y.}~\bibnamefont {Nam}},\ }\bibfield
  {title} {\bibinfo {title} {Efficient arbitrary simultaneously entangling
  gates on a trapped-ion quantum computer},\ }\href
  {https://doi.org/10.1038/s41467-020-16790-9} {\bibfield  {journal} {\bibinfo
  {journal} {Nat. Commun.}\ }\textbf {\bibinfo {volume} {11}},\ \bibinfo
  {pages} {2963} (\bibinfo {year} {2020})}\BibitemShut {NoStop}%
\bibitem [{\citenamefont {Evered}\ \emph {et~al.}(2023)\citenamefont {Evered},
  \citenamefont {Bluvstein}, \citenamefont {Kalinowski}, \citenamefont {Ebadi},
  \citenamefont {Manovitz}, \citenamefont {Zhou}, \citenamefont {Li},
  \citenamefont {Geim}, \citenamefont {Wang}, \citenamefont {Maskara},
  \citenamefont {Levine}, \citenamefont {Semeghini} \emph
  {et~al.}}]{Evered2023}%
  \BibitemOpen
  \bibfield  {author} {\bibinfo {author} {\bibfnamefont {S.~J.}\ \bibnamefont
  {Evered}}, \bibinfo {author} {\bibfnamefont {D.}~\bibnamefont {Bluvstein}},
  \bibinfo {author} {\bibfnamefont {M.}~\bibnamefont {Kalinowski}}, \bibinfo
  {author} {\bibfnamefont {S.}~\bibnamefont {Ebadi}}, \bibinfo {author}
  {\bibfnamefont {T.}~\bibnamefont {Manovitz}}, \bibinfo {author}
  {\bibfnamefont {H.}~\bibnamefont {Zhou}}, \bibinfo {author} {\bibfnamefont
  {S.~H.}\ \bibnamefont {Li}}, \bibinfo {author} {\bibfnamefont {A.~A.}\
  \bibnamefont {Geim}}, \bibinfo {author} {\bibfnamefont {T.~T.}\ \bibnamefont
  {Wang}}, \bibinfo {author} {\bibfnamefont {N.}~\bibnamefont {Maskara}},
  \bibinfo {author} {\bibfnamefont {H.}~\bibnamefont {Levine}}, \bibinfo
  {author} {\bibfnamefont {G.}~\bibnamefont {Semeghini}}, \emph {et~al.},\
  }\bibfield  {title} {\bibinfo {title} {High-fidelity parallel entangling
  gates on a neutral-atom quantum computer},\ }\href
  {https://doi.org/10.1038/s41586-023-06481-y} {\bibfield  {journal} {\bibinfo
  {journal} {Nature}\ }\textbf {\bibinfo {volume} {622}},\ \bibinfo {pages}
  {268} (\bibinfo {year} {2023})}\BibitemShut {NoStop}%
\bibitem [{\citenamefont {Kyprianidis}\ \emph {et~al.}(2024)\citenamefont
  {Kyprianidis}, \citenamefont {Rasmusson},\ and\ \citenamefont
  {Richerme}}]{Kyprianidis2024}%
  \BibitemOpen
  \bibfield  {author} {\bibinfo {author} {\bibfnamefont {A.}~\bibnamefont
  {Kyprianidis}}, \bibinfo {author} {\bibfnamefont {A.~J.}\ \bibnamefont
  {Rasmusson}},\ and\ \bibinfo {author} {\bibfnamefont {P.}~\bibnamefont
  {Richerme}},\ }\bibfield  {title} {\bibinfo {title} {Interaction graph
  engineering in trapped-ion quantum simulators with global drives},\ }\href
  {https://doi.org/10.1088/1367-2630/ad264d} {\bibfield  {journal} {\bibinfo
  {journal} {New J. Phys.}\ }\textbf {\bibinfo {volume} {26}},\ \bibinfo
  {pages} {023033} (\bibinfo {year} {2024})}\BibitemShut {NoStop}%
\bibitem [{\citenamefont {Peleg}\ \emph {et~al.}(2026)\citenamefont {Peleg},
  \citenamefont {Schwerdt}, \citenamefont {Nemirovsky}, \citenamefont
  {Shapira}, \citenamefont {Akerman}, \citenamefont {Stern}, \citenamefont
  {Ben~Kish},\ and\ \citenamefont {Ozeri}}]{Peleg2026}%
  \BibitemOpen
  \bibfield  {author} {\bibinfo {author} {\bibfnamefont {L.}~\bibnamefont
  {Peleg}}, \bibinfo {author} {\bibfnamefont {D.}~\bibnamefont {Schwerdt}},
  \bibinfo {author} {\bibfnamefont {J.}~\bibnamefont {Nemirovsky}}, \bibinfo
  {author} {\bibfnamefont {Y.}~\bibnamefont {Shapira}}, \bibinfo {author}
  {\bibfnamefont {N.}~\bibnamefont {Akerman}}, \bibinfo {author} {\bibfnamefont
  {A.}~\bibnamefont {Stern}}, \bibinfo {author} {\bibfnamefont
  {A.}~\bibnamefont {Ben~Kish}},\ and\ \bibinfo {author} {\bibfnamefont
  {R.}~\bibnamefont {Ozeri}},\ }\bibfield  {title} {\bibinfo {title} {Fast
  design and scaling of multiqubit gates in large-scale trapped-ion quantum
  computers},\ }\href {https://doi.org/10.1103/r78y-3q89} {\bibfield  {journal}
  {\bibinfo  {journal} {PRX Quantum}\ }\textbf {\bibinfo {volume} {7}},\
  \bibinfo {pages} {033021} (\bibinfo {year} {2026})}\BibitemShut {NoStop}%
\bibitem [{\citenamefont {Rather}\ \emph {et~al.}(2022)\citenamefont {Rather},
  \citenamefont {Burchardt}, \citenamefont {Bruzda}, \citenamefont
  {Rajchel-Mieldzio{\'c}}, \citenamefont {Lakshminarayan},\ and\ \citenamefont
  {{\.Z}yczkowski}}]{Rather2022}%
  \BibitemOpen
  \bibfield  {author} {\bibinfo {author} {\bibfnamefont {S.~A.}\ \bibnamefont
  {Rather}}, \bibinfo {author} {\bibfnamefont {A.}~\bibnamefont {Burchardt}},
  \bibinfo {author} {\bibfnamefont {W.}~\bibnamefont {Bruzda}}, \bibinfo
  {author} {\bibfnamefont {G.}~\bibnamefont {Rajchel-Mieldzio{\'c}}}, \bibinfo
  {author} {\bibfnamefont {A.}~\bibnamefont {Lakshminarayan}},\ and\ \bibinfo
  {author} {\bibfnamefont {K.}~\bibnamefont {{\.Z}yczkowski}},\ }\bibfield
  {title} {\bibinfo {title} {Thirty-six entangled officers of {E}uler: Quantum
  solution to a classically impossible problem},\ }\href
  {https://doi.org/10.1103/PhysRevLett.128.080507} {\bibfield  {journal}
  {\bibinfo  {journal} {Phys. Rev. Lett.}\ }\textbf {\bibinfo {volume} {128}},\
  \bibinfo {pages} {080507} (\bibinfo {year} {2022})}\BibitemShut {NoStop}%
\bibitem [{\citenamefont {Bruzda}\ and\ \citenamefont
  {{\.Z}yczkowski}(2024)}]{Bruzda2024}%
  \BibitemOpen
  \bibfield  {author} {\bibinfo {author} {\bibfnamefont {W.}~\bibnamefont
  {Bruzda}}\ and\ \bibinfo {author} {\bibfnamefont {K.}~\bibnamefont
  {{\.Z}yczkowski}},\ }\bibfield  {title} {\bibinfo {title} {Two-unitary
  complex {H}adamard matrices of order 36},\ }\href
  {https://doi.org/10.1515/spma-2024-0010} {\bibfield  {journal} {\bibinfo
  {journal} {Special Matrices}\ }\textbf {\bibinfo {volume} {12}},\ \bibinfo
  {pages} {20240010} (\bibinfo {year} {2024})}\BibitemShut {NoStop}%
\bibitem [{\citenamefont {Bruzda}(2026)}]{BruzdaGeomagic2026}%
  \BibitemOpen
  \bibfield  {author} {\bibinfo {author} {\bibfnamefont {W.}~\bibnamefont
  {Bruzda}},\ }\href {https://arxiv.org/abs/2610.16882} {\bibinfo {title}
  {Geomagic square representation of a real {AME}(4,6) state}} (\bibinfo {year}
  {2026}),\ \bibinfo {note} {arXiv:2610.16882}\BibitemShut {NoStop}%
\bibitem [{\citenamefont {Pandya}(2026)}]{Pandya2026}%
  \BibitemOpen
  \bibfield  {author} {\bibinfo {author} {\bibfnamefont {P.}~\bibnamefont
  {Pandya}},\ }\href {https://arxiv.org/abs/2610.16673} {\bibinfo {title}
  {Euler's thirty-six officers need not all be entangled: real {AME}(4,6),
  families and classification}} (\bibinfo {year} {2026}),\ \bibinfo {note}
  {arXiv:2610.16673}\BibitemShut {NoStop}%
\bibitem [{\citenamefont {P{\l}odzie{\'n}}\ \emph {et~al.}(2026)\citenamefont
  {P{\l}odzie{\'n}}, \citenamefont {Rajchel-Mieldzio{\'c}},\ and\ \citenamefont
  {{\.Z}yczkowski}}]{AMEcode}%
  \BibitemOpen
  \bibfield  {author} {\bibinfo {author} {\bibfnamefont {M.}~\bibnamefont
  {P{\l}odzie{\'n}}}, \bibinfo {author} {\bibfnamefont {G.}~\bibnamefont
  {Rajchel-Mieldzio{\'c}}},\ and\ \bibinfo {author} {\bibfnamefont
  {K.}~\bibnamefont {{\.Z}yczkowski}},\ }\href
  {https://github.com/MarcinPlodzien/analog-preparation-AME-states} {\bibinfo
  {title} {Analog preparation of absolutely maximally entangled states: code
  and data}},\ \bibinfo {howpublished}
  {\url{https://github.com/MarcinPlodzien/analog-preparation-AME-states}}
  (\bibinfo {year} {2026})\BibitemShut {NoStop}%
\bibitem [{\citenamefont {Islam}\ \emph {et~al.}(2015)\citenamefont {Islam},
  \citenamefont {Ma}, \citenamefont {Preiss}, \citenamefont {Tai},
  \citenamefont {Lukin}, \citenamefont {Rispoli},\ and\ \citenamefont
  {Greiner}}]{Islam2015}%
  \BibitemOpen
  \bibfield  {author} {\bibinfo {author} {\bibfnamefont {R.}~\bibnamefont
  {Islam}}, \bibinfo {author} {\bibfnamefont {R.}~\bibnamefont {Ma}}, \bibinfo
  {author} {\bibfnamefont {P.~M.}\ \bibnamefont {Preiss}}, \bibinfo {author}
  {\bibfnamefont {M.~E.}\ \bibnamefont {Tai}}, \bibinfo {author} {\bibfnamefont
  {A.}~\bibnamefont {Lukin}}, \bibinfo {author} {\bibfnamefont
  {M.}~\bibnamefont {Rispoli}},\ and\ \bibinfo {author} {\bibfnamefont
  {M.}~\bibnamefont {Greiner}},\ }\bibfield  {title} {\bibinfo {title}
  {Measuring entanglement entropy in a quantum many-body system},\ }\href
  {https://doi.org/10.1038/nature15750} {\bibfield  {journal} {\bibinfo
  {journal} {Nature}\ }\textbf {\bibinfo {volume} {528}},\ \bibinfo {pages}
  {77} (\bibinfo {year} {2015})}\BibitemShut {NoStop}%
\bibitem [{\citenamefont {Bradbury}\ \emph {et~al.}(2018)\citenamefont
  {Bradbury}, \citenamefont {Frostig}, \citenamefont {Hawkins}, \citenamefont
  {Johnson}, \citenamefont {Leary}, \citenamefont {Maclaurin}, \citenamefont
  {Necula}, \citenamefont {Paszke}, \citenamefont {VanderPlas}, \citenamefont
  {Wanderman-Milne},\ and\ \citenamefont {Zhang}}]{Bradbury2018}%
  \BibitemOpen
  \bibfield  {author} {\bibinfo {author} {\bibfnamefont {J.}~\bibnamefont
  {Bradbury}}, \bibinfo {author} {\bibfnamefont {R.}~\bibnamefont {Frostig}},
  \bibinfo {author} {\bibfnamefont {P.}~\bibnamefont {Hawkins}}, \bibinfo
  {author} {\bibfnamefont {M.~J.}\ \bibnamefont {Johnson}}, \bibinfo {author}
  {\bibfnamefont {C.}~\bibnamefont {Leary}}, \bibinfo {author} {\bibfnamefont
  {D.}~\bibnamefont {Maclaurin}}, \bibinfo {author} {\bibfnamefont
  {G.}~\bibnamefont {Necula}}, \bibinfo {author} {\bibfnamefont
  {A.}~\bibnamefont {Paszke}}, \bibinfo {author} {\bibfnamefont
  {J.}~\bibnamefont {VanderPlas}}, \bibinfo {author} {\bibfnamefont
  {S.}~\bibnamefont {Wanderman-Milne}},\ and\ \bibinfo {author} {\bibfnamefont
  {Q.}~\bibnamefont {Zhang}},\ }\href {https://github.com/jax-ml/jax} {\bibinfo
  {title} {{JAX}: composable transformations of {P}ython+{N}um{P}y programs}},\
  \bibinfo {howpublished} {\url{https://github.com/jax-ml/jax}} (\bibinfo
  {year} {2018})\BibitemShut {NoStop}%
\bibitem [{\citenamefont {Kingma}\ and\ \citenamefont {Ba}(2015)}]{Kingma2015}%
  \BibitemOpen
  \bibfield  {author} {\bibinfo {author} {\bibfnamefont {D.~P.}\ \bibnamefont
  {Kingma}}\ and\ \bibinfo {author} {\bibfnamefont {J.}~\bibnamefont {Ba}},\
  }\bibfield  {title} {\bibinfo {title} {Adam: A method for stochastic
  optimization},\ }in\ \href {https://arxiv.org/abs/1412.6980} {\emph {\bibinfo
  {booktitle} {International Conference on Learning Representations (ICLR)}}}\
  (\bibinfo {year} {2015})\ \bibinfo {note} {arXiv:1412.6980}\BibitemShut
  {NoStop}%
\bibitem [{\citenamefont {{DeepMind}}\ \emph {et~al.}(2020)\citenamefont
  {{DeepMind}}, \citenamefont {Babuschkin}, \citenamefont {Baumli},
  \citenamefont {Bell}, \citenamefont {Bhupatiraju}, \citenamefont {Bruce},
  \citenamefont {Buchlovsky}, \citenamefont {Budden}, \citenamefont {Cai},
  \citenamefont {Clark}, \citenamefont {Danihelka} \emph
  {et~al.}}]{DeepMind2020}%
  \BibitemOpen
  \bibfield  {author} {\bibinfo {author} {\bibnamefont {{DeepMind}}}, \bibinfo
  {author} {\bibfnamefont {I.}~\bibnamefont {Babuschkin}}, \bibinfo {author}
  {\bibfnamefont {K.}~\bibnamefont {Baumli}}, \bibinfo {author} {\bibfnamefont
  {A.}~\bibnamefont {Bell}}, \bibinfo {author} {\bibfnamefont {S.}~\bibnamefont
  {Bhupatiraju}}, \bibinfo {author} {\bibfnamefont {J.}~\bibnamefont {Bruce}},
  \bibinfo {author} {\bibfnamefont {P.}~\bibnamefont {Buchlovsky}}, \bibinfo
  {author} {\bibfnamefont {D.}~\bibnamefont {Budden}}, \bibinfo {author}
  {\bibfnamefont {T.}~\bibnamefont {Cai}}, \bibinfo {author} {\bibfnamefont
  {A.}~\bibnamefont {Clark}}, \bibinfo {author} {\bibfnamefont
  {I.}~\bibnamefont {Danihelka}}, \emph {et~al.},\ }\href
  {http://github.com/google-deepmind} {\bibinfo {title} {The {DeepMind} {JAX}
  ecosystem}} (\bibinfo {year} {2020})\BibitemShut {NoStop}%
\bibitem [{\citenamefont {Broyden}(1970)}]{Broyden1970}%
  \BibitemOpen
  \bibfield  {author} {\bibinfo {author} {\bibfnamefont {C.~G.}\ \bibnamefont
  {Broyden}},\ }\bibfield  {title} {\bibinfo {title} {The convergence of a
  class of double-rank minimization algorithms 1. {G}eneral considerations},\
  }\href {https://doi.org/10.1093/imamat/6.1.76} {\bibfield  {journal}
  {\bibinfo  {journal} {IMA J. Appl. Math.}\ }\textbf {\bibinfo {volume} {6}},\
  \bibinfo {pages} {76} (\bibinfo {year} {1970})}\BibitemShut {NoStop}%
\bibitem [{\citenamefont {Fletcher}(1970)}]{Fletcher1970}%
  \BibitemOpen
  \bibfield  {author} {\bibinfo {author} {\bibfnamefont {R.}~\bibnamefont
  {Fletcher}},\ }\bibfield  {title} {\bibinfo {title} {A new approach to
  variable metric algorithms},\ }\href
  {https://doi.org/10.1093/comjnl/13.3.317} {\bibfield  {journal} {\bibinfo
  {journal} {Comput. J.}\ }\textbf {\bibinfo {volume} {13}},\ \bibinfo {pages}
  {317} (\bibinfo {year} {1970})}\BibitemShut {NoStop}%
\bibitem [{\citenamefont {Goldfarb}(1970)}]{Goldfarb1970}%
  \BibitemOpen
  \bibfield  {author} {\bibinfo {author} {\bibfnamefont {D.}~\bibnamefont
  {Goldfarb}},\ }\bibfield  {title} {\bibinfo {title} {A family of
  variable-metric methods derived by variational means},\ }\href
  {https://doi.org/10.1090/S0025-5718-1970-0258249-6} {\bibfield  {journal}
  {\bibinfo  {journal} {Math. Comp.}\ }\textbf {\bibinfo {volume} {24}},\
  \bibinfo {pages} {23} (\bibinfo {year} {1970})}\BibitemShut {NoStop}%
\bibitem [{\citenamefont {Shanno}(1970)}]{Shanno1970}%
  \BibitemOpen
  \bibfield  {author} {\bibinfo {author} {\bibfnamefont {D.~F.}\ \bibnamefont
  {Shanno}},\ }\bibfield  {title} {\bibinfo {title} {Conditioning of
  quasi-{N}ewton methods for function minimization},\ }\href
  {https://doi.org/10.1090/S0025-5718-1970-0274029-X} {\bibfield  {journal}
  {\bibinfo  {journal} {Math. Comp.}\ }\textbf {\bibinfo {volume} {24}},\
  \bibinfo {pages} {647} (\bibinfo {year} {1970})}\BibitemShut {NoStop}%
\bibitem [{\citenamefont {Virtanen}\ \emph {et~al.}(2020)\citenamefont
  {Virtanen}, \citenamefont {Gommers}, \citenamefont {Oliphant}, \citenamefont
  {Haberland}, \citenamefont {Reddy}, \citenamefont {Cournapeau}, \citenamefont
  {Burovski}, \citenamefont {Peterson}, \citenamefont {Weckesser},
  \citenamefont {Bright} \emph {et~al.}}]{Virtanen2020}%
  \BibitemOpen
  \bibfield  {author} {\bibinfo {author} {\bibfnamefont {P.}~\bibnamefont
  {Virtanen}}, \bibinfo {author} {\bibfnamefont {R.}~\bibnamefont {Gommers}},
  \bibinfo {author} {\bibfnamefont {T.~E.}\ \bibnamefont {Oliphant}}, \bibinfo
  {author} {\bibfnamefont {M.}~\bibnamefont {Haberland}}, \bibinfo {author}
  {\bibfnamefont {T.}~\bibnamefont {Reddy}}, \bibinfo {author} {\bibfnamefont
  {D.}~\bibnamefont {Cournapeau}}, \bibinfo {author} {\bibfnamefont
  {E.}~\bibnamefont {Burovski}}, \bibinfo {author} {\bibfnamefont
  {P.}~\bibnamefont {Peterson}}, \bibinfo {author} {\bibfnamefont
  {W.}~\bibnamefont {Weckesser}}, \bibinfo {author} {\bibfnamefont
  {J.}~\bibnamefont {Bright}}, \emph {et~al.},\ }\bibfield  {title} {\bibinfo
  {title} {{SciPy} 1.0: fundamental algorithms for scientific computing in
  {Python}},\ }\href {https://doi.org/10.1038/s41592-019-0686-2} {\bibfield
  {journal} {\bibinfo  {journal} {Nat. Methods}\ }\textbf {\bibinfo {volume}
  {17}},\ \bibinfo {pages} {261} (\bibinfo {year} {2020})}\BibitemShut
  {NoStop}%
\bibitem [{\citenamefont {Liu}\ and\ \citenamefont {Nocedal}(1989)}]{Liu1989}%
  \BibitemOpen
  \bibfield  {author} {\bibinfo {author} {\bibfnamefont {D.~C.}\ \bibnamefont
  {Liu}}\ and\ \bibinfo {author} {\bibfnamefont {J.}~\bibnamefont {Nocedal}},\
  }\bibfield  {title} {\bibinfo {title} {On the limited memory {BFGS} method
  for large scale optimization},\ }\href {https://doi.org/10.1007/BF01589116}
  {\bibfield  {journal} {\bibinfo  {journal} {Math. Program.}\ }\textbf
  {\bibinfo {volume} {45}},\ \bibinfo {pages} {503} (\bibinfo {year}
  {1989})}\BibitemShut {NoStop}%
\bibitem [{\citenamefont {Byrd}\ \emph {et~al.}(1995)\citenamefont {Byrd},
  \citenamefont {Lu}, \citenamefont {Nocedal},\ and\ \citenamefont
  {Zhu}}]{Byrd1995}%
  \BibitemOpen
  \bibfield  {author} {\bibinfo {author} {\bibfnamefont {R.~H.}\ \bibnamefont
  {Byrd}}, \bibinfo {author} {\bibfnamefont {P.}~\bibnamefont {Lu}}, \bibinfo
  {author} {\bibfnamefont {J.}~\bibnamefont {Nocedal}},\ and\ \bibinfo {author}
  {\bibfnamefont {C.}~\bibnamefont {Zhu}},\ }\bibfield  {title} {\bibinfo
  {title} {A limited memory algorithm for bound constrained optimization},\
  }\href {https://doi.org/10.1137/0916069} {\bibfield  {journal} {\bibinfo
  {journal} {SIAM J. Sci. Comput.}\ }\textbf {\bibinfo {volume} {16}},\
  \bibinfo {pages} {1190} (\bibinfo {year} {1995})}\BibitemShut {NoStop}%
\bibitem [{\citenamefont {Levenberg}(1944)}]{Levenberg1944}%
  \BibitemOpen
  \bibfield  {author} {\bibinfo {author} {\bibfnamefont {K.}~\bibnamefont
  {Levenberg}},\ }\bibfield  {title} {\bibinfo {title} {A method for the
  solution of certain non-linear problems in least squares},\ }\href
  {https://doi.org/10.1090/qam/10666} {\bibfield  {journal} {\bibinfo
  {journal} {Q. Appl. Math.}\ }\textbf {\bibinfo {volume} {2}},\ \bibinfo
  {pages} {164} (\bibinfo {year} {1944})}\BibitemShut {NoStop}%
\bibitem [{\citenamefont {Marquardt}(1963)}]{Marquardt1963}%
  \BibitemOpen
  \bibfield  {author} {\bibinfo {author} {\bibfnamefont {D.~W.}\ \bibnamefont
  {Marquardt}},\ }\bibfield  {title} {\bibinfo {title} {An algorithm for
  least-squares estimation of nonlinear parameters},\ }\href
  {https://doi.org/10.1137/0111030} {\bibfield  {journal} {\bibinfo  {journal}
  {J. Soc. Ind. Appl. Math.}\ }\textbf {\bibinfo {volume} {11}},\ \bibinfo
  {pages} {431} (\bibinfo {year} {1963})}\BibitemShut {NoStop}%
\bibitem [{\citenamefont {Reck}\ \emph {et~al.}(1994)\citenamefont {Reck},
  \citenamefont {Zeilinger}, \citenamefont {Bernstein},\ and\ \citenamefont
  {Bertani}}]{Reck1994}%
  \BibitemOpen
  \bibfield  {author} {\bibinfo {author} {\bibfnamefont {M.}~\bibnamefont
  {Reck}}, \bibinfo {author} {\bibfnamefont {A.}~\bibnamefont {Zeilinger}},
  \bibinfo {author} {\bibfnamefont {H.~J.}\ \bibnamefont {Bernstein}},\ and\
  \bibinfo {author} {\bibfnamefont {P.}~\bibnamefont {Bertani}},\ }\bibfield
  {title} {\bibinfo {title} {Experimental realization of any discrete unitary
  operator},\ }\href {https://doi.org/10.1103/PhysRevLett.73.58} {\bibfield
  {journal} {\bibinfo  {journal} {Phys. Rev. Lett.}\ }\textbf {\bibinfo
  {volume} {73}},\ \bibinfo {pages} {58} (\bibinfo {year} {1994})}\BibitemShut
  {NoStop}%
\bibitem [{\citenamefont {Nielsen}\ and\ \citenamefont
  {Chuang}(2010)}]{NielsenChuang2010}%
  \BibitemOpen
  \bibfield  {author} {\bibinfo {author} {\bibfnamefont {M.~A.}\ \bibnamefont
  {Nielsen}}\ and\ \bibinfo {author} {\bibfnamefont {I.~L.}\ \bibnamefont
  {Chuang}},\ }\href {https://doi.org/10.1017/CBO9780511976667} {\emph
  {\bibinfo {title} {Quantum Computation and Quantum Information: 10th
  Anniversary Edition}}}\ (\bibinfo  {publisher} {Cambridge University Press},\
  \bibinfo {address} {Cambridge},\ \bibinfo {year} {2010})\BibitemShut
  {NoStop}%
\bibitem [{\citenamefont {Musto}\ and\ \citenamefont
  {Vicary}(2016)}]{Musto2016}%
  \BibitemOpen
  \bibfield  {author} {\bibinfo {author} {\bibfnamefont {B.}~\bibnamefont
  {Musto}}\ and\ \bibinfo {author} {\bibfnamefont {J.}~\bibnamefont {Vicary}},\
  }\bibfield  {title} {\bibinfo {title} {Quantum {L}atin squares and unitary
  error bases},\ }\href {https://doi.org/10.26421/QIC16.15-16-4} {\bibfield
  {journal} {\bibinfo  {journal} {Quantum Inf. Comput.}\ }\textbf {\bibinfo
  {volume} {16}},\ \bibinfo {pages} {1318} (\bibinfo {year}
  {2016})}\BibitemShut {NoStop}%
\bibitem [{\citenamefont {Goyeneche}\ \emph {et~al.}(2018)\citenamefont
  {Goyeneche}, \citenamefont {Raissi}, \citenamefont {Di~Martino},\ and\
  \citenamefont {{\.Z}yczkowski}}]{Goyeneche2018}%
  \BibitemOpen
  \bibfield  {author} {\bibinfo {author} {\bibfnamefont {D.}~\bibnamefont
  {Goyeneche}}, \bibinfo {author} {\bibfnamefont {Z.}~\bibnamefont {Raissi}},
  \bibinfo {author} {\bibfnamefont {S.}~\bibnamefont {Di~Martino}},\ and\
  \bibinfo {author} {\bibfnamefont {K.}~\bibnamefont {{\.Z}yczkowski}},\
  }\bibfield  {title} {\bibinfo {title} {Entanglement and quantum combinatorial
  designs},\ }\href {https://doi.org/10.1103/PhysRevA.97.062326} {\bibfield
  {journal} {\bibinfo  {journal} {Phys. Rev. A}\ }\textbf {\bibinfo {volume}
  {97}},\ \bibinfo {pages} {062326} (\bibinfo {year} {2018})}\BibitemShut
  {NoStop}%
\end{thebibliography}%

\end{document}